\documentclass[%
preprint,
 amsmath,amssymb,
 aps,showkeys
]{revtex4-2}

\usepackage{graphicx}
\usepackage{dcolumn}
\usepackage{bm}

\usepackage{graphicx}             
\usepackage{array}
\usepackage{xmpmulti}
\usepackage{amsmath,latexsym,amssymb,fixmath,tipa}
\usepackage{mathrsfs}
\usepackage{float}
\usepackage{rotating}
\usepackage{xcolor}
\usepackage{soul}
\usepackage{transparent}
\usepackage{ragged2e}
\usepackage{bigints}

\usepackage{xcolor}
\usepackage{tcolorbox}

\usepackage{caption,setspace}
\def\mixedindices#1#2{\mathstrut_{#1}^{#2}}

\def\T#1#2{\mathbf{T} \mixedindices{#1}{#2}}

\def\i{{\rm i}}

\usepackage{amsthm}

\newtheorem{remark}{Remark}

\renewcommand\thesection{\arabic{section}}
\renewcommand\thesubsection{\thesection.\arabic{subsection}}
\renewcommand\thesubsubsection{\thesubsection.\arabic{subsubsection}}

\def\andname{and}

\def\T#1#2{\mathbf{T} \mixedindices{#1}{#2}}
\def\E{\mathbf{T_{0,0}}}
\def\F{\mathbf{T_{0,1}}}
\def\G{\mathbf{T_{1,0}}}
\def\H{\mathbf{T_{0,2}}}
\def\J{\mathbf{T_{2,0}}}
\def\K{\mathbf{T_{1,1}}}
\def\L{\mathbf{T_{0,3}}}
\def\M{\mathbf{T_{1,2}}}
\def\N{\mathbf{T_{2,1}}}
\def\W{\mathbf{T_{3,0}}}
\def\Q{\mathbf{T_{0,4}}}
\def\R{\mathbf{T_{1,3}}}
\def\S{\mathbf{T_{2,2}}}
\def\T{\mathbf{T_{3,1}}}
\def\U{\mathbf{T_{4,0}}}
\def\Tone{\mathbf{T_{5,0}}}
\def\Ttwo{\mathbf{T_{4,1}}}
\def\Tthree{\mathbf{T_{3,2}}}
\def\Tfour{\mathbf{T_{2,3}}}
\def\Tfive{\mathbf{T_{1,4}}}
\def\Tsix{\mathbf{T_{0,5}}}
\def\i{{\rm i}}

\def\Vbold#1#2{\mathbf{V} \mixedindices{#1}{#2}}
\def\Qbold#1#2{\mathbf{Q} \mixedindices{#1}{#2}}

\begin{document}

\preprint{}

\title{Riemann-Hilbert problems of a non-local reverse-time  AKNS system of six-order and dynamical behaviours of $N$-soliton}

\author{Ahmed M. G. Ahmed}
\email{ahmedmgahmed@usf.edu}
\author{Alle Adjiri}
\email{aadjiri@usf.edu}
\affiliation{Department of Mathematics and Statistics,
University of South Florida, 
\\
Tampa, FL 33620-5700,
USA \looseness=-1}


\begin{abstract}
In this paper, we are going to solve nonlinear nonlocal reverse-time six-component six-order AKNS system. We used reverse-time reduction to reduce the coupled system to an integrable six-order NLS-type equation. 
Starting from the spectral problem of the AKNS system, a Riemann-Hilbert problem will be formulated. This formulation allows to generate soliton solutions by using the vectors lying in the kernel of the matrix Jost solutions. When reflection coefficients are zeros, the jump matrix is identity and the corresponding Riemann-Hilbert problem yields soliton solutions, leading to explore their dynamics.
\end{abstract}


\maketitle

\clearpage
\section{Introduction}
Integrable systems have been the interest of many branches in mathematics and physics. They have many applications in real life, for example in optical fiber, vortex filaments, ocean and water waves, and gravitational fields \cite{KangXia2019}-\cite{Osborne2010}.
The Korteweg deVries, nonlinear Schr\"{o}dinger equation and Sine-Gordon are all well known examples of integrable systems, where their intrinsic solutions are solitons.
The phenomena of solitons arise from these systems show linear and nonlinear effects. 
\\
Nonlocal PT symmetries, reverse-spacetime and reverse-time have been studied for the NLS and KdV under the inverse scattering transformation
\cite{AblowitzMusslimani2016}-\cite{AblowitzSegur1981}. This motivates us to snoop on solutions of a six-order six-component NLS-type AKNS system based on Riemann-Hilbert formulation\cite{Yang2018}-\cite{AblowitzFokas2003},
giving an extension to the dynamical behaviours of the solitons and their solutions \cite{AlleAhmedMa}. 

Throughout this paper, we formulate the AKNS hierarchy for the six-component AKNS system of sixth-order and solve the Riemann-Hilbert problem, with the contour being the real line 
and taking the jump matrix being the identity matrix \cite{KangXiaMa2019}-\cite{MaActa2022}.

This paper is outline as follows.
In section 2, we construct the AKNS hierarchy associated with the corresponding six-order six-component integrable system. In section 3, we formulate the Riemann-Hilbert problems associated with the corresponding matrix spectral problems. In section 4, 
we obtain general soliton solutions where the reflection coefficients
are zero \cite{Yang2010}-\cite{DrazinJohnson1989}, while in section 5, we explicit exact one and two soliton solutions and explore their dynamical behaviours along with three and four solitons. Finally, in the last section, we come to conclusion and remarks.
\section{Six-component AKNS Hierarchy}
\subsection{Six-component AKNS hierarchy of coupled six-order integrable equations}
Consider the $4 \times 4$ matrix spatial spectral problem
\cite{Ma2018}
\begin{align}\label{Spatialspectralproblem}
\psi_{x} &= \i U \psi,
\end{align}
where $\psi$ is the eigenfunction and $U(u,\lambda)$ the spectral matrix is given by
\begin{equation}\label{Umatrix}
U(u,\lambda)=\begin{pmatrix}
\alpha_{1} \lambda & p_{1} & p_{2} & p_{3} 
\\
r_{1} & \alpha_{2} \lambda & 0 & 0
\\
r_{2} & 0 & \alpha_{2} \lambda & 0
\\
r_{3} & 0 & 0 & \alpha_{2} \lambda
\end{pmatrix}
=\lambda \Lambda + P(u),
\end{equation}
where $\Lambda=diag(\alpha_{1},\alpha_{2},\alpha_{2},\alpha_{2})$, $\lambda$ is the spectral parameter, $\alpha_{1},\alpha_{2}$
are two real constants and 
$u= (p,r^T)^T$ 
is a vector of six potentials, where
$p=(p_1,p_2,p_3)$ and $r=(r_1,r_2,r_3)^T$ are vector functions of $(x,t)$ and $\{p_{i},r_{i} \}_{i=1,2,3} \in \mathcal{S}(\mathbb{R})$, the Schwartz space, and $p_{i},r_{i} \rightarrow 0$ as $x \rightarrow \pm \infty$, so
\begin{equation}
P=\begin{pmatrix}
0 & p_{1} & p_{2} & p_{3} 
\\
r_{1} & 0 & 0 & 0
\\
r_{2} & 0 & 0 & 0
\\
r_{3} & 0 & 0 & 0
\end{pmatrix}.
\end{equation}
Let's construct the AKNS soliton hierarchies. To do so, we need to solve the stationary zero curvature equation
\begin{equation}\label{stationaryZC}
W_{x}=\i[U,W],     
\end{equation} 
for which
\begin{equation}
W=\begin{pmatrix} 
a & b_{1} & b_{2} & b_{3} 
\\
c_{1} & d_{11} & d_{12} & d_{13}
\\
c_{2} & d_{21} & d_{22} & d_{23}
\\
c_{3} & d_{31} & d_{32} & d_{33}
\end{pmatrix},
\end{equation}
is a solution of this equation, where $a,b_{i},c_{i},d_{ij}$ are scalar components for $i,j \in \{1,2,3\}$. From the stationary zero curvature equation we get:
\begin{align}\label{recursivesystem6order}
\begin{cases}
a_{x}&=\i \big( -\sum\limits_{i=1}^{3}b_{i}r_{i}+\sum\limits_{i=1}^{3}c_{i}p_{i} \big),
\\
b_{i,x}&= \i(\alpha\lambda b_{i}-a p_{i} +d_{1i}p_{1}+d_{2i}p_{2}+d_{3i}p_{3}),
\quad 
i \in \{1,2,3\},
\\
c_{i,x}&= \i(-\alpha\lambda c_{i}
+a r_{i}-d_{i1}r_{1}-d_{i2}r_{2}-d_{i3}r_{3}),
\quad 
i \in \{1,2,3\},
\\
d_{ij,x}&=\i(b_{j}r_{i}-c_{i}p_{j}), \quad i,j \in \{1,2,3\},
\end{cases}
\end{align}
where $\alpha=\alpha_{1}-\alpha_{2}$. We expand $W$ in Laurent series:
\begin{equation}
W = \sum\limits_{m= 0}^{\infty}W_{m}\lambda^{-m} 
\quad \text{with} \quad 
W_{m}=\begin{pmatrix} 
a^{[m]}  & b_{1}^{[m]} & b_{2}^{[m]} & b_{3}^{[m]} \\
c_{1}^{[m]} & d_{11}^{[m]} & d_{12}^{[m]} & d_{13}^{[m]}  \\
c_{2}^{[m]} & d_{21}^{[m]} & d_{22}^{[m]} & d_{23}^{[m]}  \\
c_{3}^{[m]} & d_{31}^{[m]} & d_{32}^{[m]} & d_{33}^{[m]}  \\
\end{pmatrix},
\end{equation}
explicitly,
\begin{align}
a&= \sum\limits_{m= 0}^{\infty}a^{[m]}\lambda^{-m},
\hspace{0.5cm} 
b_{i}= \sum\limits_{m= 0}^{\infty}b_{i}^{[m]}\lambda^{-m}, 
\\ 
c_{i}&= \sum\limits_{m= 0}^{\infty}c_{i}^{[m]}\lambda^{-m}, \hspace{0.5cm} 
d_{ij}= \sum\limits_{m= 0}^{\infty}d_{ij}^{[m]}\lambda^{-m},
\end{align}
for $i,j \in \{1,2,3\}$ and $m \geq 0$. The system (\ref{recursivesystem6order})
generates the recursive relations:
\begin{align}\label{recursive6order}
& b_{i}^{[0]}=0, \, c_{i}^{[0]}=0, \quad \text{for} \quad i \in \{1,2,3\},
\\
& a_{x}^{[0]}=0,
\\
& d_{ij,x}^{[0]}=0, \quad \text{for} \quad i,j \in \{1,2,3\},
\\
& b_{i}^{[m+1]}= \frac{1}{\alpha} 
(-\i b_{i,x}^{[m]}+a^{[m]}p_{i}
-d_{1i}^{[m]}p_{1}-d_{2i}^{[m]}p_{2}-d_{3i}^{[m]}p_{3}),
\quad i \in \{1,2,3\},
\\
& c_{i}^{[m+1]}= \frac{1}{\alpha} 
(\i c_{i,x}^{[m]}+a^{[m]} r_{i}
-d_{i1}^{[m]}r_{1}-d_{i2}^{[m]}r_{2}
-d_{i3}^{[m]}r_{3}),
\quad i \in \{1,2,3\},
\\
& a_{x}^{[m]}=\i(-\sum\limits_{i=1}^{3}b_{i}^{[m]}r_{i}
+\sum\limits_{i=1}^{3}c_{i}^{[m]}p_{i}),
\\[3mm]
& d_{ij,x}^{[m]}=\i(b_{j}^{[m]}r_{i}-c_{i}^{[m]}p_{j}), 
\quad i,j \in \{1,2,3\},
\end{align}
where all the involved functions are defined as follows:
\begin{flalign}
&\begin{aligned}
\begin{cases}
a^{[0]} &=\beta_{1},
\quad
a^{[1]} =0,
\quad
a^{[2]} =-\frac{\beta}{\alpha^{2}} \E, 
\quad
a^{[3]} =- \i \frac{\beta}{\alpha^{3}} (\F-\G), 
\\
a^{[4]} &=\frac{\beta}{\alpha^{4}} \bigg[ 
3\E^2
+\H-\K+\J \bigg], 
\\
a^{[5]} &=\i \frac{\beta}{\alpha^{5}} \bigg[ 6 
\E ( \F -\G )
+ \L-\W+\N-\M \bigg]  ,
\\
a^{[6]} &=-\frac{\beta}{\alpha^{6}}
\Bigg[ 
10 \E^{3}
+ 10 \E (\H+\J)
+ 5 \Big( \G^{2} + \F^{2} \Big)
\\& \hspace{6cm}
+ (\Q + \U -\R -\T{}{} + \S) \Bigg],
\\
a^{[7]} &=-\i \frac{\beta}{\alpha^{7}} \bigg[
30 \E^{2} (\F-\G)
+5 \E (\N-\M)
\\& \hspace{2cm}
+10 \E (\L-\W)
+10 \K (\F-\G)
+ 5 (
\F
\J
-\G
\H
)
\\& \hspace{3cm}
+ 20 (
\F \H -\G \J
)
+ 5 (\Tone -\Ttwo +\Tthree -\Tfour +\Tfive -\Tsix) 
\bigg],
\end{cases}
\end{aligned}&
\\
&\begin{aligned}
\begin{cases}
b^{[0]}_{k} &=0,
\quad
b^{[1]}_{k} =\frac{\beta}{\alpha} p_{k}, 
\quad
b^{[2]}_{k} =-\i \frac{\beta}{\alpha^2} p_{k,x}, 
\quad
b^{[3]}_{k} =-\frac{\beta}{\alpha^3} \bigg[ p_{k,xx} + 2\E p_{k} \bigg], 
\\
b^{[4]}_{k} &=\i \frac{\beta}{\alpha^{4}} \bigg[ p_{k,xxx}
+3 \E p_{k,x}
+3 \G p_{k} \bigg] , 
\\
b^{[5]}_{k} &=\frac{\beta}{\alpha^{5}} \bigg[ p_{k,xxxx}+
4 \E p_{k,xx}+(6 \G
+2\F)p_{k,x}
+(4 \J
+2 \K
+2 \H
+6 \E^{2})p_{k} \bigg],
\\
b^{[6]}_{k} &= -\i \frac{\beta}{\alpha^{6}}
\Bigg[ p_{k,xxxxx}
+ 5\E p_{k,xxx}
+ (10 \G + 5 \F) p_{k,xx}
\\&
+ \Bigg( 10 \J
+ 5 \H
+ 10 \K
+ 10 \E^{2}
\Bigg) p_{k,x}
+ \Bigg( 5 \W
+ 5 \N
+ 5 \M
+ 20 \E \G
\Bigg) p_{k} \Bigg],
\\
b^{[7]}_{k} &= -\frac{\beta}{\alpha^{7}}
\Bigg[
p_{k,xxxxxx} + 6 \E p_{k,xxxx}
+ (9\F+15\G) p_{k,xxx}
\\&
+(15\E^{2}+11\H+20\J+25\K)p_{k,xx}
\\
&+\bigg( \E(15\F+45\G)+15\W+4\L+20\M+25\N \bigg) p_{k,x}
\\
&+ \bigg( (20\E^{3}+\E(20\H+35\J+25\K))
+10\F^{2}
\\& \hspace{3cm}
+25\G^{2}+20\G \F+2\Q+6\U+4\R+9\T{}{}+11\S
\bigg) p_{k}
\Bigg],
\end{cases}
\end{aligned}&
\end{flalign}
\begin{flalign}
&\begin{aligned}
\begin{cases}
c^{[0]}_{k} &=0 , 
\quad
c^{[1]}_{k} =\frac{\beta}{\alpha} r_{k} , 
\quad
c^{[2]}_{k} =\i \frac{\beta}{\alpha^2} r_{k,x} , 
\quad
c^{[3]}_{k} =-\frac{\beta}{\alpha^3} \bigg[ r_{k,xx} 
+ 2 \E r_{k} \bigg] , 
\\
c^{[4]}_{k} &=-\i \frac{\beta}{\alpha^{4}} \bigg[ r_{k,xxx}
+3 \E r_{k,x}
+3\F r_{k} \bigg] , 
\\
c^{[5]}_{k} &=\frac{\beta}{\alpha^{5}} \bigg[ r_{k,xxxx}
+4 \E r_{k,xx}
+(6 \F
+2\G)r_{k,x}
+(4 \H
+2 \K
+2 \J
+6 \E^{2})r_{k} \bigg],
\\
c^{[6]}_{k} &= \i \frac{\beta}{\alpha^{6}}
\Bigg[ q_{k,xxxxx}
+ 5\E q_{k,xxx}
+ (10 \F + 5 \G) q_{k,xx}
\\&
+ \Bigg( 10 \H
+ 5 \J
+ 10 \K
+ 10 \E^{2}
\Bigg) q_{k,x}
+ \Bigg( 5 \L
+ 5 \M
+ 5 \N
+ 20 \E \F
\Bigg) q_{k} \Bigg]
\\
c^{[7]}_{k} &= -\frac{\beta}{\alpha^{7}}
\Bigg[
q_{k,xxxxxx} + 6 \E q_{k,xxxx}
+ (15\F+9\G) q_{k,xxx}
\\&
+(15\E^{2}+20\H+11\J+25\K)q_{k,xx}
\\
&+\bigg( \E(45\F+15\G)+4\W+15\L+25\M+20\N \bigg) q_{k,x}
\\
&+ \bigg( (20\E^{3}+\E(35\H+20\J+25\K))
+25\F^{2}
\\& \hspace{3cm}
+10\G^{2}+20\G \F+6\Q+2\U+9\R+4\T{}{}+11\S
\bigg) q_{k}
\Bigg],
\end{cases}
\end{aligned}&
\\
&\begin{aligned}
\begin{cases}
d^{[0]}_{kj} &= \beta_{2},
\, \text{for} \, \, k=j, 
\quad \text{and} \quad 
d^{[0]}_{kj}=0, 
\, \, \text{for} \, \, k \neq j,
\quad \text{where} \quad k,j \in \{1,2,3\}
\\ 
d^{[1]}_{kj} &= 0 , 
\quad
d^{[2]}_{kj} = \frac{\beta}{\alpha^{2}} p_{j} r_{k}  , 
\quad
d^{[3]}_{kj} = -\i \frac{\beta}{\alpha^{3}}
(p_{j,x} r_{k} - p_{j} r_{k,x})  , 
\\
d^{[4]}_{kj} &= - \frac{\beta}{\alpha^{4}}
\bigg[ 3 \E p_{j} r_{k}
+p_{j,xx} r_{k} - p_{j,x} r_{k,x} + p_{j} r_{k,xx} \bigg]  , 
\\
d^{[5]}_{kj} &= \i \frac{\beta}{\alpha^{5}} \bigg[
2 (\G - \F) p_{j} r_{k}
+4 \E (p_{j,x} r_{k} - p_{j} r_{k,x})
+ p_{j,xxx} r_{k} - p_{j} r_{k,xxx} + p_{j,x} r_{k,xx}
-p_{j,xx} r_{k,x} \bigg],
\\
d^{[6]}_{kj} &= \frac{\beta}{\alpha^{6}}
\bigg[ \bigg( 
\E^{2} + 5(\H + \K + \J) \bigg) p_{j} r_{k}
+5 \F p_{j} r_{k,x}
+5 \G p_{j,x} r_{k}
\\
& +5 \E
(p_{j} r_{k,xx} -p_{j,x} r_{k,x} +p_{j,xx} r_{k})
+p_{j,xxxx} r_{k}-p_{j,xxx} r_{k,x}
+p_{j,xx} r_{k,xx}-p_{j,x} r_{k,xxx}+p_{j} r_{k,xxxx} \bigg]
\\
d^{[7]}_{kj} &= -\i \frac{\beta}{\alpha^{7}} \bigg[ \bigg(
\E (\G -\F) -4(\W-\L) +(\N-\M) \bigg) p_{j} r_{k}
\\
&+\bigg(
15 \E^{2}
+8 \H
+11 \J
+13 \K \bigg) p_{j,x} r_{k}
-\bigg( 15 \E^{2} +11 \H +8 \J +13 \K \bigg) p_{j} r_{k,x}
\\
&
+\bigg( 3 \F + 9 \G \bigg) p_{j,xx} r_{k}
+\bigg( 3 \F - 3 \G \bigg) p_{j,x} r_{k,x}
-\bigg( 9 \F + 3 \G \bigg) p_{j} r_{k,xx}
\\
&+ 6 \E
(p_{j,xxx} r_{k} -p_{j,xx} r_{k,x}+p_{j,x} r_{k,xx}
-p_{j} r_{k,xxx})
\\
&+p_{j,xxxxx} r_{k}-p_{j,xxxx} r_{k,x}+p_{j,xxx} r_{k,xx}
-p_{j,xx} r_{k,xxx}+p_{j,x} r_{k,xxxx}-p_{j} r_{k,xxxxx}
\bigg],
\end{cases}
\end{aligned}&
\end{flalign}
where $\alpha=\alpha_{1}-\alpha_{2}$, $\beta=\beta_{1}-\beta_{2}$
and
\begin{small}
\begin{align*}
\begin{cases}
&\E=\sum\limits_{i=1}^{3} p_{i} r_{i}, 
\\[5mm]
&\F=\sum\limits_{i=1}^{3} p_{i} r_{i,x}, \,
\G=\sum\limits_{i=1}^{3} p_{i,x} r_{i},
\\[5mm]
&\H=\sum\limits_{i=1}^{3} p_{i} r_{i,xx}, \,
\J=\sum\limits_{i=1}^{3} p_{i,xx} r_{i}, \,
\K=\sum\limits_{i=1}^{3} p_{i,x} r_{i,x},
\\[5mm]
&\L=\sum\limits_{i=1}^{3} p_{i} r_{i,xxx}, \,
\M=\sum\limits_{i=1}^{3} p_{i,x} r_{i,xx}, \,
\N=\sum\limits_{i=1}^{3} p_{i,xx} r_{i,x}, \,
\W=\sum\limits_{i=1}^{3} p_{i,xxx} r_{i},
\\[6mm]
&\Q=\sum\limits_{i=1}^{3} p_{i} r_{i,xxxx}, \,
\R=\sum\limits_{i=1}^{3} p_{i,x} r_{i,xxx}, \,
\S=\sum\limits_{i=1}^{3} p_{i,xx} r_{i,xx}, 
\\& 
\T{}{}=\sum\limits_{i=1}^{3} p_{i,xxx} r_{i,x} \, 
\U=\sum\limits_{i=1}^{3} p_{i,xxxx} r_{i}.
\end{cases}
\end{align*}
\end{small}
\\
We always assume that 
$b^{[m]}=(b^{[m]}_{1},b^{[m]}_{2},b^{[m]}_{3})$ 
and
$c^{[m]}=(c^{[m]}_{1},c^{[m]}_{2},c^{[m]}_{3})^{T}$, 
for $m \in \{1,2,3,4,5,6,7\}$. 
To derive the six-order six-component AKNS integrable hierarchy, we take the Lax matrices
\begin{equation}
V^{[6]}=V^{[6]}(u,\lambda)=(\lambda^{6} W)_{+}= \sum\limits_{m= 0}^{6}W_{m}\lambda^{6-m}. 
\end{equation}
By taking the modification terms to be zero.
\newline
We begin with the spatial and temporal equations of the spectral problems, with the associated Lax pair $\{U,V\}$: \cite{Ma2018}
\begin{align}\label{AKNSsystem}
\psi_{x} &= \i U \psi,
\\
\psi_{t} &= \i {V} \psi,
\end{align}
where $V=V^{[6]}$ and $\psi$ is the eigenfunction. 
\newline
The Lax matrix operator $V$ is determined by 
the compatibility condition $\psi_{xt}=\psi_{tx}$ which leads to the zero curvature equation:
\begin{equation}
U_{t} - V_{x} + \i [U,V] = 0 ,
\end{equation}
that gives the six-component system of soliton equations
\begin{equation}
u_{t}=\begin{pmatrix}
p^{T}  \\ r
\end{pmatrix}_{t} 
=\i \begin{pmatrix}
\alpha {b}^{[7]T}  \\ -\alpha c^{[7]}
\end{pmatrix}
\end{equation}
where ${b}^{[7]}$ and ${c}^{[7]}$ are defined earlier, and
\begin{equation}
V=
\begin{pmatrix}
\Vbold{11}{} & \Vbold{12}{} & \Vbold{13}{} & \Vbold{14}{}
\\
\Vbold{21}{} & \Vbold{22}{} & \Vbold{23}{} & \Vbold{24}{}
\\
\Vbold{31}{} & \Vbold{32}{} & \Vbold{33}{} & \Vbold{34}{}
\\
\Vbold{41}{} & \Vbold{42}{} & \Vbold{43}{} & \Vbold{44}{}
\end{pmatrix}
\end{equation}
where 
\begin{align} \nonumber
\Vbold{11}{} &= a^{[0]} \lambda^{6} + a^{[2]} \lambda^{4} + a^{[3]} \lambda^{3} 
+ a^{[4]} \lambda^{2} + a^{[5]} \lambda + a^{[6]},
&
\Vbold{12}{} &= b_{1}^{[1]} \lambda^{5} + b_{1}^{[2]} \lambda^{4}
+b_{1}^{[3]} \lambda^{3} 
+ b_{1}^{[4]} \lambda^{2} + b_{1}^{[5]} \lambda + b_{1}^{[6]},
\\ \nonumber
\Vbold{13}{} &= b_{2}^{[1]} \lambda^{5} + b_{2}^{[2]} \lambda^{4}
+b_{2}^{[3]} \lambda^{3} 
+ b_{2}^{[4]} \lambda^{2} + b_{2}^{[5]} \lambda + b_{2}^{[6]},
&
\Vbold{14}{} &= b_{3}^{[1]} \lambda^{5} + b_{3}^{[2]} \lambda^{4}
+b_{3}^{[3]} \lambda^{3} 
+ b_{3}^{[4]} \lambda^{2} + b_{3}^{[5]} \lambda + b_{3}^{[6]},
\\ \nonumber
\Vbold{21}{} &= c_{1}^{[1]} \lambda^{5} + c_{1}^{[2]} \lambda^{4}
+c_{1}^{[3]} \lambda^{3} 
+ c_{1}^{[4]} \lambda^{2} + c_{1}^{[5]} \lambda + c_{1}^{[6]},
&
\Vbold{22}{} &= d_{11}^{[0]} \lambda^{6} + d_{11}^{[2]} \lambda^{4} 
+ d_{11}^{[3]} \lambda^{3} 
+ d_{11}^{[4]} \lambda^{2} + d_{11}^{[5]} \lambda
+ d_{11}^{[6]},
\\ \nonumber
\Vbold{23}{} &= d_{12}^{[2]} \lambda^{4} 
+ d_{12}^{[3]} \lambda^{3} 
+ d_{12}^{[4]} \lambda^{2} + d_{12}^{[5]} \lambda
+ d_{12}^{[6]},
&
\Vbold{24}{} &= d_{13}^{[2]} \lambda^{4} 
+ d_{13}^{[3]} \lambda^{3} 
+ d_{13}^{[4]} \lambda^{2} + d_{13}^{[5]} \lambda
+ d_{13}^{[6]},
\\ \nonumber
\Vbold{31}{} &= c_{2}^{[1]} \lambda^{5} + c_{2}^{[2]} \lambda^{4}
+c_{2}^{[3]} \lambda^{3} 
+ c_{2}^{[4]} \lambda^{2} + c_{2}^{[5]} \lambda + c_{2}^{[6]},
&
\Vbold{32}{} &= d_{21}^{[2]} \lambda^{4} 
+ d_{21}^{[3]} \lambda^{3} 
+ d_{21}^{[4]} \lambda^{2} + d_{21}^{[5]} \lambda
+ d_{21}^{[6]},
\\ \nonumber
\Vbold{33}{} &= d_{22}^{[0]} \lambda^{6} + d_{22}^{[2]} \lambda^{4} 
+ d_{22}^{[3]} \lambda^{3} 
+ d_{22}^{[4]} \lambda^{2} + d_{22}^{[5]} \lambda
+ d_{22}^{[6]},
&
\Vbold{34}{} &= d_{23}^{[2]} \lambda^{4} 
+ d_{23}^{[3]} \lambda^{3} 
+ d_{23}^{[4]} \lambda^{2} + d_{23}^{[5]} \lambda
+ d_{23}^{[6]},
\\ \nonumber
\Vbold{41}{} &= c_{3}^{[1]} \lambda^{5} + c_{3}^{[2]} \lambda^{4}
+c_{3}^{[3]} \lambda^{3} 
+ c_{3}^{[4]} \lambda^{2} + c_{3}^{[5]} \lambda + c_{3}^{[6]},
&
\Vbold{42}{} &= d_{31}^{[2]} \lambda^{4} 
+ d_{31}^{[3]} \lambda^{3} 
+ d_{31}^{[4]} \lambda^{2} + d_{31}^{[5]} \lambda
+ d_{31}^{[6]},
\\ \nonumber
\Vbold{43}{} &= d_{32}^{[2]} \lambda^{4} 
+ d_{32}^{[3]} \lambda^{3} 
+ d_{32}^{[4]} \lambda^{2} + d_{32}^{[5]} \lambda
+ d_{32}^{[6]},
&
\Vbold{44}{} &= d_{33}^{[0]} \lambda^{6} + d_{33}^{[2]} \lambda^{4} 
+ d_{33}^{[3]} \lambda^{3} 
+ d_{33}^{[4]} \lambda^{2} + d_{33}^{[5]} \lambda
+ d_{33}^{[6]}.
\end{align}
Thus, we deduce the coupled AKNS system of sixth order equations:\cite{Ma2018}
\begin{align} \nonumber 
p_{k,t} &= -\i \frac{\beta}{\alpha^{6}}
\Bigg[ p_{k,xxxxxx}
+6 (\sum\limits_{i=1}^{3} p_{i} r_{i}) p_{k,xxxx}
+(9 \sum\limits_{i=1}^{3} p_{i} r_{i,x}
+15 \sum\limits_{i=1}^{3} p_{i,x} r_{i}) p_{k,xxx}
\\ \nonumber
&+ \bigg(
15 (\sum\limits_{i=1}^{3} p_{i} r_{i})^{2}
+ 11 \sum\limits_{i=1}^{3} p_{i} r_{i,xx}
+ 20 \sum\limits_{i=1}^{3} p_{i,xx} r_{i}
+ 25 \sum\limits_{i=1}^{3} p_{i,x} r_{i,x}
\bigg) p_{k,xx}
\\ \nonumber
&+ \bigg(
(\sum\limits_{i=1}^{3} p_{i} r_{i})
(15 \sum\limits_{i=1}^{3} p_{i} r_{i,x}
+45 \sum\limits_{i=1}^{3} p_{i,x} r_{i})
+ 15 \sum\limits_{i=1}^{3} p_{i,xxx} r_{i}
+ 4  \sum\limits_{i=1}^{3} p_{i} r_{i,xxx}
\\&  \nonumber
+ 20 \sum\limits_{i=1}^{3} p_{i,x} r_{i,xx}
+ 25 \sum\limits_{i=1}^{3} p_{i,xx} r_{i,x}
\bigg) p_{k,x}
\\& \nonumber
+ \bigg(
20 (\sum\limits_{i=1}^{3} p_{i} r_{i})^{3}
+(\sum\limits_{i=1}^{3} p_{i} r_{i})
(20 \sum\limits_{i=1}^{3} p_{i} r_{i,xx}
+35 \sum\limits_{i=1}^{3} p_{i,xx} r_{i}
+25 \sum\limits_{i=1}^{3} p_{i,x} r_{i,x})
\\& \nonumber
+ 10 (\sum\limits_{i=1}^{3} p_{i} r_{i,x})^{2}
+ 20 (\sum\limits_{i=1}^{3} p_{i,x} r_{i})
(\sum\limits_{i=1}^{3} p_{i} r_{i,x})
+ 25 (\sum\limits_{i=1}^{3} p_{i,x} r_{i})^{2}
\\& 
+ 2 \sum\limits_{i=1}^{3} p_{i} r_{i,xxxx}
+ 4 \sum\limits_{i=1}^{3} p_{i,x} r_{i,xxx}
+ 11 \sum\limits_{i=1}^{3} p_{i,xx} r_{i,xx}
+ 9  \sum\limits_{i=1}^{3} p_{i,xxx} r_{i,x}
+ 6  \sum\limits_{i=1}^{3} p_{i,xxxx} r_{i}
\bigg) p_{k}
\Bigg], \nonumber
\end{align}
\vspace{-1cm}
\begin{align} \nonumber
r_{k,t} &= \i \frac{\beta}{\alpha^{6}}
\Bigg[ q_{k,xxxxxx}
+6 (\sum\limits_{i=1}^{3} p_{i} r_{i}) q_{k,xxxx}
+(15 \sum\limits_{i=1}^{3} p_{i} r_{i,x}
+9 \sum\limits_{i=1}^{3} p_{i,x} r_{i}) q_{k,xxx}
\\ \nonumber
&+ \bigg(
15 (\sum\limits_{i=1}^{3} p_{i} r_{i})^{2}
+ 20 \sum\limits_{i=1}^{3} p_{i} r_{i,xx}
+ 11 \sum\limits_{i=1}^{3} p_{i,xx} r_{i}
+ 25 \sum\limits_{i=1}^{3} p_{i,x} r_{i,x}
\bigg) q_{k,xx}
\\ \nonumber
&+ \bigg(
(\sum\limits_{i=1}^{3} p_{i} r_{i})
(45 \sum\limits_{i=1}^{3} p_{i} r_{i,x}
+15 \sum\limits_{i=1}^{3} p_{i,x} r_{i})
+ 4 \sum\limits_{i=1}^{3} p_{i,xxx} r_{i}
+ 15  \sum\limits_{i=1}^{3} p_{i} r_{i,xxx}
\\& \nonumber
+ 25 \sum\limits_{i=1}^{3} p_{i,x} r_{i,xx}
+ 20 \sum\limits_{i=1}^{3} p_{i,xx} r_{i,x}
\bigg) q_{k,x}
\\& \nonumber
+ \bigg(
20 (\sum\limits_{i=1}^{3} p_{i} r_{i})^{3}
+(\sum\limits_{i=1}^{3} p_{i} r_{i})
(35 \sum\limits_{i=1}^{3} p_{i} r_{i,xx}
+20 \sum\limits_{i=1}^{3} p_{i,xx} r_{i}
+25 \sum\limits_{i=1}^{3} p_{i,x} r_{i,x})
\\& \nonumber
+ 25 (\sum\limits_{i=1}^{3} p_{i} r_{i,x})^{2}
+ 20 (\sum\limits_{i=1}^{3} p_{i,x} r_{i})
(\sum\limits_{i=1}^{3} p_{i} r_{i,x})
+ 10 (\sum\limits_{i=1}^{3} p_{i,x} r_{i})^{2}
\\& \label{coupledequs}
+ 6 \sum\limits_{i=1}^{3} p_{i} r_{i,xxxx}
+ 9 \sum\limits_{i=1}^{3} p_{i,x} r_{i,xxx}
+ 11 \sum\limits_{i=1}^{3} p_{i,xx} r_{i,xx}
+ 4  \sum\limits_{i=1}^{3} p_{i,xxx} r_{i,x}
+ 2  \sum\limits_{i=1}^{3} p_{i,xxxx} r_{i}
\bigg) q_{k}
\Bigg],
\end{align}
where $k \in \{1,2,3\}$.

\subsection{Nonlocal reverse-time six-component AKNS system}
We study the nonlocal reverse-time by considering specific reductions for the spectral matrix
\begin{equation}\label{Ureduction}
U^{T}(x,-t,-\lambda)=-CU(x,t,\lambda)C^{-1},
\end{equation}
where 
$C=
\begin{pmatrix} 
1 & 0 \\
0 & \Sigma
\end{pmatrix}$
and $\Sigma$ is a constant invertible symmetric $3 \times 3$ matrix, in other words $\det \Sigma \neq 0$ and $\Sigma^{T}=\Sigma$. 

Because $U(x,t,\lambda)=\lambda \Lambda+P(x,t)$, for $P=
\begin{pmatrix}
0 & p \\
r & 0
\end{pmatrix}$, using the reduction (\ref{Ureduction}) we can easily prove that
\begin{equation}\label{Pequ}
P^{T}(x,-t)=-C P(x,t) C^{-1}.
\end{equation}
It follows from (\ref{Pequ}) that
\begin{equation}\label{prrelation}
p^{T}(x,-t)=- \Sigma r(x,t) \quad \text{i.e.} \quad 
r(x,t) = -\Sigma^{-1} p^{T}(x,-t).
\end{equation}
Similarly from $V(x,t,\lambda)=\lambda^{6} \Omega + Q(x,t,\lambda)$
along with (\ref{prrelation}), one can prove with a tedious calculations that
\begin{equation}
Q^{T}(x,-t,-\lambda) = C Q(x,t,\lambda) C^{-1},
\end{equation}
and
\begin{equation}\label{VQequ}
V^{T}(x,-t,-\lambda) = C V(x,t,\lambda) C^{-1},
\end{equation}
where 
$\Omega=\textit{diag}(\beta_{1},\beta_{2},\beta_{2},\beta_{2})$.
\\
It is interesting that the two nonlocal Lax matrices 
$U^{T}(x,-t,-\lambda)$ and $V^{T}(x,-t,-\lambda)$ 
satisfy the equivalent zero curvature equation:
\begin{equation}
U_{t}^{T}(x,-t,-\lambda)  
+ V_{x}^{T}(x,-t,-\lambda)  
+\i \big[ U^{T}(x,-t,-\lambda), V^{T}(x,-t,-\lambda) \big] = 0.
\end{equation}
By taking $\Sigma=diag(\rho^{-1}_{1},\rho^{-1}_{2},\rho^{-1}_{3})$, where $\rho_{1}, \rho_{2}, \rho_{3}$ are non-zero real,
we deduce from (\ref{prrelation}) the nonlocal relation between
the components of the vectors $p$ and $r$, that is
\begin{equation}
r_{i}(x,t) = -\rho_{i} p_{i}(x,-t) \quad \text{for} \quad 
i \in \{1,2,3\}.
\end{equation}
Hence, we can reduce the coupled equations (\ref{coupledequs}) 
to the nonlocal reverse-time six-order equation:
\begin{align*}
p_{k,t}(x,t) &= -\i \frac{\beta}{\alpha^{6}}
\Bigg[ p_{k,xxxxxx}(x,t)
\\
& -6 \bigg( \sum\limits_{i=1}^{3} \rho_{i} p_{i}(x,t) p_{i}(x,-t) \bigg) p_{k,xxxx}
\\& - \bigg( 9 \sum\limits_{i=1}^{3} \rho_{i} p_{i}(x,t) p_{i,x}(x,-t)
+15 \sum\limits_{i=1}^{3} \rho_{i} p_{i,x}(x,t) p_{i}(x,-t) \bigg) p_{k,xxx}
\\ 
&+ \bigg(
15 \big( \sum\limits_{i=1}^{3} \rho_{i} p_{i}(x,t) p_{i}(x,-t) \big)^{2}
- 11 \sum\limits_{i=1}^{3} \rho_{i} p_{i}(x,t) p_{i,xx}(x,-t)
\\& \hspace{3cm}
- 20 \sum\limits_{i=1}^{3} \rho_{i} p_{i,xx}(x,t) p_{i}(x,-t)
- 25 \sum\limits_{i=1}^{3} \rho_{i} p_{i,x}(x,t) p_{i,x}(x,-t)
\bigg) p_{k,xx}
\\ 
&+ \Bigg(
\bigg( \sum\limits_{i=1}^{3} \rho_{i} p_{i}(x,t) p_{i}(x,-t) \bigg)
\bigg( 15 \sum\limits_{i=1}^{3} \rho_{i} p_{i}(x,t) p_{i,x}(x,-t)
+45 \sum\limits_{i=1}^{3} \rho_{i} p_{i,x}(x,t) p_{i}(x,-t) \bigg)
\\& \hspace{2cm}
- 15 \sum\limits_{i=1}^{3} \rho_{i} p_{i,xxx}(x,t) p_{i}(x,-t)
- 4  \sum\limits_{i=1}^{3} \rho_{i} p_{i}(x,t) p_{i,xxx}(x,-t)
\\& \hspace{3cm} 
- 20 \sum\limits_{i=1}^{3} \rho_{i} p_{i,x}(x,t) p_{i,xx}(x,-t)
- 25 \sum\limits_{i=1}^{3} \rho_{i} p_{i,xx}(x,t) p_{i,x}(x,-t)
\Bigg) p_{k,x}
\\& 
+ \Bigg(
-20 \bigg( \sum\limits_{i=1}^{3} \rho_{i} p_{i}(x,t) p_{i}(x,-t) \bigg)^{3}
+\bigg( \sum\limits_{i=1}^{3} \rho_{i} p_{i}(x,t) p_{i}(x,-t) \bigg)
\bigg( 20 \sum\limits_{i=1}^{3} \rho_{i} p_{i}(x,t) p_{i,xx}(x,-t)
\\&
+35 \sum\limits_{i=1}^{3} \rho_{i} p_{i,xx}(x,t) p_{i}(x,-t)
+25 \sum\limits_{i=1}^{3} \rho_{i} p_{i,x}(x,t) p_{i,x}(x,-t) \bigg)
\\& 
+ 10 \big(\sum\limits_{i=1}^{3} \rho_{i} p_{i}(x,t) p_{i,x}(x,-t) \big)^{2}
+ 20 \bigg( \sum\limits_{i=1}^{3} \rho_{i}p_{i,x}(x,t) p_{i}(x,-t) \bigg)
\bigg( \sum\limits_{i=1}^{3} \rho_{i} p_{i}(x,t) p_{i,x}(x,-t) \bigg)
\\&
+ 25 \big( \sum\limits_{i=1}^{3} \rho_{i} p_{i,x}(x,t) p_{i}(x,-t)\big)^{2}
- 2 \sum\limits_{i=1}^{3} \rho_{i} p_{i}(x,t) p_{i,xxxx}(x,-t)
\\&
- 4 \sum\limits_{i=1}^{3} \rho_{i} p_{i,x}(x,t) p_{i,xxx}(x,-t)
- 11 \sum\limits_{i=1}^{3} \rho_{i} p_{i,xx}(x,t) p_{i,xx}(x,-t)
\\&
- 9  \sum\limits_{i=1}^{3} \rho_{i} p_{i,xxx}(x,t) p_{i,x}(x,-t)
- 6  \sum\limits_{i=1}^{3} \rho_{i} p_{i,xxxx}(x,t) p_{i}(x,-t)
\Bigg) p_{k}
\Bigg] 
\end{align*}
for $k \in \{1,2,3\}$. 
\\[3mm]
We can see that when all $\rho_{i} < 0$ for $i \in \{1,2,3\}$,
the dispersive term and the nonlinear terms attract.
Hence, we obtain the focusing nonlocal reverse-time six-component six-order equation. Otherwise, if $\rho_{i}$'s are not all negative
for $i \in \{1,2,3\}$, then we have combined focussing and defocussing cases.
\section{Riemann-Hilbert problems}
The spatial and temporal spectral problem of the six-component six-order AKNS equations can be written:
\begin{equation}\label{spatialequ}
\psi_{x}=i U \psi = i (\lambda \Lambda + P) \psi,
\end{equation}
\begin{equation}\label{temporalequ}
\psi_{t}=i V \psi = i (\lambda^{6} \Omega + Q) \psi,
\end{equation}
where 
$\Lambda=\textit{diag}(\alpha_{1},\alpha_{2},\alpha_{2},\alpha_{2})$, 
$\Omega=\textit{diag}(\beta_{1},\beta_{2},\beta_{2},\beta_{2})$,
and 
\begin{equation}
P=
\begin{pmatrix}\label{Pequation}
0 & p_{1} & p_{2} & p_{3} \\
r_{1} & 0 & 0 & 0 \\
r_{2} & 0 & 0 & 0 \\
r_{3} & 0 & 0 & 0 
\end{pmatrix},
\quad
Q=
\begin{pmatrix}
\Qbold{11}{} & \Qbold{12}{} & \Qbold{13}{} & \Qbold{14}{}
\\
\Qbold{21}{} & \Qbold{22}{} & \Qbold{23}{} & \Qbold{24}{}
\\
\Qbold{31}{} & \Qbold{32}{} & \Qbold{33}{} & \Qbold{34}{}
\\
\Qbold{41}{} & \Qbold{42}{} & \Qbold{43}{} & \Qbold{44}{}
\end{pmatrix},
\end{equation}
\begin{align*}
\Qbold{11}{} &= a^{[2]} \lambda^{4} + a^{[3]} \lambda^{3} 
+ a^{[4]} \lambda^{2} + a^{[5]} \lambda + a^{[6]},
&
\Qbold{12}{} &= b_{1}^{[1]} \lambda^{5} + b_{1}^{[2]} \lambda^{4}
+b_{1}^{[3]} \lambda^{3} 
+ b_{1}^{[4]} \lambda^{2} + b_{1}^{[5]} \lambda + b_{1}^{[6]},
\\
\Qbold{13}{} &= b_{2}^{[1]} \lambda^{5} + b_{2}^{[2]} \lambda^{4}
+b_{2}^{[3]} \lambda^{3} 
+ b_{2}^{[4]} \lambda^{2} + b_{2}^{[5]} \lambda + b_{2}^{[6]},
&
\Qbold{14}{} &= b_{3}^{[1]} \lambda^{5} + b_{3}^{[2]} \lambda^{4}
+b_{3}^{[3]} \lambda^{3} 
+ b_{3}^{[4]} \lambda^{2} + b_{3}^{[5]} \lambda + b_{3}^{[6]},
\\
\Qbold{21}{} &= c_{1}^{[1]} \lambda^{5} + c_{1}^{[2]} \lambda^{4}
+c_{1}^{[3]} \lambda^{3} 
+ c_{1}^{[4]} \lambda^{2} + c_{1}^{[5]} \lambda + c_{1}^{[6]},
&
\Qbold{22}{} &= d_{11}^{[2]} \lambda^{4} 
+ d_{11}^{[3]} \lambda^{3} 
+ d_{11}^{[4]} \lambda^{2} + d_{11}^{[5]} \lambda
+ d_{11}^{[6]},
\\
\Qbold{23}{} &= d_{12}^{[2]} \lambda^{4} 
+ d_{12}^{[3]} \lambda^{3} 
+ d_{12}^{[4]} \lambda^{2} + d_{12}^{[5]} \lambda,
+ d_{12}^{[6]},
&
\Qbold{24}{} &= d_{13}^{[2]} \lambda^{4} 
+ d_{13}^{[3]} \lambda^{3} 
+ d_{13}^{[4]} \lambda^{2} + d_{13}^{[5]} \lambda
+ d_{13}^{[6]},
\\
\Qbold{31}{} &= c_{2}^{[1]} \lambda^{5} + c_{2}^{[2]} \lambda^{4}
+c_{2}^{[3]} \lambda^{3} 
+ c_{2}^{[4]} \lambda^{2} + c_{2}^{[5]} \lambda + c_{2}^{[6]},
&
\Qbold{32}{}  &= d_{21}^{[2]} \lambda^{4} 
+ d_{21}^{[3]} \lambda^{3} 
+ d_{21}^{[4]} \lambda^{2} + d_{21}^{[5]} \lambda
+ d_{21}^{[6]},
\\
\Qbold{33}{} &= d_{22}^{[2]} \lambda^{4} 
+ d_{22}^{[3]} \lambda^{3} 
+ d_{22}^{[4]} \lambda^{2} + d_{22}^{[5]} \lambda
+ d_{22}^{[6]},
&
\Qbold{34}{} &= d_{23}^{[2]} \lambda^{4} 
+ d_{23}^{[3]} \lambda^{3} 
+ d_{23}^{[4]} \lambda^{2} + d_{23}^{[5]} \lambda
+ d_{23}^{[6]},
\\
\Qbold{41}{} &= c_{3}^{[1]} \lambda^{5} + c_{3}^{[2]} \lambda^{4}
+c_{3}^{[3]} \lambda^{3} 
+ c_{3}^{[4]} \lambda^{2} + c_{3}^{[5]} \lambda + c_{3}^{[6]},
&
\Qbold{42}{} &= d_{31}^{[2]} \lambda^{4} 
+ d_{31}^{[3]} \lambda^{3} 
+ d_{31}^{[4]} \lambda^{2} + d_{31}^{[5]} \lambda
+ d_{31}^{[6]},
\\
\Qbold{43}{} &= d_{32}^{[2]} \lambda^{4} 
+ d_{32}^{[3]} \lambda^{3} 
+ d_{32}^{[4]} \lambda^{2} + d_{32}^{[5]} \lambda
+ d_{32}^{[6]},
&
\Qbold{44}{} &= d_{33}^{[2]} \lambda^{4} 
+ d_{33}^{[3]} \lambda^{3} 
+ d_{33}^{[4]} \lambda^{2} + d_{33}^{[5]} \lambda
+ d_{33}^{[6]}.
\end{align*}
Throughout the presentation of this paper, we assume that 
$\alpha=\alpha_{1}-\alpha_{2}<0$ and 
$\beta=\beta_{1}-\beta_{2}<0$.
\newline
To find soliton solutions we start with an initial condition 
$(p(x,0),r^{T}(x,0))^{T}$ and evolute in time to reach $(p(x,t),r^{T}(x,t))^{T}$. Taking $p_{i}$ and $r_{i}$ in Schwartz space, they will decay exponentially,  i.e., $p_{i} \rightarrow 0$ and 
$r_{i} \rightarrow 0$  as
$x,t \rightarrow \pm \infty$ 
for $i \in \{1,2,3\}$.
Therefore from the spectral problems (\ref{spatialequ})
and (\ref{temporalequ}), the asymptotic behaviour of the fundamental matrix $\psi$ 
can be written as
\begin{equation}
\psi(x,t) \leadsto e^{i \lambda \Lambda x+i \lambda^{6}\Omega t}.
\end{equation}
Hence, the solution of the spectral problems can be written in the form:
\begin{equation}\label{psiequ}
\psi(x,t) = \phi(x,t) e^{i \lambda \Lambda x+i \lambda^{6}\Omega t}.
\end{equation}
The Jost solution of the eigenfunction (\ref{psiequ}) requires that \cite{Yang2010,DrazinJohnson1989}
\begin{equation}\label{boundary}
\quad \phi(x,t) \rightarrow I_{4},\quad \text{as}\quad x,t \rightarrow \pm \infty,  
\end{equation}
where $I_{4}$ is the $4\times 4$ identity matrix. 
The Lax pair (\ref{spatialequ}) and (\ref{temporalequ}) can be 
rewritten in terms of $\phi$ using equation (\ref{psiequ}), giving the equivalent expression of the spectral problems
\begin{equation}\label{spatialphi}
\phi_{x} = i \lambda [\Lambda,\phi] +i P \phi,
\end{equation}
\begin{equation}\label{temporalphi}
\phi_{t} = i \lambda^{6} [\Omega,\phi] +i Q \phi.
\end{equation}
To construct the Riemann-Hilbert problems and their solutions in the reflectionless case, we are going to use the adjoint scattering equations of the spectral problems $\psi_{x}=iU\psi$
and $\psi_{t}=iV^{[6]}\psi$. Their adjoints are
\begin{equation}\label{psiadjointequation}
\tilde{\psi}_{x} = -i \tilde{\psi} U,
\end{equation}
\begin{equation}\label{psiadjointtemporalequation}
\tilde{\psi}_{t} = -i \tilde{\psi} V^{[6]},
\end{equation}
and the equivalent spectral adjoint equations read
\begin{equation}\label{adjointspatialphi}
\tilde{\phi}_{x} = -i \lambda [\tilde{\phi},\Lambda]-i\tilde{\phi} P,
\end{equation}
\begin{equation}\label{adjointtemporalphi}
\tilde{\phi}_{t} = -i \lambda^{6} [\tilde{\phi},\Omega]-i\tilde{\phi} Q.
\end{equation}
Because $tr(iP)=0$ and $tr(iQ)=0$, using Liouvilles's formula \cite{Yang2010}, it is easy to see that the $(det(\phi))_{x}=0$, 
that is, $det(\phi)$ is a constant, and utilizing the boundary condition (\ref{boundary}), we conclude
\begin{equation}
det(\phi)=1,
\end{equation}
hence the Jost matrix $\phi$ is invertible.
\newline
Furthermore, as $\phi^{-1}_{x}=-\phi^{-1} \phi_{x} \phi^{-1}$, 
we can derive from (\ref{spatialphi}),
\begin{equation}\label{adjointspatialphiinv}
\phi^{-1}_{x} = -i \lambda [\phi^{-1},\Lambda]-i\phi^{-1} P.
\end{equation}
Thus, we can see that both $(\phi^{+})^{-1}$
and $(\phi^{-})^{-1}$
satisfies the spatial adjoint equation (\ref{adjointspatialphi}). 
We can also show that both satisfies
the temporal adjoint equation (\ref{adjointtemporalphi}) as well.
\par
Notice that if the eigenfunction $\phi(x,t,\lambda)$ is a solution of the spectral problem (\ref{spatialphi}), then 
$\phi^{-1}(x,t,\lambda)$ is a solution of the adjoint spectral problem (\ref{adjointspatialphi}), implying that
$C \phi^{-1}(x,t,\lambda)$
is also a solution of (\ref{adjointspatialphi}) with the same eigenvalue because
$\phi_{x}^{-1}=- \phi^{-1} \phi_{x} \phi^{-1}$. 
In a similar way, the nonlocal
$\phi^{T}(x,-t,-\lambda) C$ is also a solution of the spectral adjoint problem (\ref{adjointspatialphi}). 
Since the boundary condition is the same for both solutions as $x \rightarrow \pm \infty$,
this guarantees the uniqueness of the solution, so
\begin{equation}\label{phiTphiminus}
\phi^{T}(x,-t,-\lambda) = C \phi^{-1}(x,t,\lambda) C^{-1}.
\end{equation}
As a result, if $\lambda$
is an eigenvalue of the spectral problems, then $-\lambda$ is also an eigenvalue and
the relation (\ref{phiTphiminus}) holds. 
\newline
Now, we are going to work with the spatial spectral problem (\ref{spatialphi}), assuming that the time is $t=0$.
\newline
For notation simplicity, we denote 
$Y^{+}$ and $Y^{-}$ to indicate the boundary conditions are set as $x \rightarrow \infty$
and $x \rightarrow -\infty$, respectively. 
\newline
We know that
\begin{equation}\label{phiboundary}
\phi^{\pm} \rightarrow I_{4} \quad \text{when} \quad x \rightarrow \pm \infty.
\end{equation}
From (\ref{psiequ}), this allows us to write
\begin{equation}\label{psiphirelation}
\psi^{\pm} = \phi^{\pm} e^{i \lambda \Lambda x}.
\end{equation}
Both $\psi^{+}$ and $\psi^{-}$ satisfy 
the spectral spatial differential equation (\ref{spatialequ}),
i.e. both are two solutions of that equation. 
Thus, they are linearly dependent, hence there exists
a scattering matrix $S(\lambda)$, such that
\begin{equation}\label{psiminusplus}
\psi^{-} = \psi^{+} S(\lambda).
\end{equation}
Substituting (\ref{psiphirelation}) into (\ref{psiminusplus}),
leads to
\begin{equation}\label{psiplusminus}
\phi^{-} = \phi^{+} e^{i \lambda \Lambda x} S(\lambda) e^{-i \lambda \Lambda x}, \quad \text{for}
\quad \lambda \in \mathbb{R}, 
\end{equation}
where 
\begin{equation}
S(\lambda)=(s_{ij})_{4 \times 4}=
\begin{pmatrix}
s_{11} & s_{12} & s_{13} & s_{14} \\
s_{21} & s_{22} & s_{23} & s_{24} \\
s_{31} & s_{32} & s_{33} & s_{34} \\
s_{41} & s_{42} & s_{43} & s_{44} 
\end{pmatrix}.
\end{equation}
Given that $det(\phi^{\pm})=1$, we obtain
\begin{equation}
det(S(\lambda))=1.
\end{equation}
In addition, we can show from
(\ref{psiplusminus}) and (\ref{phiTphiminus})
that $S(\lambda)$ possess the involution relation
\begin{equation}\label{Sequ}
S^{T}(-\lambda) = C S^{-1}(\lambda) C^{-1}.
\end{equation}
we deduce from (\ref{Sequ}) that
\begin{equation}\label{s11hats11relation}
\hat{s}_{11}({\lambda})=s_{11}(-\lambda),
\end{equation}
where the inverse scattering data matrix 
$S^{-1}=(\hat{s}_{ij})_{4 \times 4}$ for $i,j \in \{1,2,3,4\}$. 
\\[3mm] 
From $\phi^{-} = \phi^{+} e^{i \lambda \Lambda x} S(\lambda) e^{-i \lambda \Lambda x}$,
$\phi^{\pm} \rightarrow I_{4}$ when 
$x \rightarrow \pm \infty$. In order to formulate Riemann-Hilbert problems we need to analyse the analyticity of the Jost matrix $\phi^{\pm}$. 
\newline
To do so, we can use the Volterra integral equations to write the solutions $\phi^{\pm}$ in a uniquely manner by using the spatial spectral problem (\ref{spatialequ}):
\begin{align}\label{Volt1}
\phi^{-}(x,\lambda) &= I_{4} + i \int\limits^{x}_{- \infty} 
e^{i \lambda \Lambda (x-y)} P(y) \phi^{-}(y,\lambda) 
e^{i \lambda \Lambda (y-x)} dy,
\\
\phi^{+}(x,\lambda) &= I_{4} - i \int\limits^{+ \infty}_{x} 
e^{i \lambda \Lambda (x-y)} P(y) \phi^{+}(y,\lambda) 
e^{i \lambda \Lambda (y-x)} dy.
\end{align} 
We denote the matrix $\phi^{-}$ to be
\begin{equation}
\phi^{-}=
\begin{pmatrix}
\phi^{-}_{11} & \phi^{-}_{12} & \phi^{-}_{13} & \phi^{-}_{14} \\
\phi^{-}_{21} & \phi^{-}_{22} & \phi^{-}_{23} & \phi^{-}_{24} \\
\phi^{-}_{31} & \phi^{-}_{32} & \phi^{-}_{33} & \phi^{-}_{34} \\
\phi^{-}_{41} & \phi^{-}_{42} & \phi^{-}_{43} & \phi^{-}_{44} 
\end{pmatrix}.
\end{equation}
and $\phi^{+}$ is denoted similarly.
So from (\ref{Volt1}) the components of the first column of $\phi^{-}$ are
\begin{align}
\phi^{-}_{11} &= 1 + i \int_{- \infty}^{x} 
(p_{1}(y) \phi^{-}_{21}(y,\lambda)+p_{2}(y) \phi^{-}_{31}(y,\lambda)
+p_{3}(y) \phi^{-}_{41}(y,\lambda)) dy,
\\
\phi^{-}_{21} &= i \int^{x}_{- \infty} 
r_{1}(y) \phi^{-}_{11}(y,\lambda) e^{-i \lambda \alpha (x-y)} dy,
\\
\phi^{-}_{31} &= i \int^{x}_{- \infty} 
r_{2}(y) \phi^{-}_{11}(y,\lambda) e^{-i \lambda \alpha (x-y)} dy,
\\
\phi^{-}_{41} &= i \int^{x}_{- \infty} 
r_{3}(y) \phi^{-}_{11}(y,\lambda) e^{-i \lambda \alpha (x-y)} dy.
\end{align}
\\
Similarly, the components of the second column of $\phi^{-}$ are
\begin{align}
\phi^{-}_{12} &= i \int_{- \infty}^{x} 
\bigg( p_{1}(y) \phi^{-}_{22}(y,\lambda)+p_{2}(y) \phi^{-}_{32}(y,\lambda)
+p_{3}(y) \phi^{-}_{42}(y,\lambda) \bigg) e^{i \lambda \alpha (x-y)} dy,
\\
\phi^{-}_{22} &=1 + i \int^{x}_{- \infty} 
r_{1}(y) \phi^{-}_{12}(y,\lambda) dy , 
\\
\phi^{-}_{32} &= i \int^{x}_{- \infty} 
r_{2}(y) \phi^{-}_{12}(y,\lambda) dy ,
\\
\phi^{-}_{42} &= i \int^{x}_{- \infty} 
r_{3}(y) \phi^{-}_{12}(y,\lambda) dy,
\end{align}
and the components of the third column of $\phi^{-}$ are
\begin{align}
\phi^{-}_{13} &= i \int_{- \infty}^{x} 
\bigg( p_{1}(y) \phi^{-}_{23}(y,\lambda)+p_{2}(y) \phi^{-}_{33}(y,\lambda)
+p_{3}(y) \phi^{-}_{43}(y,\lambda) \bigg) e^{i \lambda \alpha (x-y)} dy , 
\\
\phi^{-}_{23} &= i \int^{x}_{- \infty} 
r_{1}(y) \phi^{-}_{13}(y,\lambda) dy , 
\\
\phi^{-}_{33} &= 1 + i \int^{x}_{- \infty} 
r_{2}(y) \phi^{-}_{13}(y,\lambda) dy , 
\\
\phi^{-}_{43} &= i \int^{x}_{- \infty} 
r_{3}(y) \phi^{-}_{13}(y,\lambda) dy,
\end{align}
and finally the components of the fourth column of $\phi^{-}$ are
\begin{align}
\phi^{-}_{14} &= i \int_{- \infty}^{x} 
\bigg( p_{1}(y) \phi^{-}_{24}(y,\lambda)+p_{2}(y) \phi^{-}_{34}(y,\lambda)
+p_{3}(y) \phi^{-}_{44}(y,\lambda) \bigg) e^{i \lambda \alpha (x-y)} dy , 
\\
\phi^{-}_{24} &= i \int^{x}_{- \infty} 
r_{1}(y) \phi^{-}_{14}(y,\lambda) dy  ,
\\
\phi^{-}_{34} &= i \int^{x}_{- \infty} 
r_{2}(y) \phi^{-}_{14}(y,\lambda) dy  ,
\\
\phi^{-}_{44} &= 1 + i \int^{x}_{- \infty} 
r_{3}(y) \phi^{-}_{14}(y,\lambda) dy.
\end{align}
Recall that $\alpha < 0$. If $Im(\lambda)>0$  
and $y<x$ then, $Re(e^{-i \lambda \alpha (x-y)})$ decays exponentially and so each integral of the first column of $\phi^{-}$ converges.
As a result, the components of the first column of $\phi^{-}$, are analytic in the upper half complex plane for 
$\lambda \in \mathbb{C}_{+}$, and 
continuous for $\lambda \in \mathbb{C}_{+} \cup \mathbb{R}$. 
\newline
In the same way, for $y > x$, the components of the last three 
columns of $\phi^{+}$ are analytic in the upper half plane for 
$\lambda \in \mathbb{C}_{+}$ and continuous for 
$\lambda \in \mathbb{C}_{+} \cup \mathbb{R}$.
\\[3mm]
It is worth mentioning the case when $Im(\lambda) < 0$, then
the first column $\phi^{+}$ is analytic in the lower half plane for $\lambda \in \mathbb{C}_{-}$ and continuous for 
$\lambda \in \mathbb{C}_{-} \cup \mathbb{R}$, and the components of the last three columns of $\phi^{-}$ are analytic in the lower half plane for $\lambda \in \mathbb{C}_{-}$ and continuous for $\lambda \in \mathbb{C}_{-} \cup \mathbb{R}$.
\\[3mm]
Now, let us construct the Riemann-Hilbert problems.
To construct the upper-half plane we
note that \begin{equation}\label{psiphiplusminus}
\phi^{\pm} = \psi^{\pm} e^{-i \lambda \Lambda x}.
\end{equation}
Let $\phi^{\pm}_{j}$ be the $j$th column of $\phi^{\pm}$ for 
$j \in \{1,2,3,4\}$, hence the first Jost matrix solution can be taken as
\begin{equation}\label{Pplusequ}
P^{+}(x,\lambda)=
(\phi_{1}^{-},\phi_{2}^{+},\phi_{3}^{+},\phi_{4}^{+})
=\phi^{-} H_{1} + \phi^{+} H_{2},
\end{equation}
where $H_{1}=diag(1,0,0,0)$ and $H_{2}=diag(0,1,1,1)$.
\\[3mm]
Therefore, $P^{+}$ is then analytic for $\lambda \in \mathbb{C}_{+}$ and continuous 
for $\lambda \in \mathbb{C}_{+} \cup \mathbb{R}$.
\\[3mm]
For the lower-half plane, we can construct $P^{-} \in \mathbb{C}_{-}$ which is the analytic counterpart of $P^{+} \in \mathbb{C}_{+}$.
To do so, we utilize the equivalent spectral adjoint equation (\ref{adjointspatialphiinv}).
Because $\tilde{\phi}^{\pm}=(\phi^{\pm})^{-1}$ and  
$\psi^{\pm} = \phi^{\pm} e^{i \lambda \Lambda x}$, we have
\begin{equation}
(\phi^{\pm})^{-1} = e^{i \lambda \Lambda x} (\psi^{\pm})^{-1}.
\end{equation}
Let $\tilde{\phi}_{j}^{\pm}$ be the $j$th row of $\tilde{\phi}^{\pm}$
for $j \in \{1,2,3,4\}$. As above, we can get
\begin{equation}\label{Pminusequ}
P^{-}(x,\lambda)=\bigg( \tilde{\phi}_{1}^{-},\tilde{\phi}_{2}^{+},\tilde{\phi}_{3}^{+},\tilde{\phi}_{4}^{+} 
\bigg)^{T}=
H_{1}(\phi^{-})^{-1}+H_{2}(\phi^{+})^{-1}.
\end{equation}
Hence, $P^{-}$ is analytic for $\lambda \in \mathbb{C}_{-}$ and continuous 
for $\lambda \in \mathbb{C}_{-} \cup \mathbb{R}$. 
\\[3mm]
Since both $\phi^{-}$ and $\phi^{+}$ satisfy
\begin{equation}\label{phiTinvequation}
\phi^{T}(x,-t,-\lambda) = C \phi^{-1}(x,t,\lambda) C^{-1},  
\end{equation}
using (\ref{Pplusequ}), we have
\begin{equation}
P^{+}(x,-t,-\lambda) = \phi^{-}(x,-t,-\lambda) H_{1}
+ \phi^{+}(x,-t,-\lambda) H_{2}
\end{equation}
or equivalently
\begin{equation}\label{Pplustranspose}
(P^{+})^{T}(x,-t,-\lambda) = H_{1}^{T} (\phi^{-})^{T}(x,-t,-\lambda) 
+ H_{2}^{T} (\phi^{+})^{T}(x,-t,-\lambda) 
\end{equation}
substituting (\ref{phiTinvequation}) in (\ref{Pplustranspose}) we have the nonlocal involution property
\begin{equation}\label{PplusPminusrelation}
(P^{+})^{T}(x,-t,-\lambda) = C P^{-} (x,t,\lambda) C^{-1}.
\end{equation}
Employing analyticity of both $P^{+}$ and $P^{-}$, we can construct the Riemann-Hilbert problems 
\begin{equation}
P^{-}P^{+}=J,    
\end{equation}
where $J=e^{i \lambda \Lambda x} (H_{1}+H_{2}S)(H_{1}+S^{-1}H_{2}) e^{-i \lambda \Lambda x}$ for\quad $\lambda \in \mathbb{R}$, and
\\
$S^{-1}=(\hat{s}_{ij})_{4 \times 4}$ for $i,j \in \{1,2,3,4\}$ is the inverse scattering data matrix.

Replacing (\ref{psiplusminus}) in (\ref{Pplusequ}), we have
\begin{equation}\label{Pplussimplified}
P^{+}(x,\lambda) = \phi^{+} (e^{i \lambda \Lambda x} S 
e^{-i \lambda \Lambda x} H_{1} +H_{2}).
\end{equation}
Because $\phi^{+}(x,\lambda) \rightarrow I_{4}$ when $x \rightarrow + \infty$, we get
\begin{equation}
\lim_{x \rightarrow + \infty} P^{+} = 
\begin{pmatrix}
s_{11}(\lambda) & 0 & 0 & 0 \\
0 & 1 & 0 & 0 \\
0 & 0 & 1 & 0 \\
0 & 0 & 0 & 1 
\end{pmatrix},
\quad
\text{for}
\quad \lambda \in \mathbb{C}_{+} \cup \mathbb{R}.
\end{equation}
In the same way, 
\begin{equation}
\lim_{x \rightarrow - \infty} P^{-} = 
\begin{pmatrix}
\hat{s}_{11}(\lambda) & 0 & 0 & 0 \\
0 & 1 & 0 & 0 \\
0 & 0 & 1 & 0 \\
0 & 0 & 0 & 1 
\end{pmatrix},
\quad
\text{for}
\quad \lambda \in \mathbb{C}_{-} \cup \mathbb{R}.
\end{equation}
Thus if we choose
\begin{equation}
\begin{array}{cc}\label{GplusPplusrelation}
G^{+}(x,\lambda) = P^{+}(x,\lambda) 
\begin{pmatrix}
s_{11}^{-1} (\lambda) & 0 & 0 & 0
\\
0 & 1 & 0 & 0
\\
0 & 0 & 1 & 0
\\
0 & 0 & 0 & 1
\end{pmatrix}
\quad
\text{and}
\quad
(G^{-})^{-1}(x,\lambda) = 
\begin{pmatrix}
\hat{s}_{11}^{-1} (\lambda) & 0 & 0 & 0
\\
0 & 1 & 0 & 0
\\
0 & 0 & 1 & 0
\\
0 & 0 & 0 & 1
\end{pmatrix}
P^{-}(x,\lambda) 
\end{array},
\end{equation}
the two generalized matrices
$G^{+}(x,\lambda)$ and $G^{-}(x,\lambda)$ generate the matrix Riemann-Hilbert problems on the real line for the 
six-component AKNS system of sixth-order given by
\begin{equation}
G^{+}(x,\lambda) = G^{-}(x,\lambda) G_0(x,\lambda),
\quad
\text{for}
\quad \lambda \in \mathbb{R},
\end{equation}
where the jump matrix $G_0(x,\lambda)$ can be cast as
\begin{equation}\label{Gequ}
G_0(x,\lambda) = 
\begin{pmatrix}
\hat{s}_{11}^{-1} (\lambda) & 0 & 0 & 0
\\
0 & 1 & 0 & 0
\\
0 & 0 & 1 & 0
\\
0 & 0 & 0 & 1
\end{pmatrix}
J
\begin{pmatrix}
s_{11}^{-1} (\lambda) & 0 & 0 & 0
\\
0 & 1 & 0 & 0
\\
0 & 0 & 1 & 0
\\
0 & 0 & 0 & 1
\end{pmatrix},
\end{equation}
this reads
\begin{equation}
G_0^{}(x,\lambda) =
\begin{pmatrix}
s_{11}^{-1} \hat{s}_{11}^{-1} 
& 
\hat{s}_{12}\hat{s}_{11}^{-1} e^{i \lambda \alpha x}
& 
\hat{s}_{13}\hat{s}_{11}^{-1} e^{i \lambda \alpha x}
& 
\hat{s}_{14}\hat{s}_{11}^{-1} e^{i \lambda \alpha x}
\\[2mm]
s_{21} s_{11}^{-1} e^{-i \lambda \alpha x} 
& 
1
& 
0
& 
0
\\[4mm]
s_{31} s_{11}^{-1} e^{-i \lambda \alpha x}
& 
0
& 
1
&
0
\\[4mm]
s_{41} s_{11}^{-1} e^{-i \lambda \alpha x}
& 
0 
& 
0
& 
1
\end{pmatrix},
\end{equation}
and its canonical normalization conditions:
\begin{align}
G^{+}(x,\lambda) \rightarrow I_{4} \quad \text{as} \quad \lambda \in \mathbb{C}_{+} \cup \mathbb{R} \rightarrow \infty,
\\
G^{-}(x,\lambda) \rightarrow I_{4} \quad \text{as} \quad \lambda \in \mathbb{C}_{-} \cup \mathbb{R} \rightarrow \infty.
\end{align}
From (\ref{PplusPminusrelation}) along with (\ref{GplusPplusrelation}) and using (\ref{s11hats11relation}), we deduce the nonlocal involution property
\begin{equation}
(G^{+})^{T}(x,-t,-\lambda) = C (G^{-})^{-1} (x,t,\lambda) C^{-1}.
\end{equation}
Furthermore, we derive the following nonlocal involution property for $G_0$
\begin{equation}
G_0^{T}(x,-t,-\lambda) = C G_0(x,t,\lambda) C^{-1},
\end{equation}
from (\ref{Gequ}) and (\ref{s11hats11relation}).

\subsection{Time evolution of the scattering data}
Reaching this point, we need to determine the scattering data
as they evolute in time. In order to do that , we differentiate
equation (\ref{psiplusminus}) with respect to 
time $t$ and applying (\ref{temporalphi}) gives 
\begin{equation}
S_{t} = i \lambda^{6} [\Omega,S],
\end{equation}
and thus
\begin{equation}
S_{t}=
\begin{pmatrix}
0 & i \beta \lambda^{6} s_{12} & 
i \beta \lambda^{6} s_{13} & i \beta \lambda^{6} s_{14} 
\\
-i \beta \lambda^{6} s_{21} & 0 & 
0 & 0
\\
-i \beta \lambda^{6} s_{31} & 0 & 
0 & 0
\\
-i \beta \lambda^{6} s_{41} & 0 & 
0 & 0
\end{pmatrix}.
\end{equation}
As a result, we have
\begin{equation}
\begin{cases}
s_{12}(t,\lambda) = s_{12}(0,\lambda) e^{i \beta \lambda^{6} t},
\\
s_{13}(t,\lambda) = s_{13}(0,\lambda) e^{i \beta \lambda^{6} t} ,
\\
s_{14}(t,\lambda) = s_{14}(0,\lambda) e^{i \beta \lambda^{6} t} ,
\\
s_{21}(t,\lambda) = s_{21}(0,\lambda) e^{-i \beta \lambda^{6} t} ,
\\
s_{31}(t,\lambda) = s_{31}(0,\lambda) e^{-i \beta \lambda^{6} t} , 
\\
s_{41}(t,\lambda) = s_{41}(0,\lambda) e^{-i \beta \lambda^{6} t} ,
\end{cases}
\end{equation}
and $s_{11},s_{22},s_{23},s_{24},s_{32},s_{33},s_{34},s_{42},s_{43},s_{44}$ are constants.

\section{Soliton solutions}
\subsection{General case}
The determinant of the matrix $G^{\pm}$ determines the type of soliton solutions generated using the Riemann-Hilbert problems. In the regular case, when
$det(G^{\pm}) \neq 0$, we obtain unique soliton solution. In the non-regular case, that is to say when $det(G^{\pm})=0$, it could generate discrete eigenvalues in the spectral plane. This non-regular case can be transformed into the regular case to solve for soliton solutions \cite{Yang2010}.

From (\ref{Pplussimplified}) and $det(\phi^{\pm})=1$, we can show that
\begin{equation}\label{ppluss11}
det(P^{+}(x,\lambda))=s_{11}(\lambda),
\end{equation}
in the same way,
\begin{equation}\label{pminuss11hat}
det(P^{-}(x,\lambda))=\hat{s}_{11}(\lambda).
\end{equation}
Because $det(S(\lambda))=1$, this implies that
$S^{-1}(\lambda)=\bigg( cof(S(\lambda)) \bigg)^{T}$ and
\begin{equation}
\hat{s}_{11}=
\begin{vmatrix}
s_{22} & s_{23} & s_{24} \\
s_{32} & s_{33} & s_{34} \\
s_{42} & s_{43} & s_{44}
\end{vmatrix},
\end{equation}
which should be zero for the non-regular case.
\\[3mm]
To give rise to soliton solutions, we need the solutions of
$det(P^{+}(x,\lambda))=det(P^{-}(x,\lambda))=0$ to be simple.
When
$det(P^{+}(x,\lambda))=s_{11}(\lambda)=0$, we assume
$s_{11}(\lambda)$ has simple zeros with
discrete eigenvalues $\lambda_{k} \in \mathbb{C}_{+}$ for 
$k \in \{1,2,...,N\}$, while for $det(P^{-}(x,\lambda))=\hat{s}_{11}(\lambda)=0$, 
we assume
$\hat{s}_{11}(\lambda)$ has simple zeros with
discrete eigenvalues $\hat{\lambda}_{k} \in \mathbb{C}_{-}$ for 
$k \in \{1,2,...,N\}$, which are the poles of the transmission coefficients \cite{DrazinJohnson1989}.
\\[3mm]
From  
$\hat{s}_{11}(\lambda)=s_{11}(-\lambda)$
and $det(P^{\pm}(x,\lambda))=0$, we have the
nonlocal involution relation
\begin{equation}\label{lambdahatlambdarelation}
\hat{\lambda}=- \lambda.
\end{equation}
Each $Ker(P^{+}(x,\lambda_{k}))$
contains only a single column vector $v_{k}$, similarly each $Ker(P^{-}(x,\hat{\lambda}_{k}))$ contains only a single row vector $\hat{v}_{k}$ such that:
\begin{equation}\label{Pplus}
P^{+}(x,\lambda_{k}) v_{k}=0 \quad \text{for} \quad k \in \{1,2,...,N\},
\end{equation}
and
\begin{equation}\label{Pminus}
\hat{v}_{k} P^{-}(x,\hat{\lambda}_{k})=0 \quad \text{for} \quad k \in \{1,2,...,N\}.
\end{equation}
To obtain explicit soliton solutions, we take $G_0=I_{4}$ in the
Riemann-Hilbert problems. This will force the reflection coefficients 
$s_{21}=s_{31}=s_{41}=0$ and $\hat{s}_{12}=\hat{s}_{13}=\hat{s}_{14}=0$.
\\[2mm]
In that case, the Riemann-Hilbert problems can be presented as follows \cite{MaAugust2020}:
\begin{equation}\label{Pplussum}
G^{+}(x,\lambda)=I_{4}-\sum\limits_{k,j=1}^{N} 
\frac{v_{k}(M^{-1})_{kj}\hat{v}_{j}}{\lambda-\hat{\lambda}_{j}},
\end{equation}
and
\begin{equation}\label{Pminussum}
(G^{-})^{-1}(x,\lambda)=I_{4}+\sum\limits_{k,j=1}^{N} 
\frac{v_{k}(M^{-1})_{kj}\hat{v}_{j}}{\lambda-\lambda_{k}},
\end{equation}
where $M=(m_{kj})_{N \times N}$ is a matrix defined by
\begin{equation}
m_{kj}=
\begin{cases}
\frac{\hat{v}_{k} v_{j}}{\lambda_{j}-\hat{\lambda}_{k}}
& \text{if} \quad
\lambda_{j} \neq \hat{\lambda}_{k}
\\
0 & \text{if} \quad
\lambda_{j} = \hat{\lambda}_{k}
\end{cases}
\quad
k,j \in \{1,2,...,N\}.
\end{equation}

Since the zeros $\lambda_{k}$ and $\hat{\lambda}_{k}$ are constants, they are independent of space and time.
We can explore the spatial and temporal evolution of the scattering vectors $v_{k}(x,t)$ and $\hat{v}_{k}(x,t)$
for $1\leq k \leq N$.
\\[3mm]
Taking the $x$-derivative of both sides of the equation
\begin{equation}
P^{+}(x,\lambda_{k})v_{k}=0, \quad 1\leq k \leq N
\end{equation}
and knowing that $P^{+}$ satisfies the spectral
spatial equivalent equation (\ref{spatialphi}), along with
(\ref{Pplus}), we obtain
\begin{equation}
P^{+}(x,\lambda_{k}) 
\Bigg( \frac{dv_{k}}{dx}-i\lambda_{k} \Lambda v_{k} \Bigg)=0
\quad
\text{for}
\quad
k,j \in \{1,2,...,N\}.
\end{equation}
In a similar manner, taking the $t$-derivative and using the temporal equation (\ref{temporalphi}) and (\ref{Pplus}),
we acquire
\begin{equation}
P^{+}(x,\lambda_{k}) 
\Bigg( \frac{dv_{k}}{dt}-i\lambda^{6}_{k} \Omega v_{k} \Bigg)=0
\quad
\text{for}
\quad
k,j \in \{1,2,...,N\}.
\end{equation}
For the adjoint spectral equations (\ref{adjointspatialphi}) and (\ref{adjointtemporalphi})
, we can obtain the following similar results
\begin{equation}
\Bigg( \frac{d\hat{v}_{k}}{dx}+i\hat{\lambda}_{k} \hat{v}_{k} \Lambda \Bigg) P^{-}(x,\hat{\lambda}_{k})=0,
\end{equation}
and
\begin{equation}
\Bigg( \frac{d\hat{v}_{k}}{dt}+i\hat{\lambda}^{6}_{k} \hat{v}_{k} \Omega \Bigg) P^{-}(x,\hat{\lambda}_{k})=0.
\end{equation}
Because $v_{k}$ is a single vector in the kernel of $P^{+}$,
so
$\frac{d v_{k}}{dx}-i\lambda_{k} \Lambda v_{k}$
and 
$\frac{d v_{k}}{dt}-i\lambda^{6}_{k} \Omega v_{k}$
are scalar multiples of $v_{k}$. 
\newline 
Hence without loss of generality, we can take the space dependence of $v_{k}$ to be:
\begin{equation}\label{vkxderivative}
\frac{d v_{k}}{dx}=i\lambda_{k} \Lambda v_{k}, \quad 1\leq k \leq N
\end{equation}
and the time dependence of $v_{k}$ as:
\begin{equation}\label{vktderivative}
\frac{d v_{k}}{dt}=i\lambda^{6}_{k} \Omega v_{k}, \quad 1\leq k \leq N.
\end{equation}
Thus, we can conclude that
\begin{equation}\label{vkequation}
v_{k}(x,t)=e^{i\lambda_{k}\Lambda x + i\lambda^{6}_{k} \Omega t} w_{k}
\quad \text{for} \quad k \in \{1,2,...,N\},
\end{equation}
by solving equations (\ref{vkxderivative}) and (\ref{vktderivative}), where $w_{k}$ is a constant column vector. Likewise, we get
\begin{equation}\label{vkhatequation}
\hat{v}_{k}(x,t)=\hat{w}_{k} e^{-i\hat{\lambda}_{k}\Lambda x - i\hat{\lambda}^{6}_{k} \Omega t} 
\quad \text{for} \quad k \in \{1,2,...,N\},
\end{equation}
where $\hat{w}_{k}$ is a constant row vector.
\\[3mm]
From (\ref{Pplus}) and using the formula (\ref{PplusPminusrelation}),
it is easy to see 
\begin{equation}
v_{k}^{T}(x,-t,-\lambda_{k}) (P^{+})^{T}(x,-t,-\lambda_{k})=
v_{k}^{T}(x,-t,-\lambda_{k}) C P^{-}(x,t,\lambda_{k}) C^{-1} = 0.
\end{equation}
Because $v_{k}^{T}(x,-t,-\lambda_{k}) C P^{-}(x,t,\lambda_{k})$
can be zero and using (\ref{Pminus}), this leads to 
\begin{align}
v_{k}^{T}(x,-t,-\lambda_{k}) C P^{-}(x,t,\lambda_{k}) 
&=
\hat{v}_{k}(x,t,\hat{\lambda}_{k}) P^{-}(x,t,\hat{\lambda}_{k})
\\
&=\hat{v}_{k}(x,t,-\hat{\lambda}_{k}) P^{-}(x,t,-\hat{\lambda}_{k})
=0.
\end{align}
From (\ref{lambdahatlambdarelation}), we have $\hat{\lambda}_{k}=-\lambda_{k}$ for $k \in \{1,2,...,N\}$, then we can take
\begin{equation}
\hat{v}_{k}(x,t,-\hat{\lambda}_{k})=
v_{k}^{T}(x,-t,-\lambda_{k}) C.
\end{equation}
Thus, the involution relations (\ref{vkequation}) and (\ref{vkhatequation}) give
\begin{equation}\label{vknonlocalequ}
v_{k}(x,t)=e^{i\lambda_{k} \Lambda x + i\lambda_{k}^{6} \Omega t} w_{k},
\end{equation}
\begin{equation}\label{vkhatnonlocalequ}
\hat{v}_{k}(x,t)= w^{T}_{k} e^{-i\hat{\lambda}_{k} \Lambda x - i\hat{\lambda}_{k}^{6} \Omega t} C.
\end{equation}
Because the jump matrix $G_{0}=I_{4}$, we can solve the Riemann-Hilbert problem precisely. As a result,
we can determine the potentials by computing the matrix $P^{+}$.
Because $P^{+}$ is analytic, we can expand $G^{+}$ as follows:
\begin{equation}\label{Gexpansion}
G^{+}(x,\lambda)=I_{4}+\frac{1}{\lambda} G^{+}_{1}(x)+O(\frac{1}{\lambda^{2}}),
\quad
\text{when}
\quad
\lambda \rightarrow \infty.
\end{equation}
Because $G^{+}$ satisfies the spectral problem, substituting
it in (\ref{spatialphi}) and matching the coefficients of the same power of $\frac{1}{\lambda}$, at order $O(1)$,
we get
\begin{equation}\label{G1plusequs}
P=-[\Lambda,G^{+}_{1}].
\end{equation}
If we denote 
\begin{equation}
G_{1}^{+}=
\begin{pmatrix}
(G_{1}^{+})_{11} & (G_{1}^{+})_{12} & (G_{1}^{+})_{13} & (G_{1}^{+})_{14}  \\
(G_{1}^{+})_{21} & (G_{1}^{+})_{22} & (G_{1}^{+})_{23} & (G_{1}^{+})_{24} \\
(G_{1}^{+})_{31} & (G_{1}^{+})_{32} & (G_{1}^{+})_{33} & (G_{1}^{+})_{34} \\
(G_{1}^{+})_{41} & (G_{1}^{+})_{42} & (G_{1}^{+})_{43} & (G_{1}^{+})_{44}
\end{pmatrix}
\end{equation}
then
\begin{equation}
P=-[\Lambda,G_{1}^{+}]=
\begin{pmatrix}
0 & -\alpha (G_{1}^{+})_{12} & -\alpha (G_{1}^{+})_{13} & -\alpha (G_{1}^{+})_{14}  \\
\alpha(G_{1}^{+})_{21} & 0 & 0 & 0 \\
\alpha(G_{1}^{+})_{31} & 0 & 0 & 0 \\
\alpha(G_{1}^{+})_{41} & 0 & 0 & 0
\end{pmatrix}.
\end{equation}
Consequently, we can recover the potentials $p_{i}$ and $r_{i}$ for
$i \in \{1,2,3\}$:
\begin{align}\label{p123equations}
p_{1} &= -\alpha (G_{1}^{+})_{12}, \quad r_{1}= \alpha (G_{1}^{+})_{21}, \nonumber
\\
p_{2} &= -\alpha (G_{1}^{+})_{13}, \quad r_{2}= \alpha (G_{1}^{+})_{31},
\\
p_{3} &= -\alpha (G_{1}^{+})_{14}, \quad r_{3}= \alpha (G_{1}^{+})_{41}. \nonumber
\end{align}
It can be seen from (\ref{Gexpansion}) that
\begin{equation}
G_{1}^{+} = \lambda \lim_{\lambda \rightarrow \infty} (G^{+}(x,\lambda)-I_{4}),
\end{equation}
then using equation (\ref{Pplussum}), we deduce
\begin{equation}\label{G1plussummation}
G_{1}^{+} = - \sum\limits_{k,j=1}^{N} v_{k} (M^{-1})_{k,j} \hat{v}_{j}.
\end{equation}
In addition, by the use of equations (\ref{Pequ}) and (\ref{G1plusequs}), we can easily prove the following
nonlocal involution property
\begin{equation}\label{P1plusequ}
(G_{1}^{+})^{T}(x,-t) = C G_{1}^{+}(x,t) C^{-1}.
\end{equation}
By substituting (\ref{G1plussummation}) into (\ref{p123equations})
and using (\ref{vknonlocalequ}) and (\ref{vkhatnonlocalequ}), 
we generate the $N$-soliton solution to the nonlocal reverse-time six-component AKNS system of six-order
\begin{equation}
p_{i} = \alpha \sum\limits_{k,j=1}^{N} v_{k1} (M^{-1})_{kj}
\hat{v}_{j,i+1}
\quad \textrm{for} \quad i \in \{1,2,3\},
\end{equation}
where $w_{k}$ 
is an arbitrary constant column vector in $\mathbb{C}^{4}$, and
\[v_{k}=(v_{k1},v_{k2},v_{k3},...,v_{kn+1})^{T},\ \hat{v}_{k}=(\hat{v}_{k1},\hat{v}_{k2},\hat{v}_{k3},...,\hat{v}_{kn+1}).\]

\section{Exact soliton solutions and dynamics}
\subsection{Explicit one-soliton solution and its dynamics}
\begin{figure}[H]
\begin{center}
\includegraphics[width=0.45\textwidth,height=4.5cm]{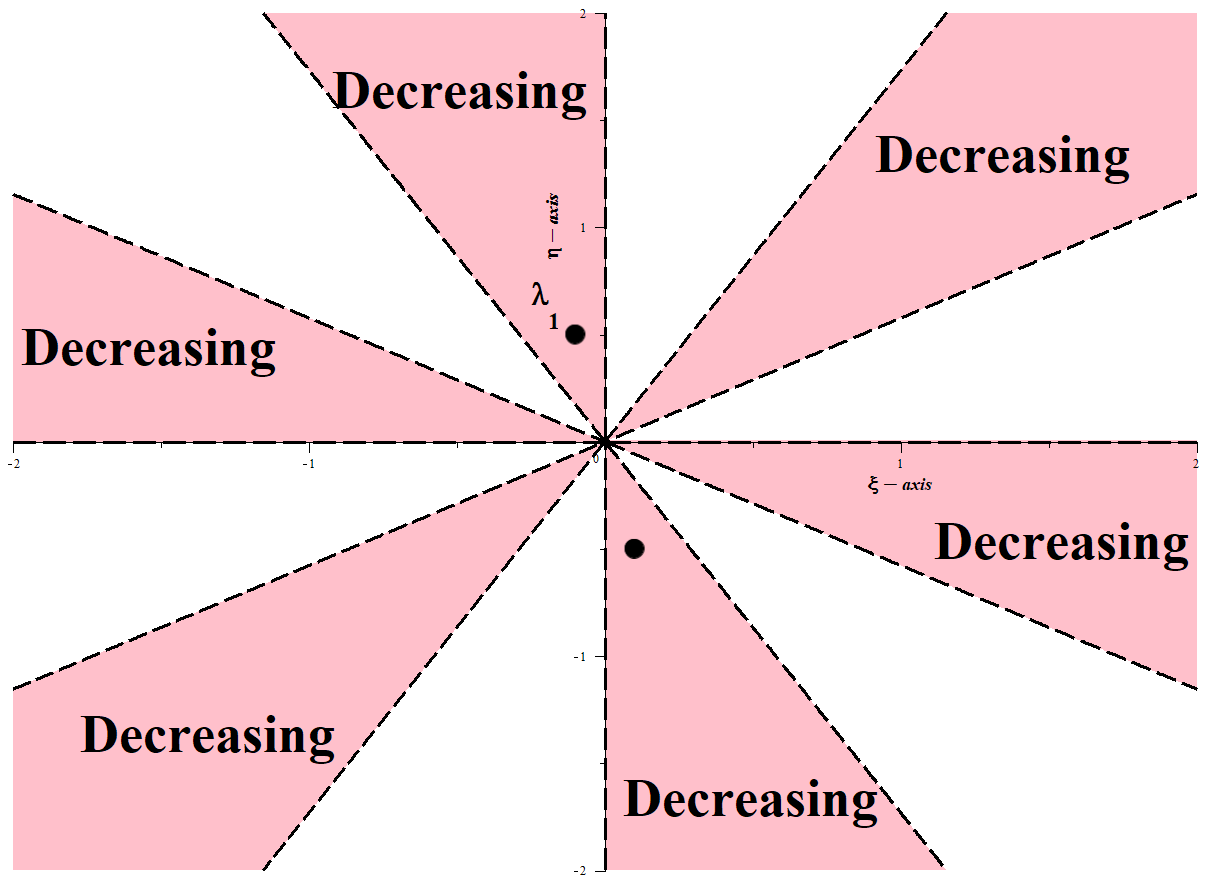}%
\includegraphics[width=0.45\textwidth,height=4.5cm]{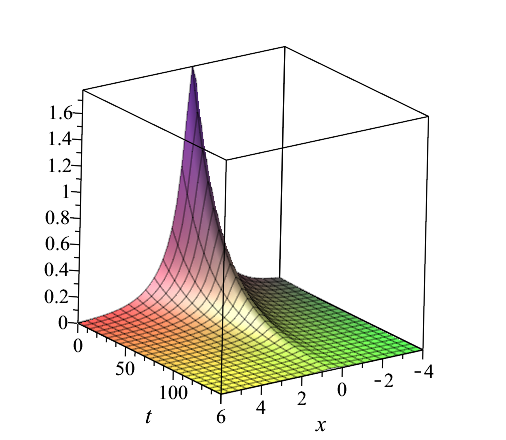}%
\\
\includegraphics[width=0.45\textwidth,height=4.5cm]{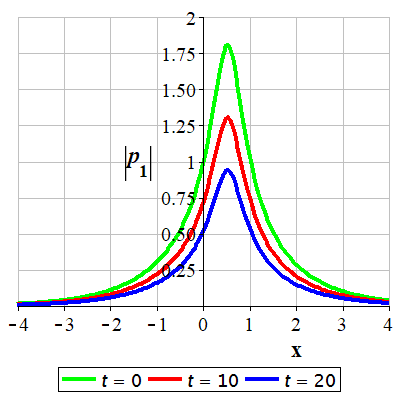}%
\includegraphics[width=0.45\textwidth,height=4.5cm]{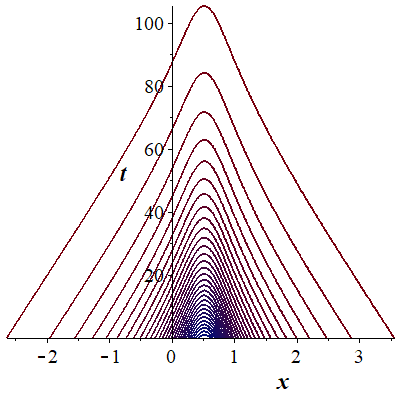}%
\caption{Spectral plane along with 3D, 2D and contour plots of $|p_{1}|$ in the focussing case of the one-soliton with parameters  $(\rho_{1},\rho_{2},\rho_{3},\alpha_{1},\alpha_{2},\beta_{1},\beta_{2})=(-1,-2,-1,-1,1,-1,1)$, $\lambda_{1}=-0.1+0.5i$, $w_{1}=(1,i,2+i,1)$.}%
\label{1solitonplot1}%
\end{center}
\begin{center}
\includegraphics[width=0.45\textwidth,height=4.5cm]{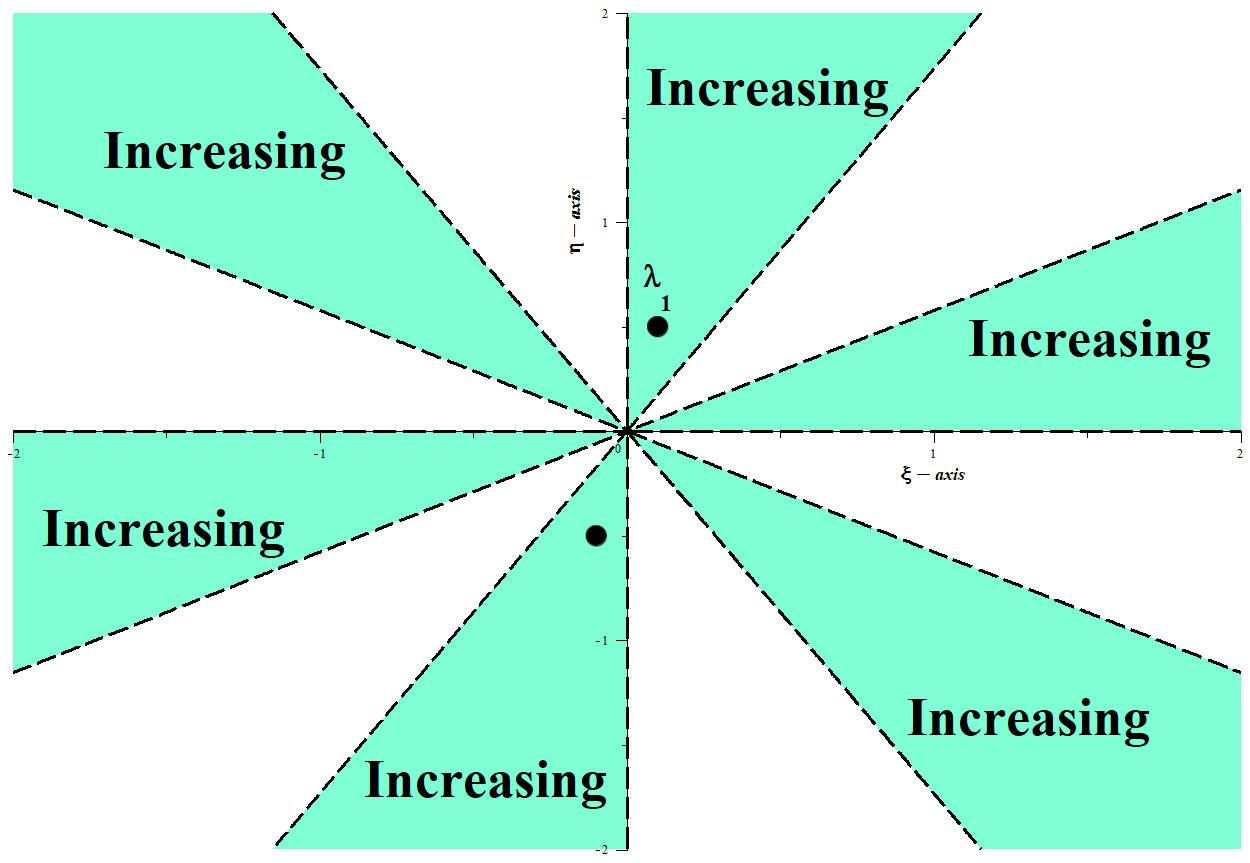}%
\includegraphics[width=0.45\textwidth,height=4.5cm]{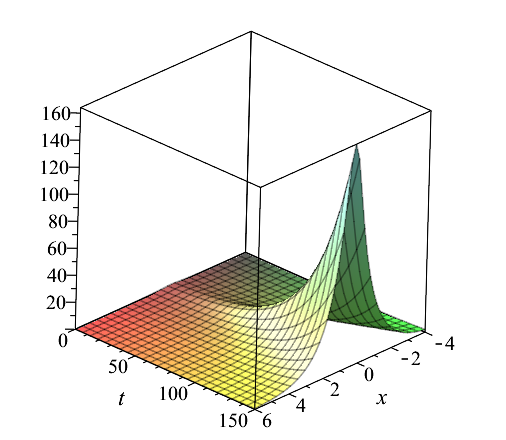}%
\\
\includegraphics[width=0.45\textwidth,height=4.5cm]{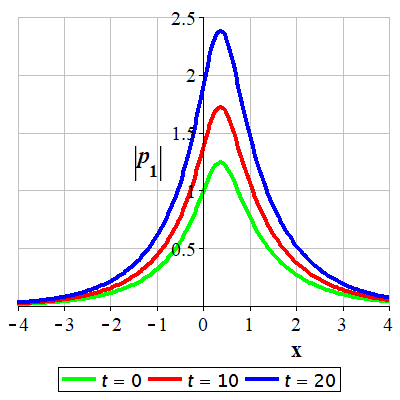}%
\includegraphics[width=0.45\textwidth,height=4.5cm]{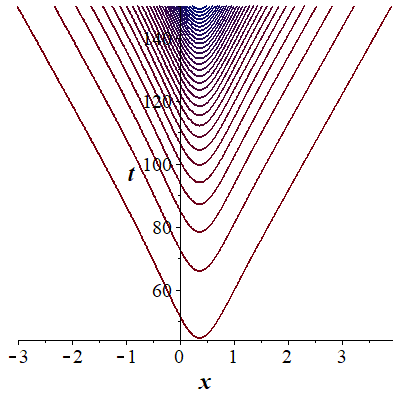}%
\caption{Spectral plane along with 3D, 2D  and contour plots of $|p_{1}|$ in the focussing case of the one-soliton with parameters $(\rho_{1},\rho_{2},\rho_{3},\alpha_{1},\alpha_{2},\beta_{1},\beta_{2})=(-1,-2,-1,-1,1,-1,1)$, $\lambda_{1}=0.1+0.5i$, $w_{1}=(1,i,2+i,1)$.}%
\label{1solitonplot2}%
\end{center}
\end{figure}

A general explicit solution for a single soliton in the reverse-time case
when $N=1$, 
$w_{1}=(w_{11},w_{12},w_{13},w_{14})^{T}$, $\lambda_{1} \in \mathbb{C}$ is arbitrary,
and $\hat{\lambda}_{1}=-\lambda_{1}$ is given by
\begin{align}\label{p1general}
p_{1}(x,t)= \frac{
2 \rho_{2} \rho_{3} \lambda_{1} (\alpha_{1}-\alpha_{2}) w_{11} w_{12}
e^{i \lambda_{1} (\alpha_{1}+\alpha_{2})x+i\lambda_{1}^{6}(\beta_{1}-\beta_{2})t}}{\rho_{1} \rho_{2} \rho_{3} w_{11}^{2} e^{2i\lambda_{1} \alpha_{1}x}
+(\rho_{2} \rho_{3} w_{12}^{2}+ \rho_{1} \rho_{3} w_{13}^{2}+ \rho_{1} \rho_{2} w_{14}^{2}) e^{2i\lambda_{1}\alpha_{2}x}},
\\
p_{2}(x,t)= \frac{
2 \rho_{1} \rho_{3} \lambda_{1} (\alpha_{1}-\alpha_{2}) w_{11} w_{13}
e^{i \lambda_{1} (\alpha_{1}+\alpha_{2})x+i\lambda_{1}^{6}(\beta_{1}-\beta_{2})t}}{\rho_{1} \rho_{2} \rho_{3} w_{11}^{2} e^{2i\lambda_{1} \alpha_{1}x}
+(\rho_{2} \rho_{3} w_{12}^{2}+ \rho_{1} \rho_{3} w_{13}^{2}+ \rho_{1} \rho_{2} w_{14}^{2}) e^{2i\lambda_{1}\alpha_{2}x}},
\\
p_{3}(x,t)= \frac{
2 \rho_{1} \rho_{2} \lambda_{1} (\alpha_{1}-\alpha_{2}) w_{11} w_{14}
e^{i \lambda_{1} (\alpha_{1}+\alpha_{2})x+i\lambda_{1}^{6}(\beta_{1}-\beta_{2})t}}{\rho_{1} \rho_{2} \rho_{3} w_{11}^{2} e^{2i\lambda_{1} \alpha_{1}x}
+(\rho_{2} \rho_{3} w_{12}^{2}+ \rho_{1} \rho_{3} w_{13}^{2}+ \rho_{1} \rho_{2} w_{14}^{2}) e^{2i\lambda_{1}\alpha_{2}x}}.
\end{align}
We can get the amplitude of $p_{1}$:
\begin{equation}\label{p1magnitude}
|p_{1}(x,t)|= 2 A e^{-\beta t Im(\lambda_{1}^{6})} 
\end{equation}
where
\begin{equation}
A= 
\Bigg| \frac{
2 \lambda_{1} \rho_{2} \rho_{3} (\alpha_{1}-\alpha_{2}) w_{11} w_{12}e^{-Im(\lambda_{1} (\alpha_{1}+\alpha_{2})x)}}{\rho_{1} \rho_{2} \rho_{3} w_{11}^{2} e^{2i\lambda_{1} \alpha_{1}x}
+(\rho_{2} \rho_{3} w_{12}^{2}+ \rho_{1} \rho_{3} w_{13}^{2}+ \rho_{1} \rho_{2} w_{14}^{2}) e^{2i\lambda_{1}\alpha_{2}x}} \Bigg|.
\end{equation}
We can see from $p_{1}$, since the real part of the phase is zero, i.e
$Re(\i 2 \lambda_{1} \alpha x)=0$, then the phase velocity is zero.
Hence the one-soliton is not a travelling wave, and it is
stationary in space.
\\[3mm]
Fixing $x=x_{0}$, the amplitude is 
$|p_{1}(x,t)|= 2 A|_{x=x_{0}} e^{-\beta t Im(\lambda_{1}^{6})}$. 
If $Im(\lambda_{1}^{6})<0$ the amplitude decays exponentially, while 
it grows exponentially for $Im(\lambda_{1}^{6})>0$
and when
$Im(\lambda_{1}^{6})=0$, the amplitude remains constant over the time.
\\[3mm]
In this reverse-time case,  any one-soliton does not collapse, either it strictly increases, decreases or stays constant.
\newline
From the spectral plane, let $\lambda_{1}=\xi + i \eta=|\lambda_{1}| e^{i \theta}$, where $|\lambda_{1}|>0$, and $0<\theta<2\pi$ then:
\begin{equation}
\text{if}
\begin{cases}
\theta \in \big\{ (\frac{n}{6} \pi,\frac{n+1}{6} \pi)\big\}, 
\text{then the amplitude of the soliton is increasing for $n=\{0,2,4,\ldots \}$,}
\\
\theta \in \big\{ (\frac{n}{6} \pi,\frac{n+1}{6} \pi)\big\},  
\text{the amplitude of the soliton is decreasing for 
$n=\{1,3,5,\ldots \}$,}
\\
\theta \in \big( \frac{n}{6}\, \text{mod} \{ n\} \big) \pi, 
\text{the amplitude of the soliton is constant for 
$n=\{0,1,2,3,4,5,\ldots \}$,}
\\
\theta \in \{ n \pi \}, \, \text{we obtain one breather with constant amplitude  for 
$n=\{0,1,2,3,4,5,\ldots \}$.}
\end{cases}
\end{equation}
This illustration is shown by the figure below.
\begin{figure}[H]
\centering
\includegraphics[width=12cm,height=9cm]{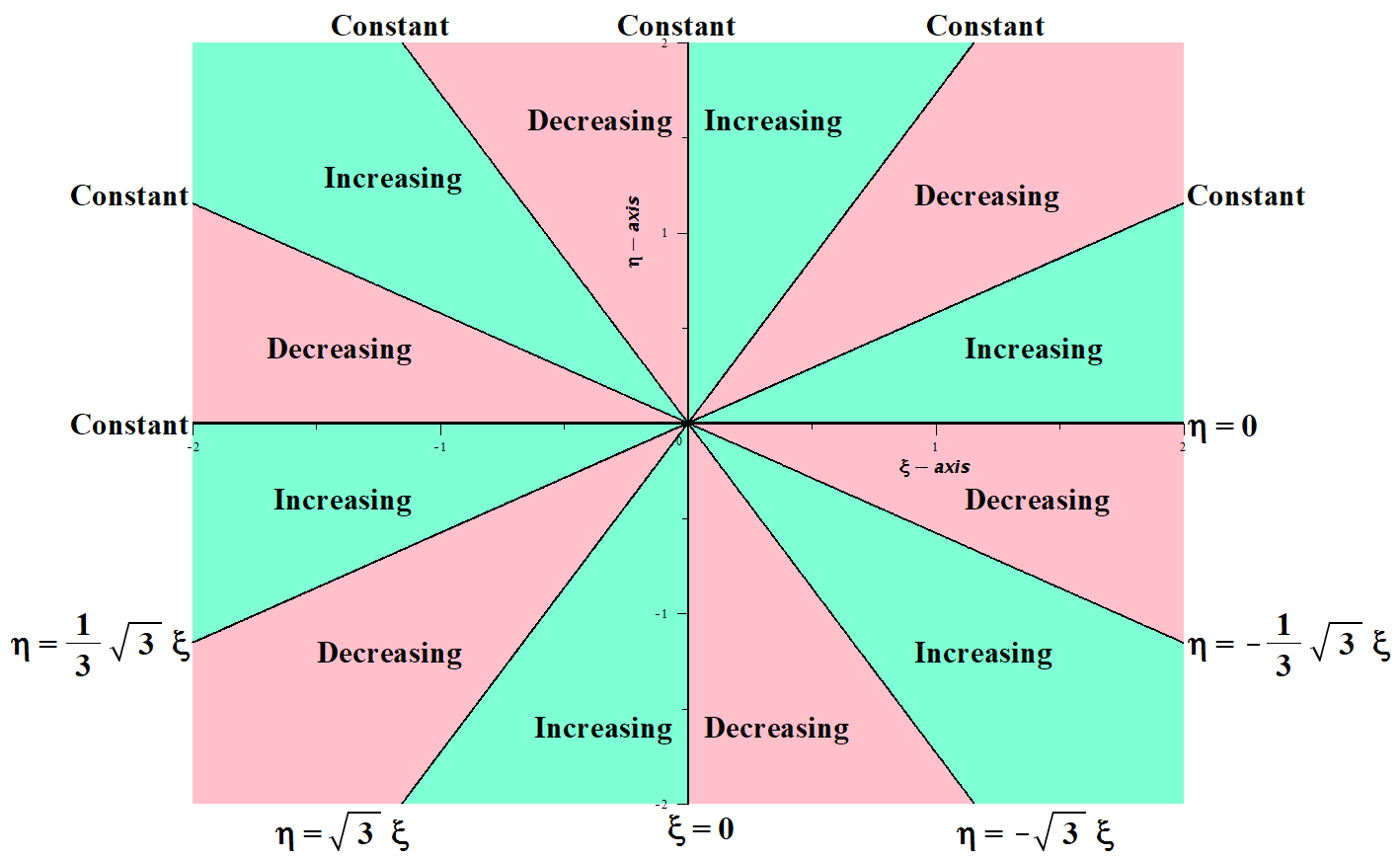}
\caption{Spectral plane of eigenvalues.}
\label{1solitonsymmetry1}
\end{figure}
\vspace{-0.5cm}
For the one-soliton solution, when 
$\lambda_{1}$ does not lie on the real axis, imaginary axis or the trisectors, i.e.
$\lambda \notin \{\xi, \i \eta, (1\pm \i \sqrt{3}) \xi,(1\pm \i \frac{1}{\sqrt{3}}) \xi\}$, the amplitude of the potential grows or decays exponentially, if $Im(\lambda_{1}^{6})>0$ or $Im(\lambda_{1}^{6})<0$ respectively. 
In Figure \ref{1solitonplot1} and Figure \ref{1solitonplot2}, we have two examples where the amplitude grows and decays exponentially.
\par
\begin{figure}[H]
\begin{center}
\includegraphics[width=0.45\textwidth,height=4.5cm]{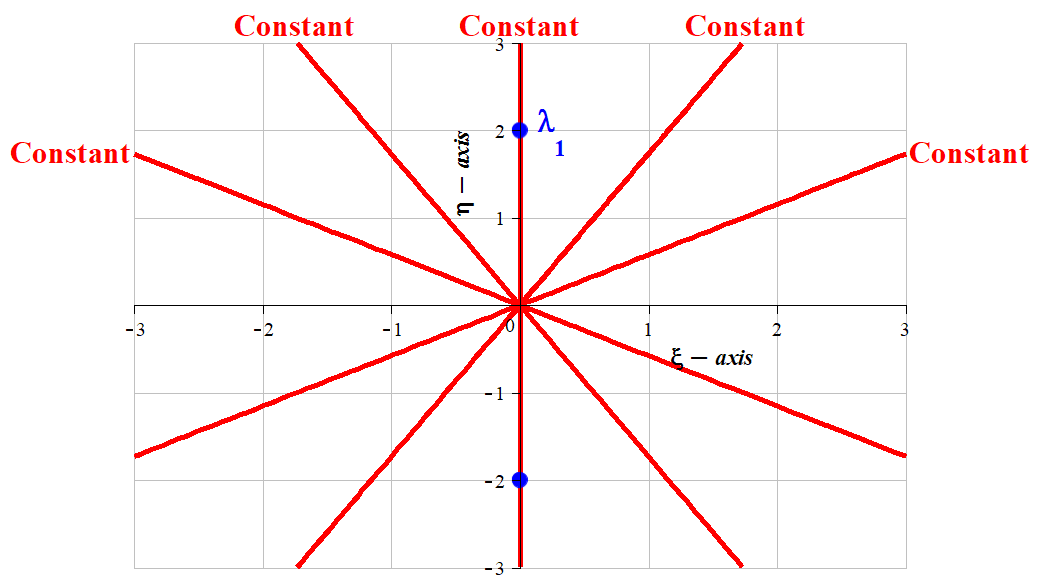}%
\includegraphics[width=0.45\textwidth,height=4.5cm]{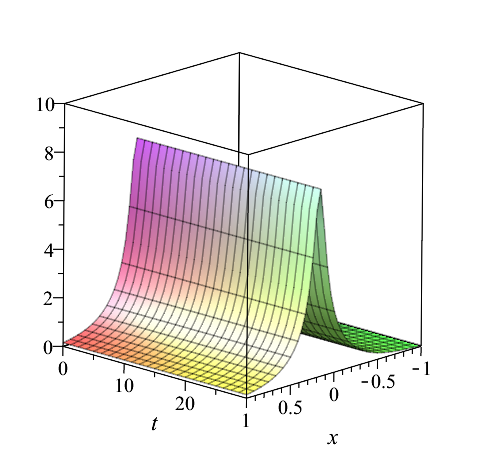}%
\\
\includegraphics[width=0.45\textwidth,height=4.5cm]{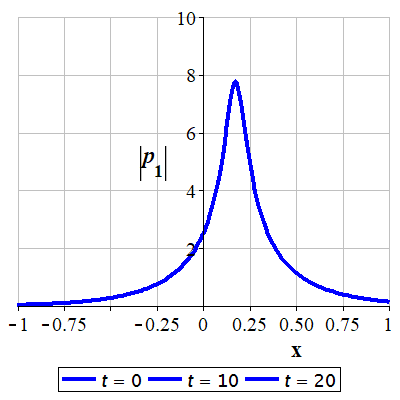}%
\includegraphics[width=0.45\textwidth,height=4.5cm]{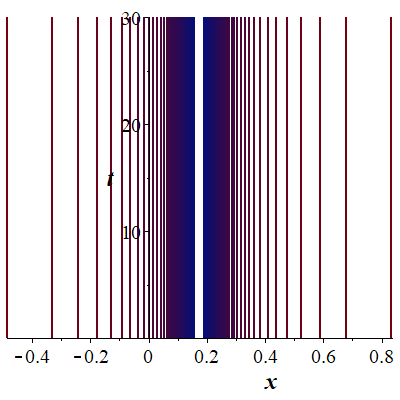}%
\caption{Spectral plane along with 3D, 2D and contour plots of $|p_{1}|$ of the one-soliton with parameters $(\rho_{1},\rho_{2},\rho_{3},\alpha_{1},\alpha_{2},\beta_{1},\beta_{2})=(1,-2,-1,-1,1,-1,1)$, $\lambda_{1}=2i$, $w_{1}=(1,i,2+i,1)$.}%
\label{1solitonplot3}%
\end{center}
\begin{center}
\includegraphics[width=0.45\textwidth,height=4.5cm]{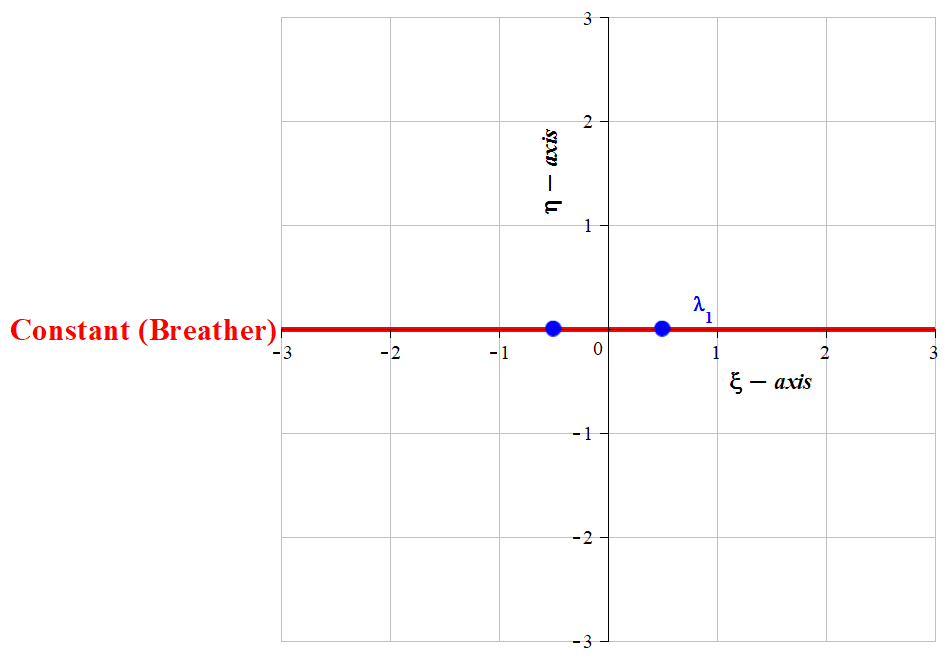}%
\includegraphics[width=0.45\textwidth,height=4.5cm]{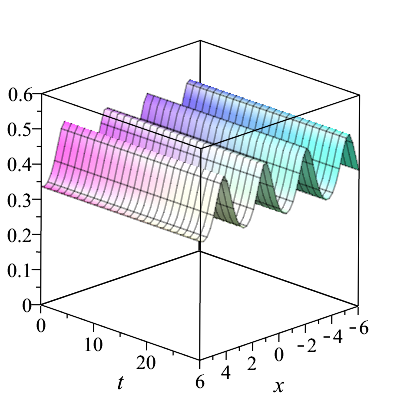}%
\\
\includegraphics[width=0.45\textwidth,height=4.5cm]{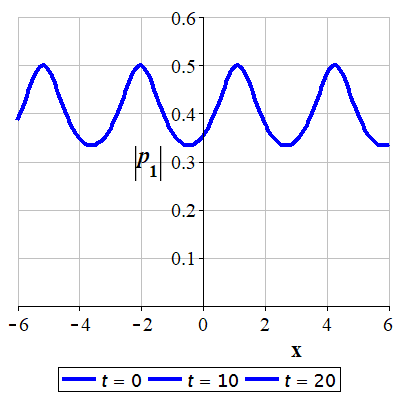}%
\includegraphics[width=0.45\textwidth,height=4.5cm]{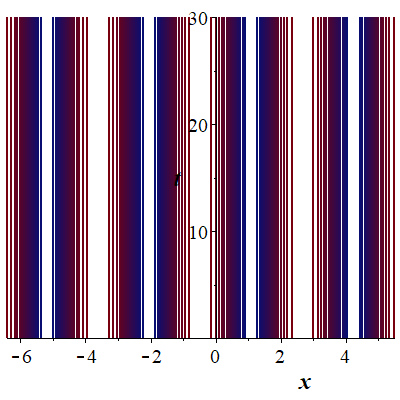}%
\caption{Spectral plane along with 3D, 2D and contour plots of $|p_{1}|$ of the one-soliton with parameters $(\rho_{1},\rho_{2},\rho_{3},\alpha_{1},\alpha_{2},\beta_{1},\beta_{2})=(1,1,1,-1,1,-1,1)$, $\lambda_{1}=0.5$, $w_{1}=(1,i,2+i,1)$.}%
\label{1solitonplot4}%
\end{center}
\end{figure}

The amplitude does not change when $Im(\lambda_{1}^{6})=0$,
i.e.
that means when $\lambda_{1}$ belongs to the real axis, imaginary axis 
or the trisectors. If $\lambda_{1}$ lies on the imaginary axis or the trisectors, then we have a fundamental soliton as seen in figure \ref{1solitonplot3}. If $\lambda_{1}$ is purely imaginary, then
the Lax matrix $U(u,\lambda)$ is a skew-Hermitian matrix. On the other hand, if $\lambda_{1}$ lies on the real axis,
we have a breather which is a periodic one-soliton with period 
$\frac{\pi}{|\alpha \lambda_{1} |}$ as seen in figure \ref{1solitonplot4}.
This is a consequence of the Lax matrix $U(u,\lambda)$ being a Hermitian matrix. 
\newpage

\newpage
\begin{remark}
In the case of one-soliton, when the AKNS system (\ref{AKNSsystem})
has higher even-orders i.e. $\lambda_{1}^{2m}=|\lambda_{1}|^{2m} e^{i2m \theta}$,
$m \in \mathbb{N}$,
the amplitude of $p_{1}$ can be written in the form:
\begin{equation}\label{p1magnitudegeneral}
|p_{1}(x,t)|= A e^{-Im(\lambda_{1}^{2m}(\beta_{1}-\beta_{2})t+ \lambda_{1} (\alpha_{1}+\alpha_{2})x)} 
\end{equation}
where $A$ is a constant. From (\ref{p1magnitudegeneral})
if $Im(\lambda_{1}^{2m})=|\lambda_{1}|^{2m} sin(2m\theta)=0$,
that gives the partition of the complex plane by $2m$-sectors.
\\
If $0<|\lambda_{1}|<1$ then $\lim\limits_{m \rightarrow \infty}Im(\lambda_{1}^{2m})=0$, this means for any $\lambda_{1}$
lying inside the disk of radius 1, the soliton has a constant amplitude. \\
If $|\lambda_{1}|=1$ it lies on the circle of radius 1, then the amplitude $|p_{1}(x,t)|$ will be bounded by
$A e^{\beta t}\leq |p_{1}(x,t)| \leq A e^{-\beta t}$, where $\beta = \beta_{1}-\beta_{2}<0$.
\\
If $|\lambda_{1}|>1$ then $\lim\limits_{m \rightarrow \infty}Im(\lambda_{1}^{2m}) \rightarrow \pm \infty$, so the amplitude will grow exponentially or it will decay to zero exponentially.
\end{remark}

\subsection{Explicit two-soliton solution and its dynamics}
\begin{figure}[H]
\begin{center}
\includegraphics[width=0.45\textwidth,height=4.5cm]{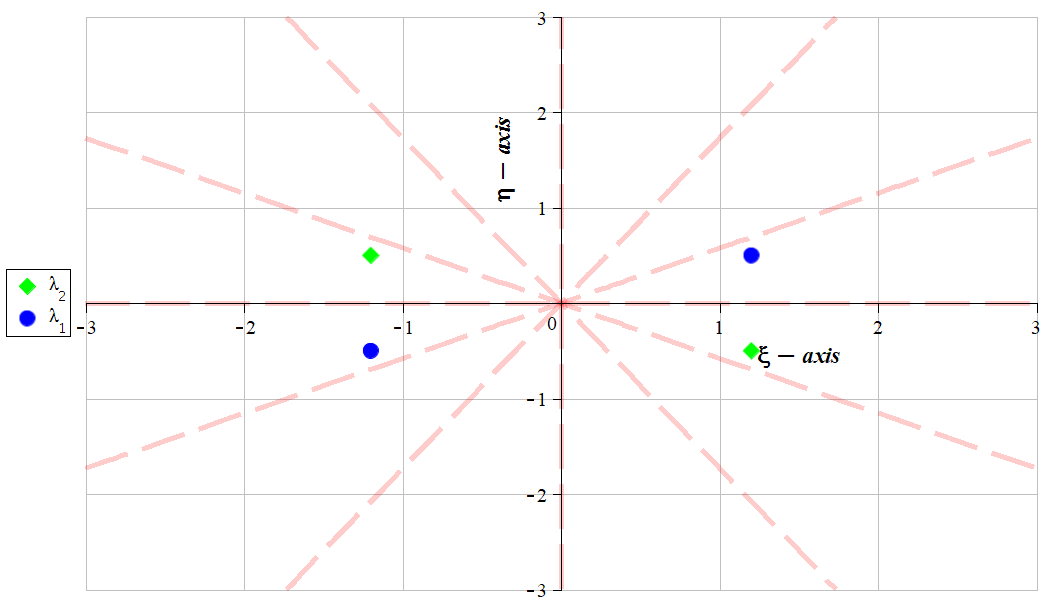}%
\includegraphics[width=0.45\textwidth,height=4.5cm]{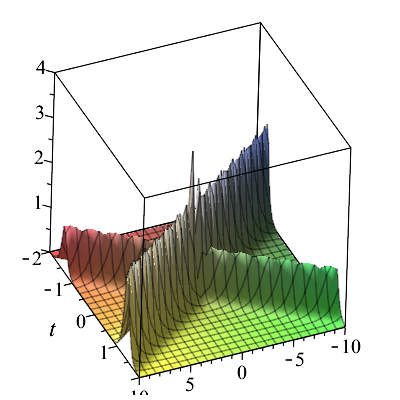}%
\\
\includegraphics[width=0.45\textwidth,height=4.5cm]{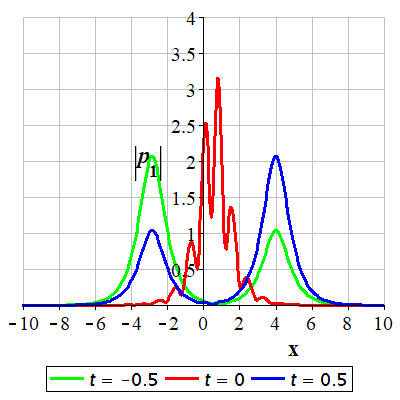}%
\includegraphics[width=0.45\textwidth,height=4.5cm]{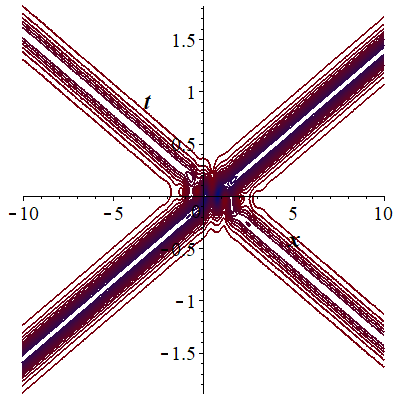}%
\caption{Spectral plane along with 3D, 2D and contour plots of $|p_{1}|$ of the two-soliton with parameters $(\rho_{1},\rho_{2},\rho_{3},\alpha_{1},\alpha_{2},\beta_{1},\beta_{2})=(-1,-1,1,-2,1,-1,2)$, $(\lambda_{1},\lambda_{2})=(1.2+0.5i,-1.2+0.5i)$, $w_{1}=(1,1-3i,-i,1+i)$, $w_{2}=(2,1-3i,-i,1+i)$.}%
\label{2solitonplot20}%
\end{center}
\begin{center}
\includegraphics[width=0.45\textwidth,height=4.5cm]{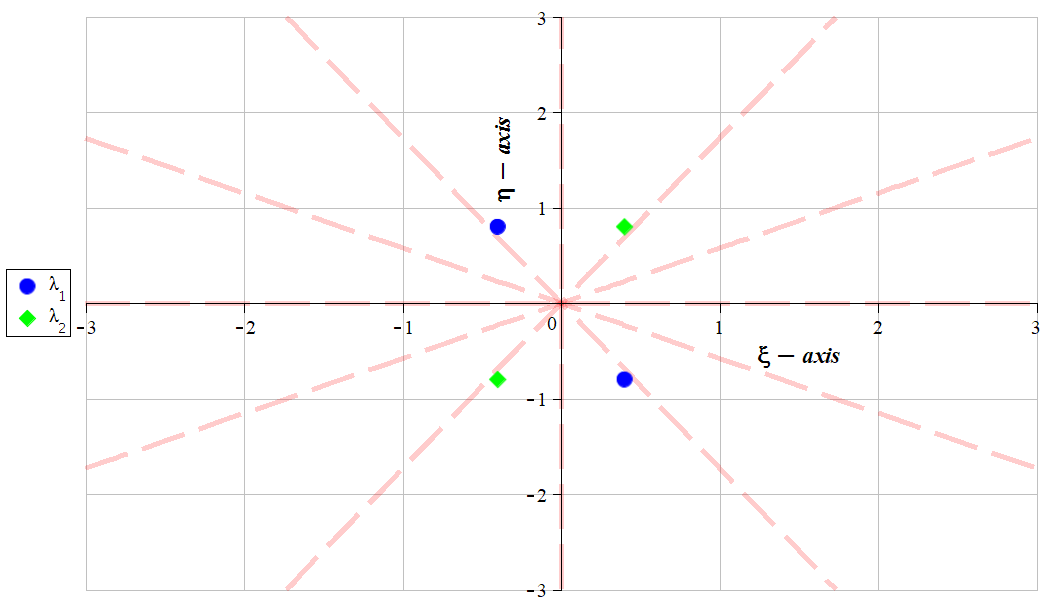}%
\includegraphics[width=0.45\textwidth,height=4.5cm]{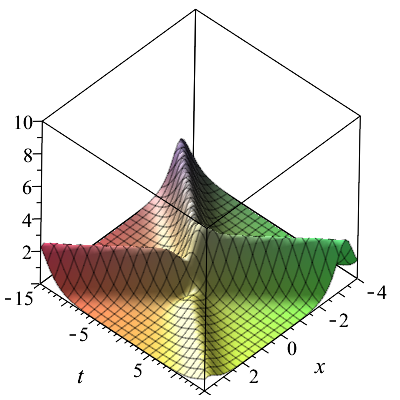}%
\\
\includegraphics[width=0.45\textwidth,height=4.5cm]{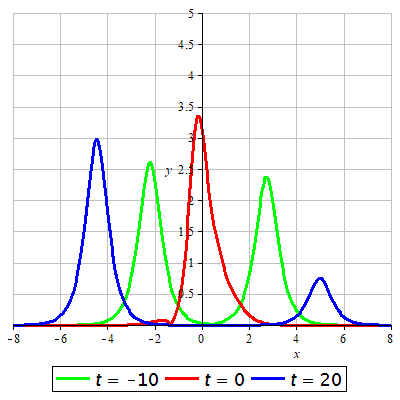}%
\includegraphics[width=0.45\textwidth,height=4.5cm]{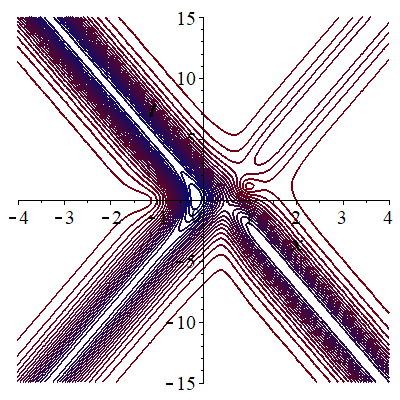}%
\caption{Spectral plane along with 3D, 2D and contour plots of $|p_{1}|$ of the two-soliton with parameters $(\rho_{1},\rho_{2},\rho_{3},\alpha_{1},\alpha_{2},\beta_{1},\beta_{2})=(-1,1,-1,-2,1,-2,1)$, $(\lambda_{1},\lambda_{2})=(-0.4+0.8i,0.4+0.8i)$, $w_{1}=(1,1-i,-0.1+i,1+i)$, $w_{2}=(-1+2i,1-0.1i,3+i,0)$.}%
\label{2solitonplot2}%
\end{center}
\end{figure}

\begin{figure}[H]
\begin{center}
\includegraphics[width=0.45\textwidth,height=4.5cm]{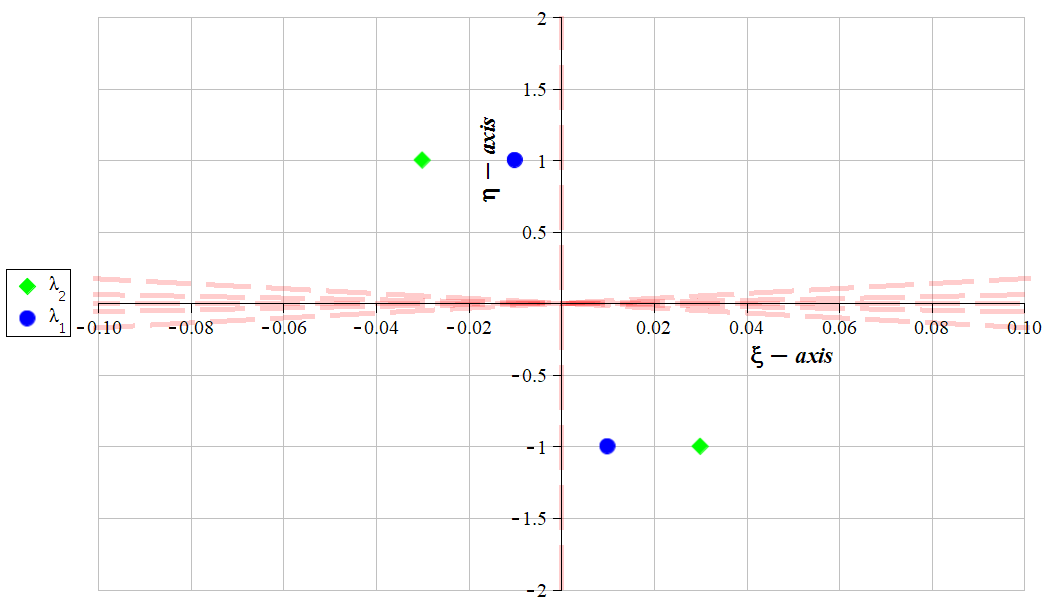}%
\includegraphics[width=0.45\textwidth,height=4.5cm]{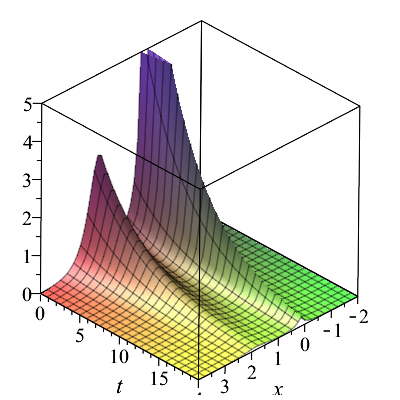}%
\\
\includegraphics[width=0.45\textwidth,height=4.5cm]{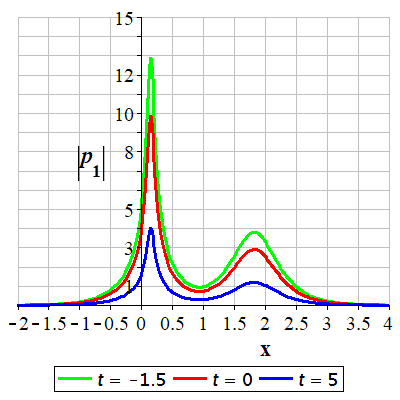}%
\includegraphics[width=0.45\textwidth,height=4.5cm]{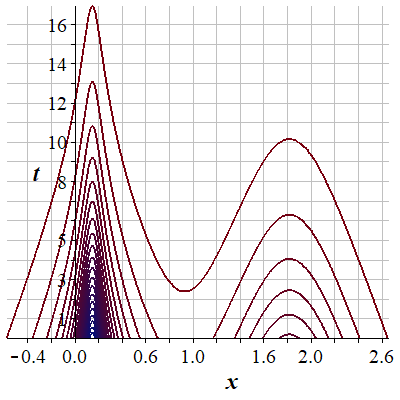}%
\caption{Spectral plane along with 3D, 2D and contour plots of $|p_{1}|$ in the focussing case of the two-soliton with parameters $(\rho_{1},\rho_{2},\rho_{3},\alpha_{1},\alpha_{2},\beta_{1},\beta_{2})=(-1,-2,-3,-2,1,-1,2)$, $(\lambda_{1},\lambda_{2})=(-0.01+i,-0.03+i)$, $w_{1}=(1,0,2+i,0)$, $w_{2}=(1,2-i,0,1)$.}%
\label{2solitonplot24}%
\end{center}
\begin{center}
\includegraphics[width=0.45\textwidth,height=4.5cm]{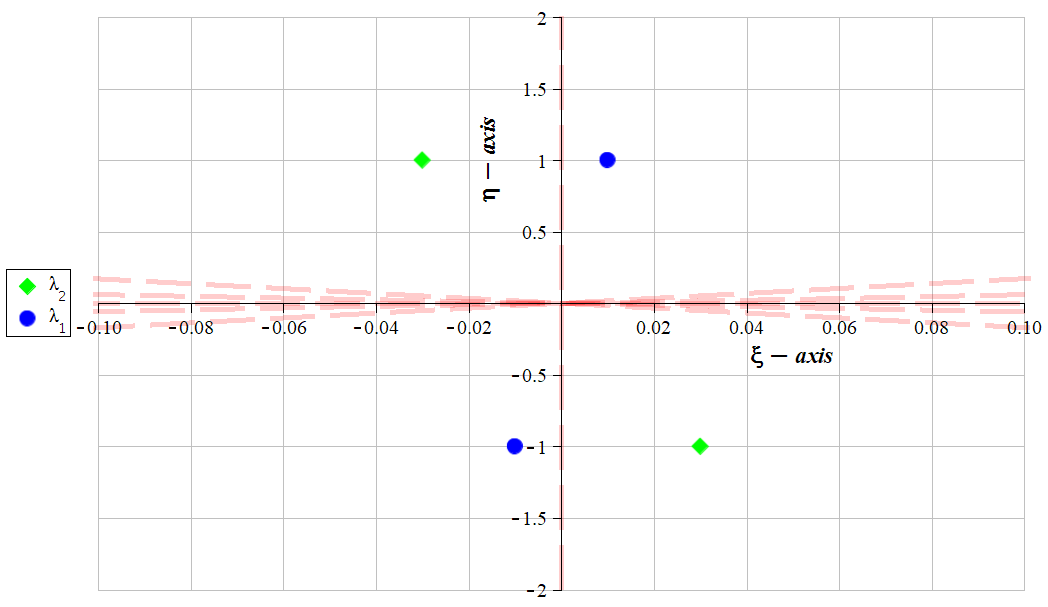}%
\includegraphics[width=0.45\textwidth,height=4.5cm]{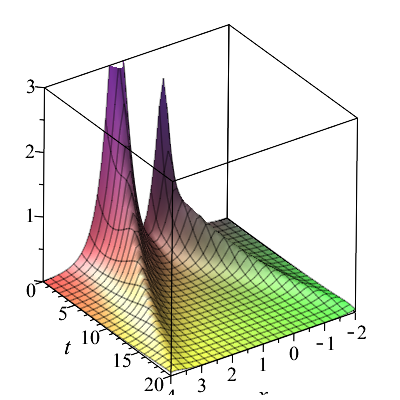}%
\\
\includegraphics[width=0.45\textwidth,height=4.5cm]{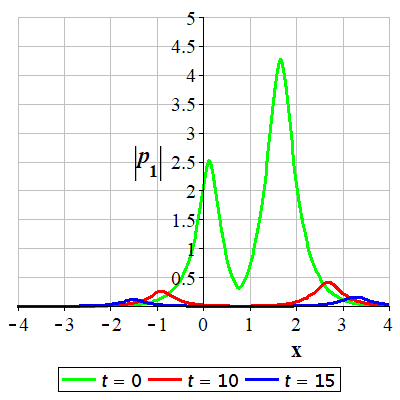}%
\includegraphics[width=0.45\textwidth,height=4.5cm]{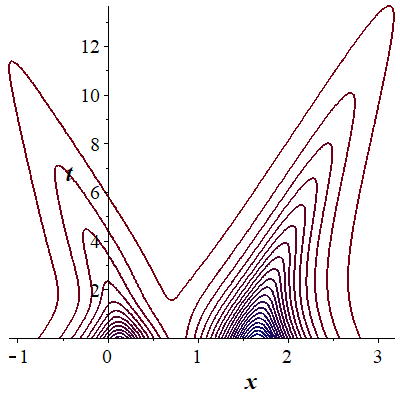}%
\caption{Spectral plane along with 3D, 2D and contour plots of $|p_{1}|$ in the focussing case of the two-soliton with parameters $(\rho_{1},\rho_{2},\rho_{3},\alpha_{1},\alpha_{2},\beta_{1},\beta_{2})=(-1,-2,-3,-2,1,-1,2)$, $(\lambda_{1},\lambda_{2})=(0.01+i,-0.03+i)$, $w_{1}=(1,i,2+i,1)$, $w_{2}=(1,2-i,i,1)$.}%
\label{2solitonplot25}%
\end{center}
\end{figure}

\begin{figure}[H]
\begin{center}
\includegraphics[width=0.45\textwidth,height=4.5cm]{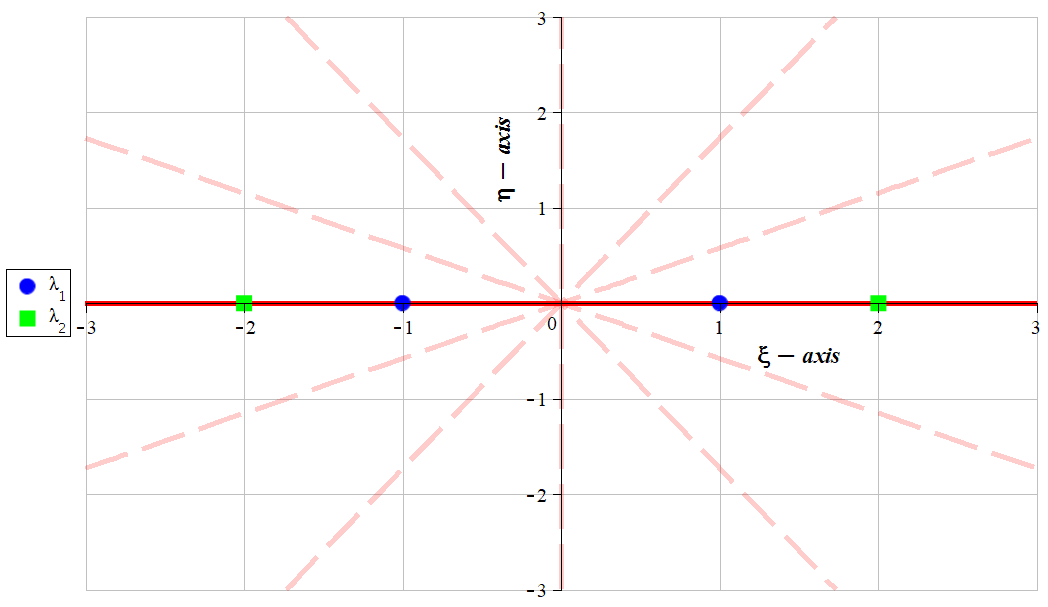}%
\includegraphics[width=0.45\textwidth,height=4.5cm]{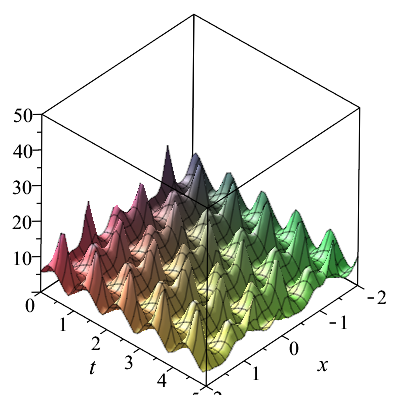}%
\\
\includegraphics[width=0.45\textwidth,height=4.5cm]{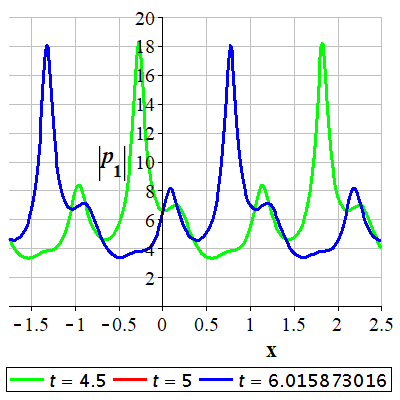}%
\includegraphics[width=0.45\textwidth,height=4.5cm]{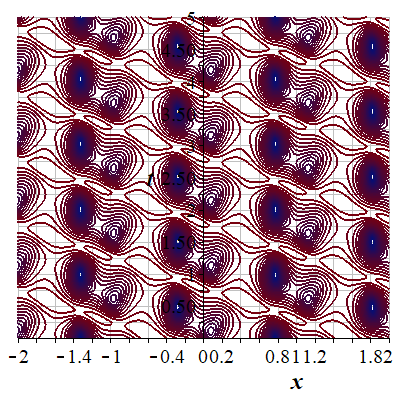}%
\caption{Spectral plane along with 3D, 2D and contour plots  of $|p_{1}|$ in the focussing case of the two-soliton with parameters $(\rho_{1},\rho_{2},\rho_{3},\alpha_{1},\alpha_{2},\beta_{1},\beta_{2})=(-1,-1,-1,-2,1,-1/64 \pi,1/64 \pi)$, $(\lambda_{1},\lambda_{2})=(1,2)$, $w_{1}=(1,0,2+i,1-i)$, $w_{2}=(-1,1-2i,-i,0)$.}%
\label{2solitonplot23}%
\end{center}
\end{figure}

A general explicit two-soliton solution in the reverse-time case when $N=2$,
$w_{1}=(w_{11},w_{12},w_{13},w_{14})^{T}$,
$w_{2}=(w_{21},w_{22},w_{23},w_{24})^{T}$,
$(\lambda_{1},\lambda_{2}) \in \mathbb{C}^{2}$
are arbitrary,
and $\hat{\lambda}_{1}=-\lambda_{1}$, $\hat{\lambda}_{2}=-\lambda_{2}$,
is given if $\lambda_{1} \neq - \lambda_{2}$ by
\begin{align}
p_{1}(x,t) = 2 \rho_{2} \rho_{3} (\lambda_{1}+\lambda_{2}) (\alpha_{1}-\alpha_{2}) \frac{A(x,t)}{B(x,t)},
\\
p_{2}(x,t) = 2 \rho_{1} \rho_{3} (\lambda_{1}+\lambda_{2}) (\alpha_{1}-\alpha_{2}) \frac{C(x,t)}{B(x,t)},
\\
p_{3}(x,t) = 2 \rho_{1} \rho_{2} (\lambda_{1}+\lambda_{2}) (\alpha_{1}-\alpha_{2}) \frac{D(x,t)}{B(x,t)},
\end{align}
where
\begin{flalign}
&\begin{aligned}
A(x,t) = 
e^{i[\lambda_{2}^{6} (\beta_{1}-\beta_{2})  t 
+\lambda_{2} (\alpha_{1}+\alpha_{2}) x]} \cdot
& \bigg[ \bigg( w_{22} M (\lambda_{1}+\lambda_{2}) 
- 2 w_{12} K \lambda_{1} \bigg)
w_{21} \lambda_{2} e^{i 2\alpha_{2} \lambda_{1} x}
\\[-2mm]
& - \rho_{1} \rho_{2} \rho_{3} (\lambda_{1}-\lambda_{2}) w_{11}^{2} w_{21} w_{22} \lambda_{2}
e^{i 2 \alpha_{1} \lambda_{1} x} \bigg]
\\[-2mm]
+ e^{i[\lambda_{1}^{6} (\beta_{1}-\beta_{2})  t 
+\lambda_{1} (\alpha_{1}+\alpha_{2}) x]} \cdot
& \bigg[ \bigg( w_{12} N (\lambda_{1}+\lambda_{2}) 
- 2 w_{22} K \lambda_{2} \bigg)
w_{11} \lambda_{1} e^{i 2\alpha_{2} \lambda_{2} x}
\\[-2mm]
& + \rho_{1} \rho_{2} \rho_{3} (\lambda_{1}-\lambda_{2}) w_{11} w_{12} w_{21}^{2} \lambda_{1}
e^{i 2 \alpha_{1} \lambda_{2} x} \bigg],
\end{aligned}&
\end{flalign}

\begin{flalign}
&\begin{aligned}
B(x,t) &= -4 \rho_{1} \rho_{2} \rho_{3} \lambda_{1} \lambda_{2} w_{11} w_{21} K
e^{i (\lambda_{1}+\lambda_{2})(\alpha_{1}+\alpha_{2}) x}
\cdot
\bigg[ 
e^{i (\lambda_{1}^{6}-\lambda_{2}^{6})(\beta_{1}-\beta_{2}) t}
+ e^{-i (\lambda_{1}^{6}-\lambda_{2}^{6})(\beta_{1}-\beta_{2}) t}
\bigg]
\\
&+ \rho_{1} \rho_{2} \rho_{3} w_{21}^{2} M (\lambda_{1}+\lambda_{2})^{2}
e^{i2 (\alpha_{1} \lambda_{2}+\alpha_{2} \lambda_{1}) x}
+ \rho_{1} \rho_{2} \rho_{3} w_{11}^{2} N (\lambda_{1}+\lambda_{2})^{2}
e^{i2 (\alpha_{1} \lambda_{1}+\alpha_{2} \lambda_{2}) x}
\\
&+
\rho_{1}^2 \rho_{2}^2 \rho_{3}^2 w_{11}^{2} w_{21}^{2} (\lambda_{1}-\lambda_{2})^{2}
e^{i2 \alpha_{1} (\lambda_{1}+\lambda_{2}) x}
+ \bigg[ (\lambda_{1}^{2}+\lambda_{2}^{2}) MN 
+ (2MN-4K^{2}) \lambda_{1} \lambda_{2}\bigg] 
e^{i2 \alpha_{2} (\lambda_{1}+\lambda_{2}) x},
\end{aligned}&
\end{flalign}
\vspace{-2cm}
\begin{flalign}
&\begin{aligned}
C(x,t) = 
e^{i[\lambda_{2}^{6} (\beta_{1}-\beta_{2})  t 
+\lambda_{2} (\alpha_{1}+\alpha_{2}) x]} \cdot
&\bigg[ \bigg( w_{23} M (\lambda_{1}+\lambda_{2}) 
- 2 w_{13} K \lambda_{1} \bigg)
w_{21} \lambda_{2} e^{i 2\alpha_{2} \lambda_{1} x}
\\[-2mm]
& - \rho_{1} \rho_{2} \rho_{3} (\lambda_{1}-\lambda_{2}) w_{11}^{2} w_{21} w_{23} \lambda_{2}
e^{i 2 \alpha_{1} \lambda_{1} x} \bigg]
\\[-2mm]
+ e^{i[\lambda_{1}^{6} (\beta_{1}-\beta_{2})  t 
+\lambda_{1} (\alpha_{1}+\alpha_{2}) x]} \cdot
& \bigg[ \bigg( w_{13} N (\lambda_{1}+\lambda_{2}) 
- 2 w_{23} K \lambda_{2} \bigg)
w_{11} \lambda_{1} e^{i 2\alpha_{2} \lambda_{2} x}
\\[-2mm]
& + \rho_{1} \rho_{2} \rho_{3} (\lambda_{1}-\lambda_{2}) w_{11} w_{13} w_{21}^{2} \lambda_{1}
e^{i 2 \alpha_{1} \lambda_{2} x} \bigg],
\end{aligned}&
\end{flalign}
\vspace{-1cm}
\begin{flalign}
&\begin{aligned}
D(x,t) = 
e^{i[\lambda_{2}^{6} (\beta_{1}-\beta_{2})  t 
+\lambda_{2} (\alpha_{1}+\alpha_{2}) x]} \cdot
& \bigg[ \bigg( w_{24} M (\lambda_{1}+\lambda_{2}) 
- 2 w_{14} K \lambda_{1} \bigg)
w_{21} \lambda_{2} e^{i 2\alpha_{2} \lambda_{1} x}
\\[-2mm]
& - \rho_{1} \rho_{2} \rho_{3} (\lambda_{1}-\lambda_{2}) w_{11}^{2} w_{21} w_{24} \lambda_{2}
e^{i 2 \alpha_{1} \lambda_{1} x} \bigg]
\\[-2mm]
+ e^{i[\lambda_{1}^{6} (\beta_{1}-\beta_{2})  t 
+\lambda_{1} (\alpha_{1}+\alpha_{2}) x]} \cdot
& \bigg[ \bigg( w_{14} N (\lambda_{1}+\lambda_{2}) 
- 2 w_{24} K \lambda_{2} \bigg)
w_{11} \lambda_{1} e^{i 2\alpha_{2} \lambda_{2} x}
\\[-2mm]
& + \rho_{1} \rho_{2} \rho_{3} (\lambda_{1}-\lambda_{2}) w_{11} w_{14} w_{21}^{2} \lambda_{1}
e^{i 2 \alpha_{1} \lambda_{2} x} \bigg],
\end{aligned}&
\end{flalign}
and 
$M=\rho_{2} \rho_{3} w_{12}^{2}+ \rho_{1} \rho_{3} w_{13}^{2}+ \rho_{1} \rho_{2} w_{14}^{2}$, 
$N=\rho_{2} \rho_{3} w_{22}^{2}+ \rho_{1} \rho_{3} w_{23}^{2}+ \rho_{1} \rho_{2} w_{24}^{2}$ and
$K=\rho_{2} \rho_{3} w_{12} w_{22} + \rho_{1} \rho_{3} w_{13} w_{23} + \rho_{1} \rho_{2} w_{14} w_{24}$.
\\
For the two-soliton dynamics, we notice that either both the two solitons are moving (repeatedly or not) in opposite directions or both are stationary, i.e., they don't move with respect to space. 
In figure \ref{2solitonplot20}, we have two travelling waves that move in opposite directions, keeping the same amplitude before and after interaction in an elastic collision, that is no energy radiation emitted \cite{Yang2010}. Whereas  in figure \ref{2solitonplot2}, the amplitudes of the two waves change after interaction to new constant amplitudes resembling Manakov waves \cite{Manakov1974}.
\\[3mm]
In figure \ref{2solitonplot24}, we have two solitons with exponentially decaying amplitude and stationary over the time, i.e. they do not move in space. On the other hand, we can have as in 
figure \ref{2solitonplot25}, two solitons with exponentially decaying amplitude but moving apart over the time.
\\
Aside, if both $\lambda_{1}$ and $\lambda_{2}$ lies on the real axis, 
then we will obtain breather solitons with time period 
$\frac{2 \pi}{\beta (\lambda_{1}^{6}-\lambda_{2}^{6})}$. 
An example is shown in figure \ref{2solitonplot23}, where the breather
waves coincide for $t=5$ and $t=6.015873016$.
\newpage
\subsection{The dynamics of the three-soliton solution}
\begin{figure}[H]
\begin{center}
\includegraphics[width=0.45\textwidth,height=4.5cm]{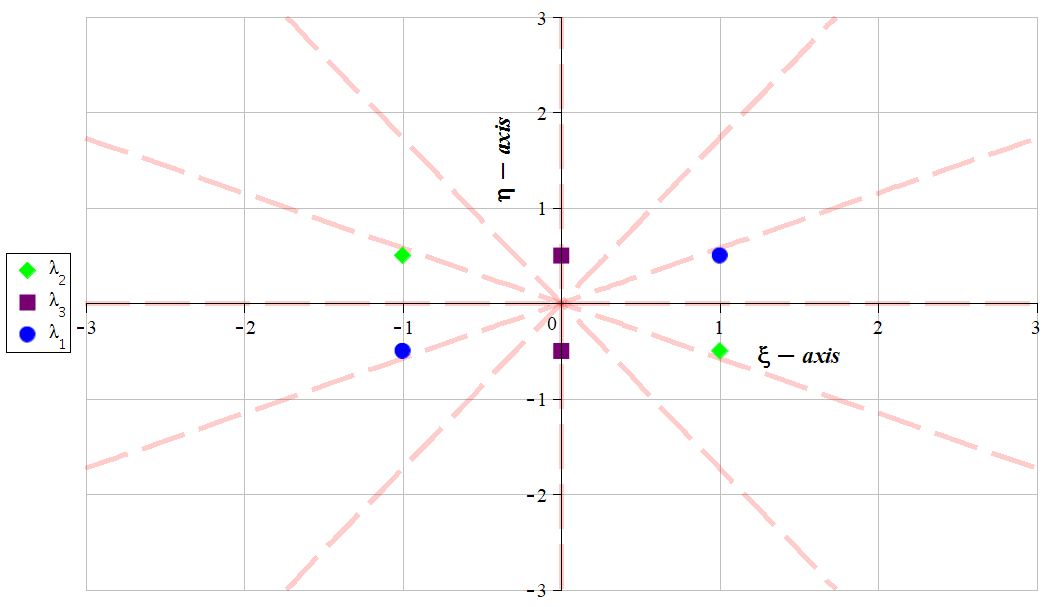}%
\includegraphics[width=0.45\textwidth,height=4.5cm]{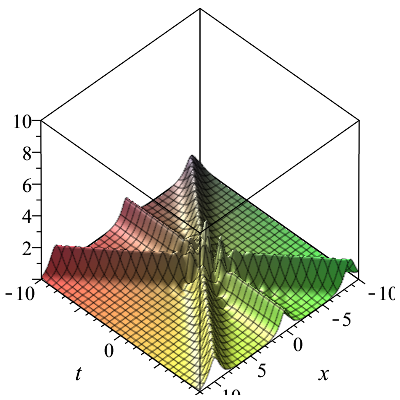}%
\\
\includegraphics[width=0.45\textwidth,height=4.5cm]{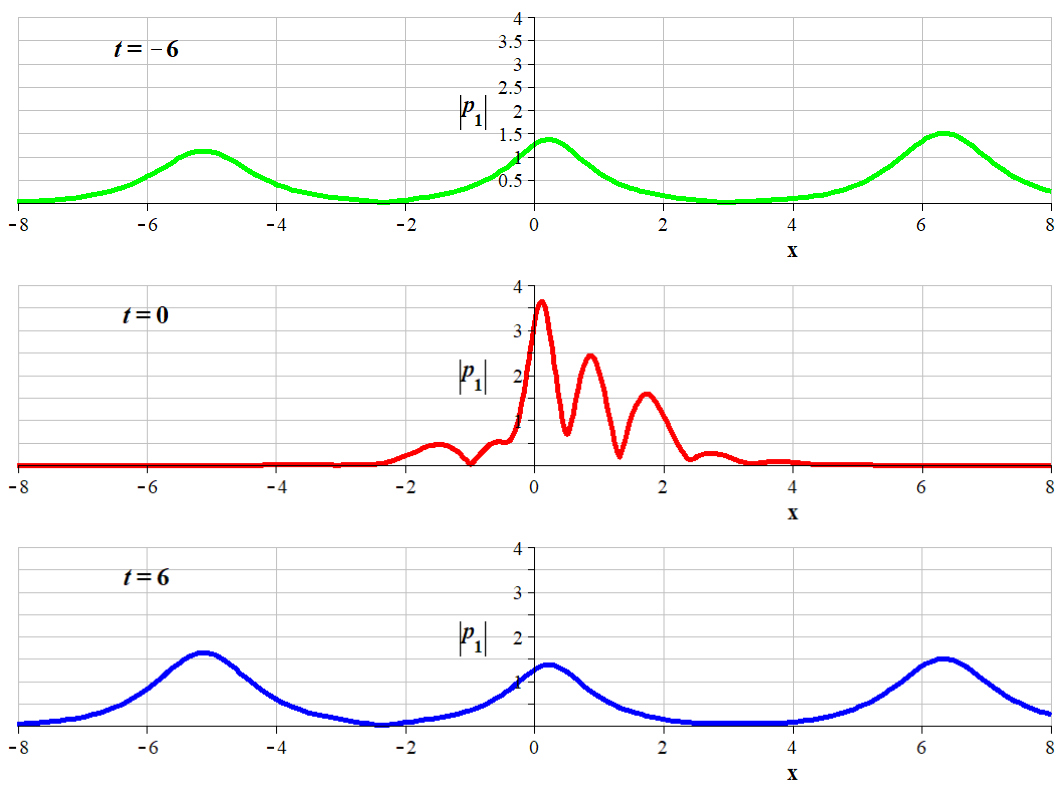}%
\includegraphics[width=0.45\textwidth,height=4.5cm]{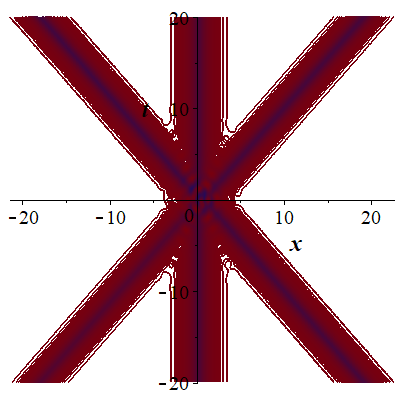}%
\caption{Spectral plane along with 3D, 2D and contour plots of $|p_{1}|$ in the focussing case of the three-soliton with parameters $(\rho_{1},\rho_{2},\rho_{3},\alpha_{1},\alpha_{2},\beta_{1},\beta_{2})=(-1,-1,-1,-2,1,-1,1)$, $(\lambda_{1},\lambda_{2},\lambda_{3})=(1+0.5i,-1+0.5i,0.5i)$, $w_{1}=(1,1+2i,0,0)$, $w_{2}=(-1,1-2i,0,0)$, $w_{3}=(2+i,1+2i,1,2i)$.}%
\label{3solitonplot7}%
\end{center}
\begin{center}
\includegraphics[width=0.45\textwidth,height=4.5cm]{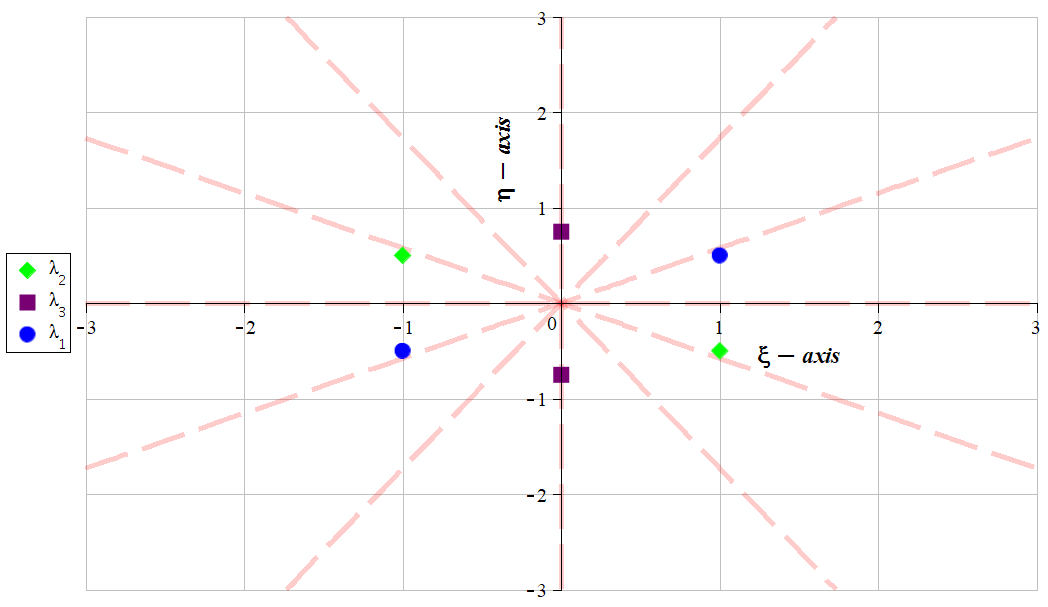}%
\includegraphics[width=0.45\textwidth,height=4.5cm]{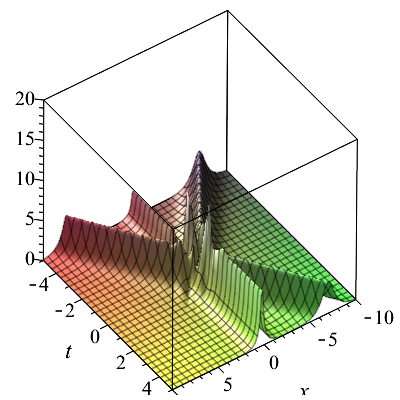}%
\\
\includegraphics[width=0.45\textwidth,height=4.5cm]{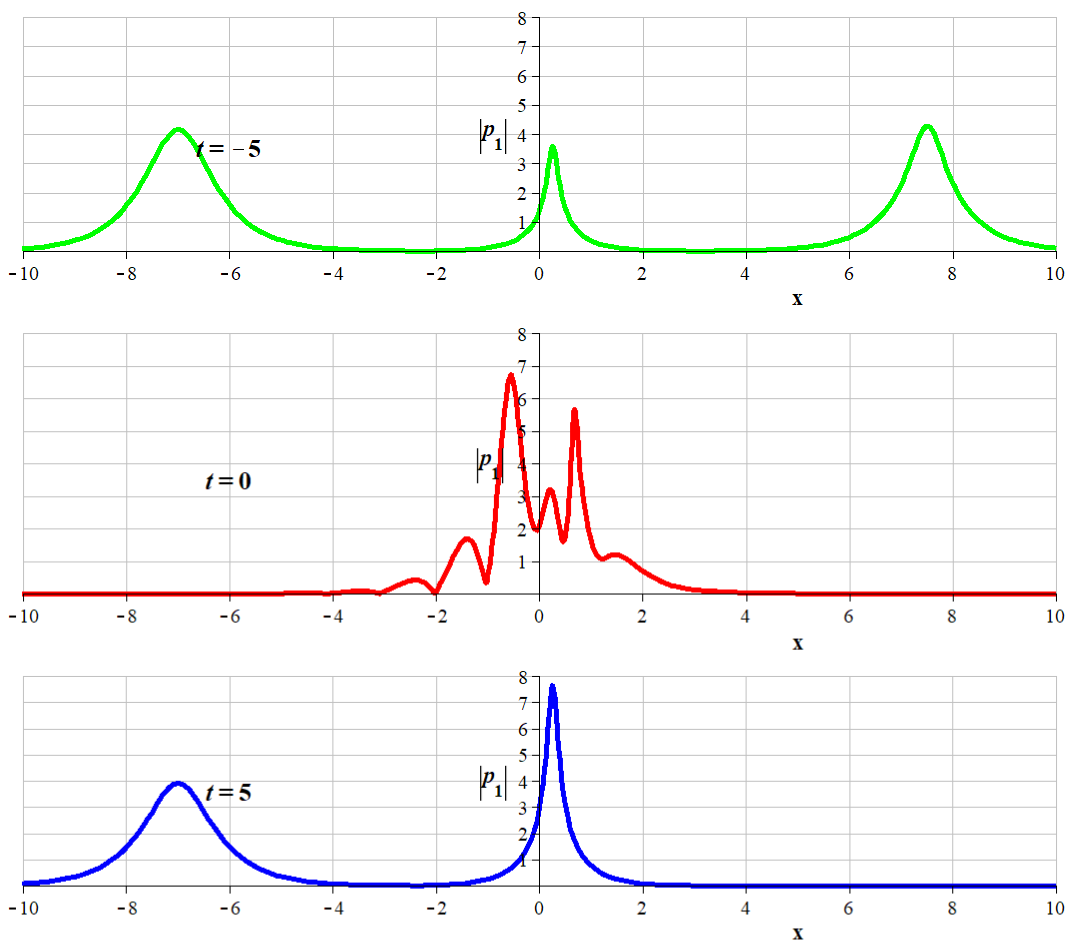}%
\includegraphics[width=0.45\textwidth,height=4.5cm]{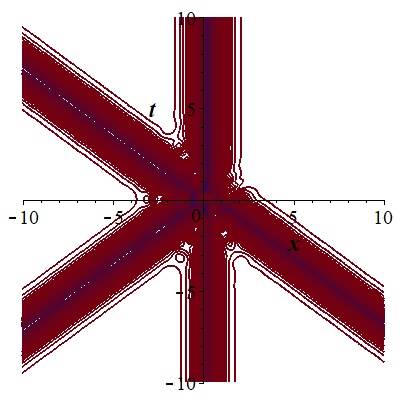}%
\caption{Spectral plane along with 3D, 2D and contour plots of $|p_{1}|$ of the three-soliton with parameters $(\rho_{1},\rho_{2},\rho_{3},\alpha_{1},\alpha_{2},\beta_{1},\beta_{2})=(1,1,1,-2,1,-2,1)$, $(\lambda_{1},\lambda_{2},\lambda_{3})=(1+0.5i,-1+0.5i,0.75i)$, $w_{1}=(1,0,2+i,1-i)$, $w_{2}=(-1,1-2i,-i,0)$, $w_{3}=(2+i,1+2i,1,2i)$.}%
\label{3solitonplot14}%
\end{center}
\end{figure}

\begin{figure}[H]
\begin{center}
\includegraphics[width=0.45\textwidth,height=4.5cm]{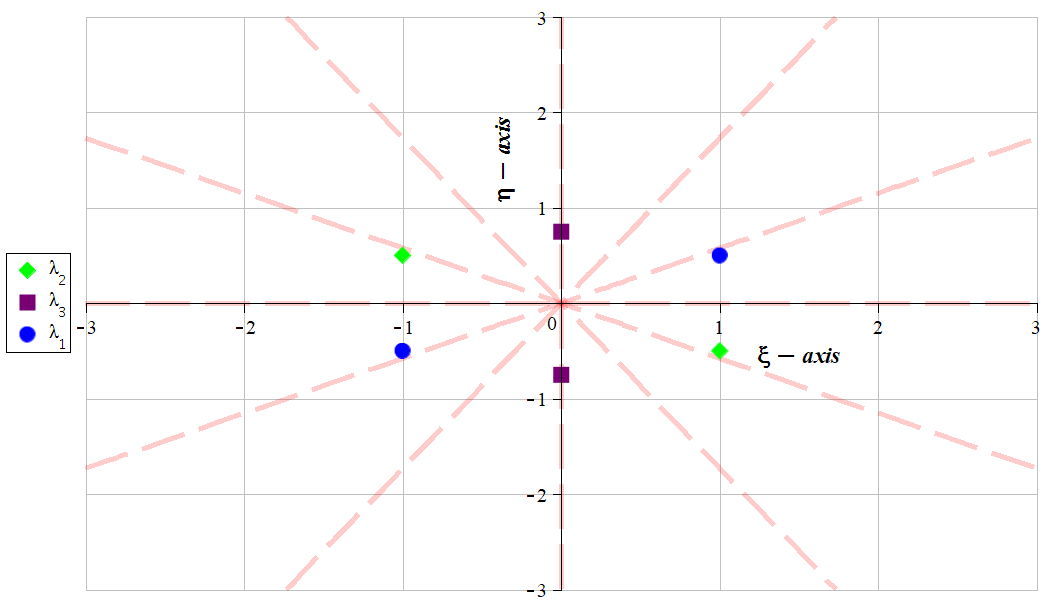}%
\includegraphics[width=0.45\textwidth,height=4.5cm]{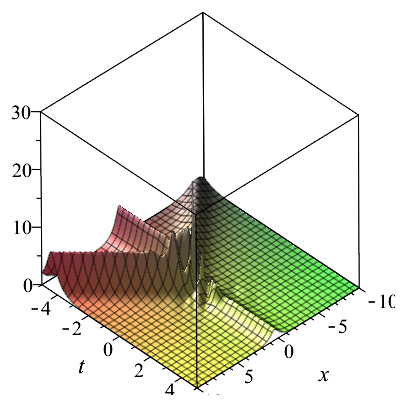}%
\\
\includegraphics[width=0.45\textwidth,height=4.5cm]{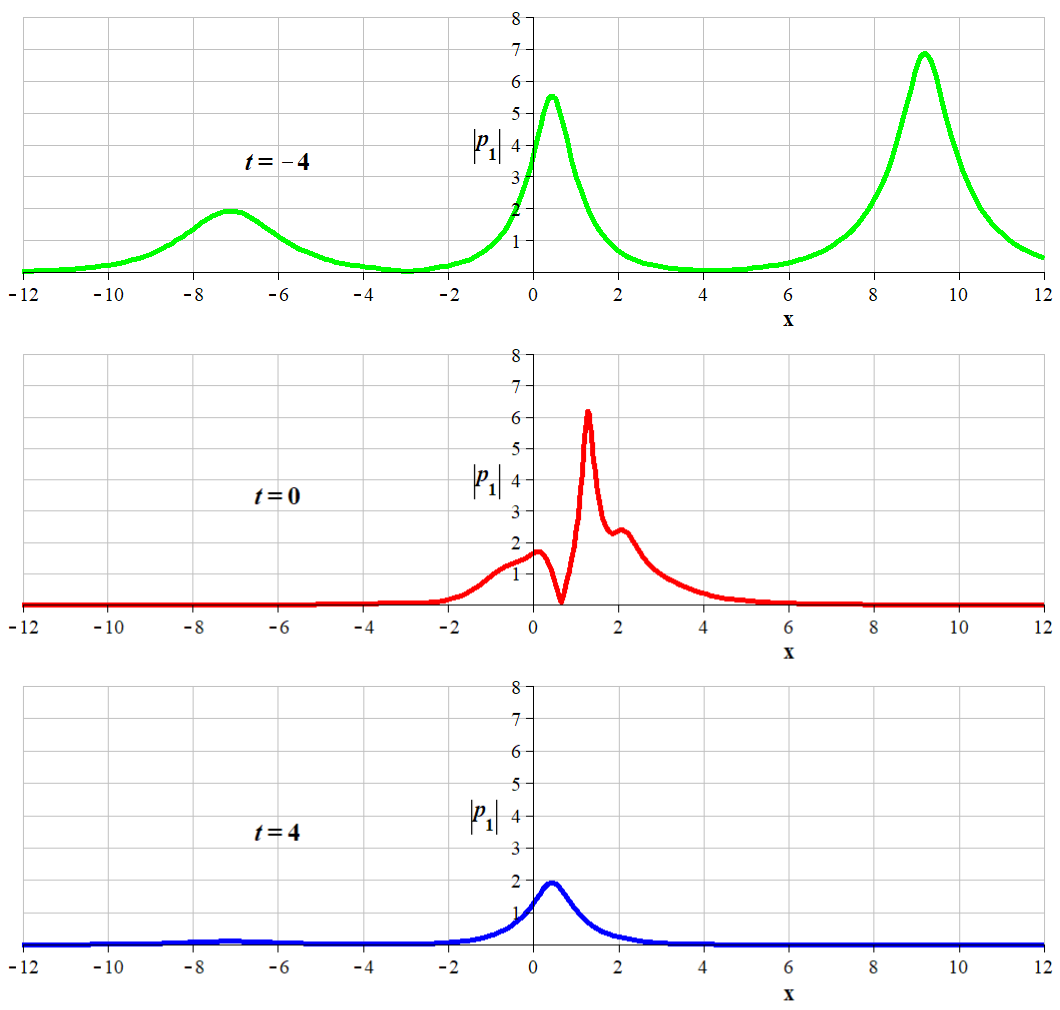}%
\includegraphics[width=0.45\textwidth,height=4.5cm]{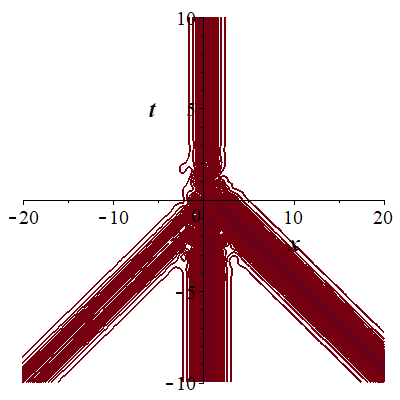}%
\caption{Spectral plane along with 3D, 2D and contour plots of $|p_{1}|$ of the three-soliton with parameters $(\rho_{1},\rho_{2},\rho_{3},\alpha_{1},\alpha_{2},\beta_{1},\beta_{2})=(1,1,1,-1,1,-2,1)$, $(\lambda_{1},\lambda_{2},\lambda_{3})=(1+0.5i,-1+0.5i,0.75i)$, $w_{1}=(1,0,2+i,1-i)$, $w_{2}=(-1,5-2i,-i,0)$, $w_{3}=(2+i,1+2i,1,2i)$.}%
\label{3solitonplot12}%
\end{center}
\begin{center}
\includegraphics[width=0.45\textwidth,height=4.5cm]{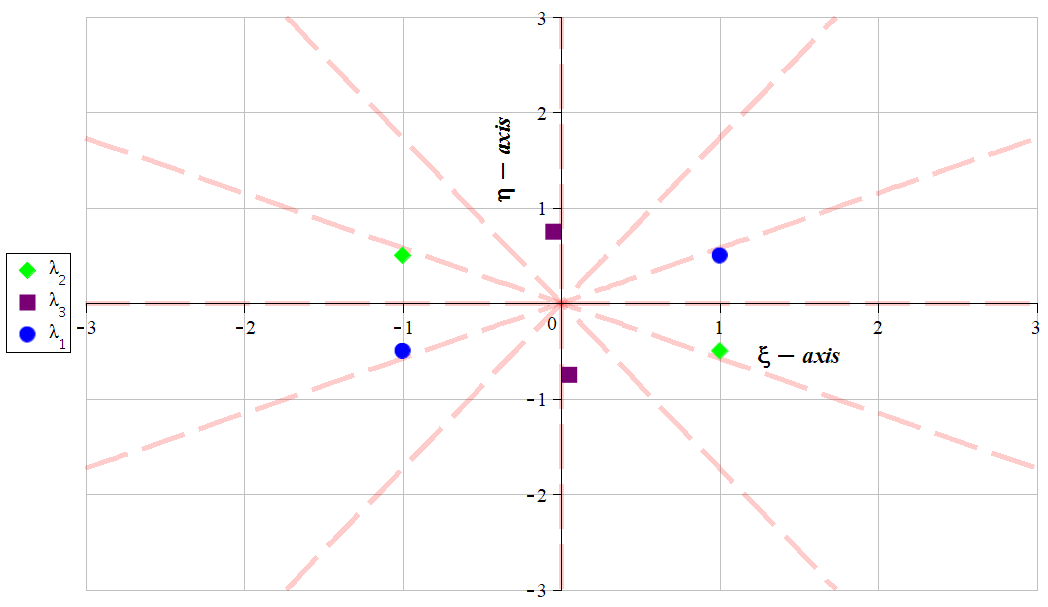}%
\includegraphics[width=0.45\textwidth,height=4.5cm]{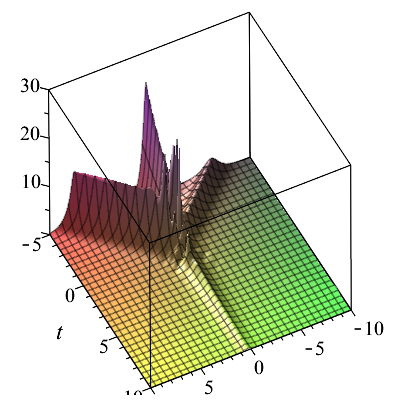}%
\\
\includegraphics[width=0.45\textwidth,height=4.5cm]{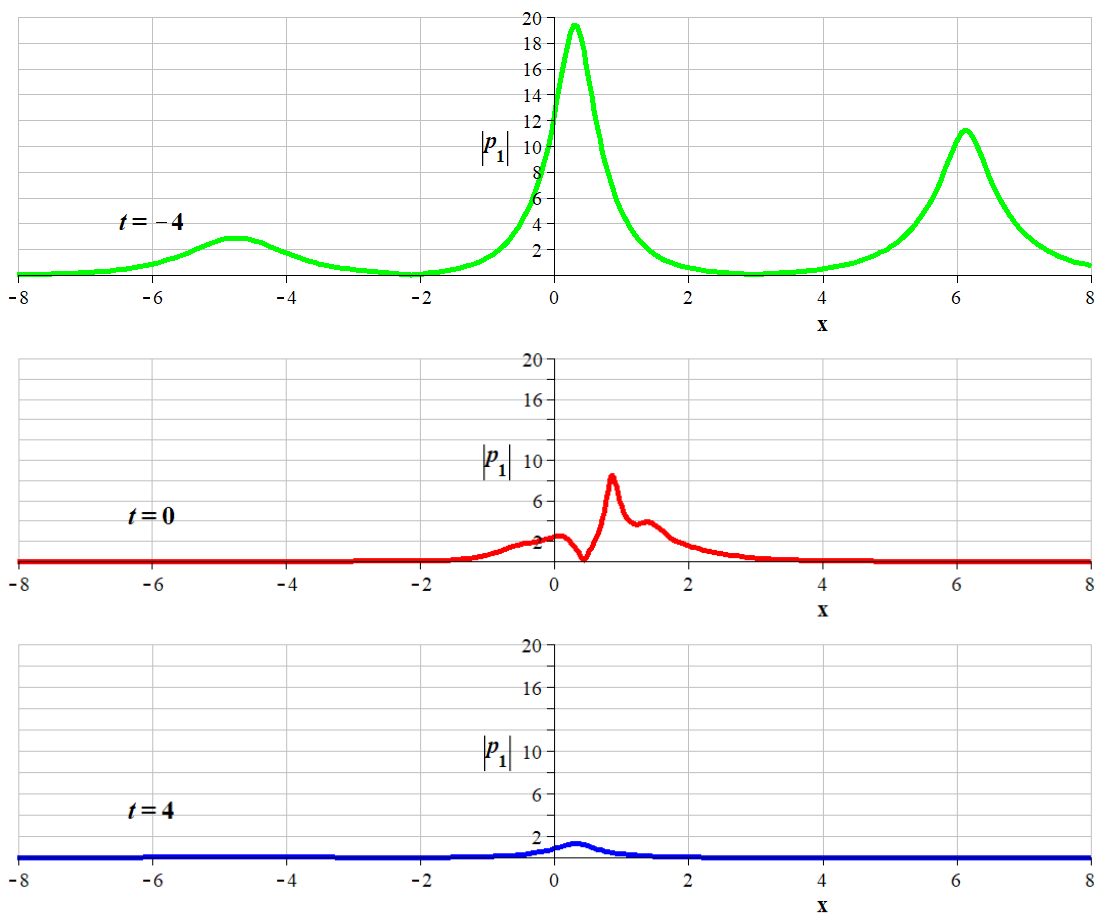}%
\includegraphics[width=0.45\textwidth,height=4.5cm]{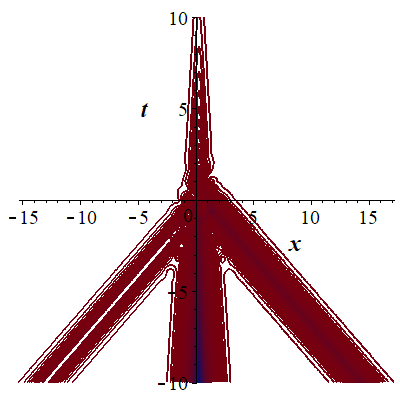}%
\caption{Spectral plane along with 3D, 2D and contour plots of $|p_{1}|$ of the three-soliton with parameters $(\rho_{1},\rho_{2},\rho_{3},\alpha_{1},\alpha_{2},\beta_{1},\beta_{2})=(1,1,1,-2,1,-2,1)$, $(\lambda_{1},\lambda_{2},\lambda_{3})=(1+0.5i,-1+0.5i,-0.05+0.75i)$, $w_{1}=(1,0,2+i,1-i)$, $w_{2}=(-1,5-2i,-i,0)$, $w_{3}=(2+i,1+2i,1,2i)$.}%
\label{3solitonplot13}%
\end{center}
\end{figure}

\begin{figure}[H]
\begin{center}
\includegraphics[width=0.45\textwidth,height=4.5cm]{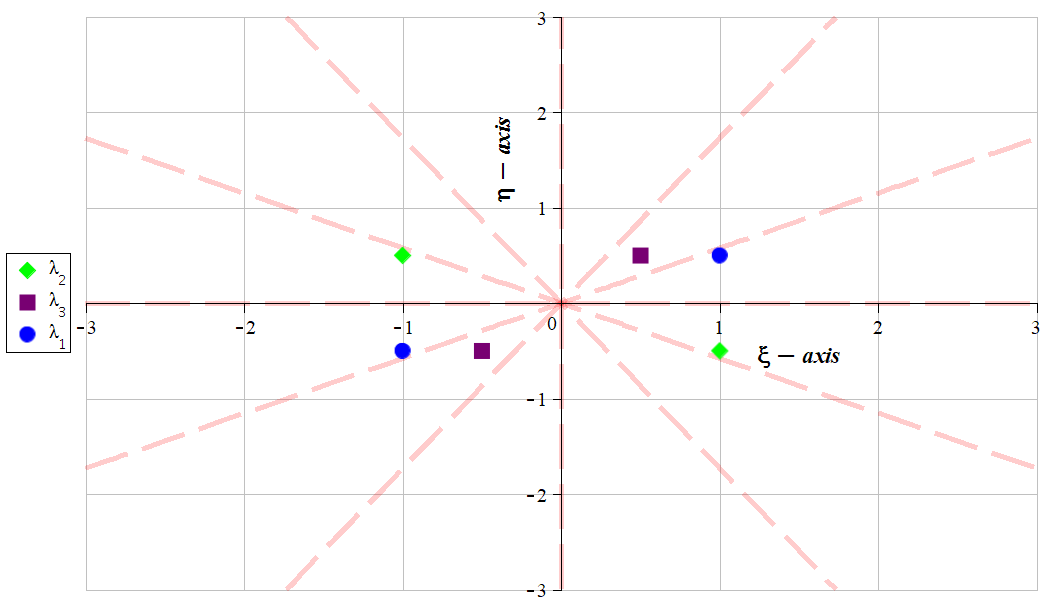}%
\includegraphics[width=0.45\textwidth,height=4.5cm]{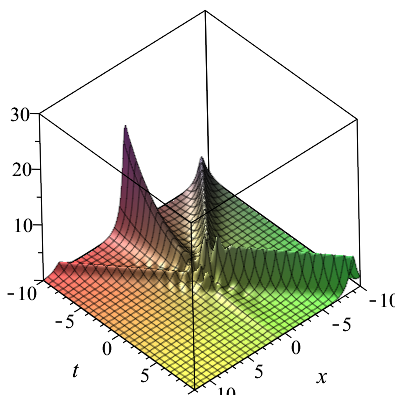}%
\\
\includegraphics[width=0.45\textwidth,height=4.5cm]{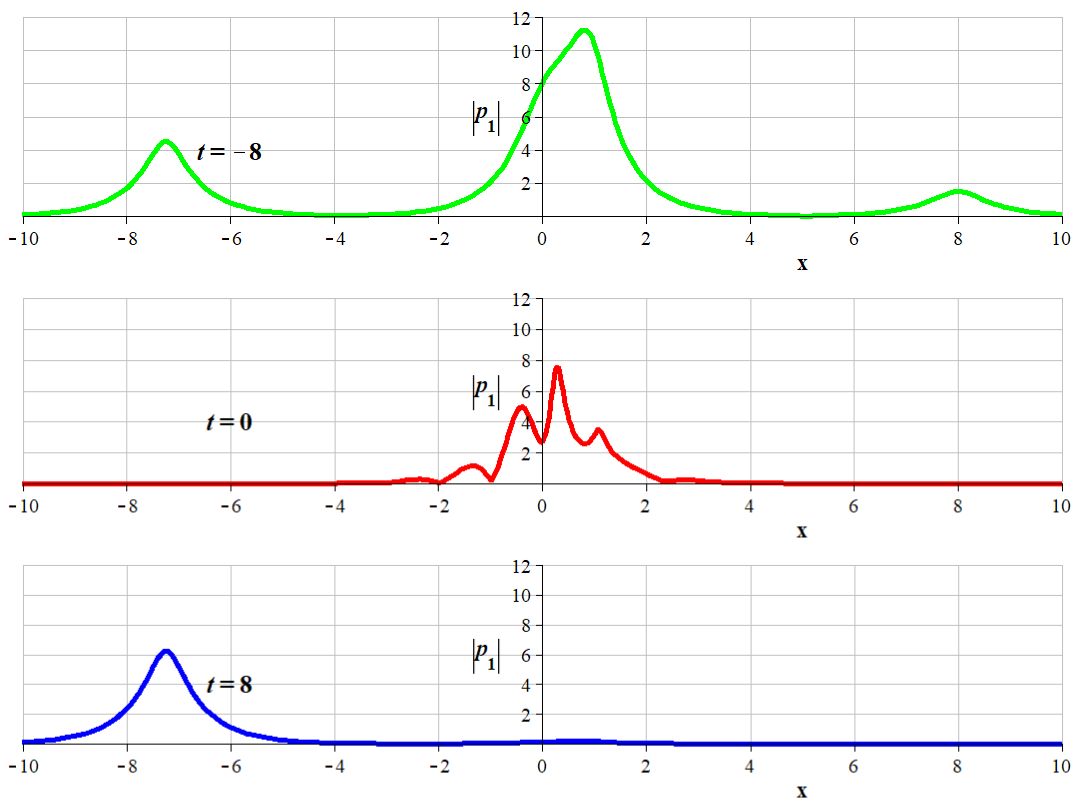}%
\includegraphics[width=0.45\textwidth,height=4.5cm]{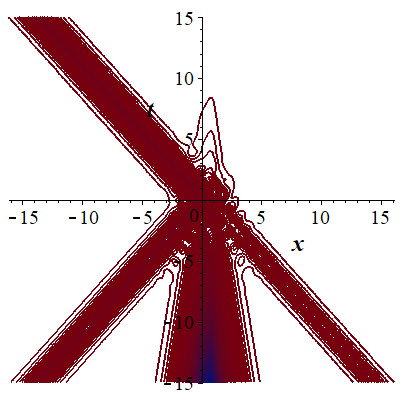}%
\caption{Spectral plane along with 3D, 2D and contour plots of $|p_{1}|$ in the focussing case of the three-soliton with parameters $(\rho_{1},\rho_{2},\rho_{3},\alpha_{1},\alpha_{2},\beta_{1},\beta_{2})=(-1,-1,-1,-2,1,-1,1)$, $(\lambda_{1},\lambda_{2},\lambda_{3})=(1+0.5i,-1+0.5i,0.5+0.5i)$, $w_{1}=(1,0,2+i,1-i)$, $w_{2}=(-1,1-2i,-i,0)$, $w_{3}=(2+i,1+2i,1,2i)$.}%
\label{3solitonplot6}%
\end{center}
\end{figure}

The three-soliton solution is given, for which $N=3$,
$w_{1}=(w_{11},w_{12},w_{13},w_{14})^{T}$,
$w_{2}=(w_{21},w_{22},w_{23},w_{24})^{T}$,
$w_{3}=(w_{31},w_{32},w_{33},w_{34})^{T}$,
$(\lambda_{1},\lambda_{2},\lambda_{3}) \in \mathbb{C}^{3}$, and
$\hat{\lambda}_{1}=-\lambda_{1}$, $\hat{\lambda}_{2}=-\lambda_{2}$, $\hat{\lambda}_{3}=-\lambda_{3}$ by
\begin{align}
p_{1} = \alpha \sum\limits_{k,j=1}^{3} v_{k1} (M^{-1})_{kj},
\hat{v}_{j,2}
\\
p_{2} = \alpha \sum\limits_{k,j=1}^{3} v_{k1} (M^{-1})_{kj},
\hat{v}_{j,3}
\\
p_{3} = \alpha \sum\limits_{k,j=1}^{3} v_{k1} (M^{-1})_{kj}.
\hat{v}_{j,4}
\end{align}
Without loss of generality, for the three-soliton,
we take all three eigenvalues in the upper-half plane in such a way that 
$\lambda_{i} \neq \lambda_{j}$ for $i,j \in \{1,2,3\}$.
\\[3mm]
Here, we can look at some of the three-soliton solution dynamics. We have
two solitons moving in opposite directions interacting with one stationary soliton. After the interaction 
either the three solitons keep their amplitudes or the amplitudes
change to new constant amplitudes, an example is shown in figure 
\ref{3solitonplot7}.
\\[3mm]
Another behaviour could be the interaction of three solitons
that are embedded into two solitons after the interaction, where
the stationary soliton keeps or changes it amplitude after collision. We may have the opposite case where the 
two solitons unfold to three solitons.
\\[3mm]
A different class of behaviour shows that three solitons can interact
and embedded in a single soliton after the interaction.
\\[2mm]
In figure \ref{3solitonplot12}, we have three solitons, two of which
are moving in opposite directions and one is stationary, all of them with constant amplitudes before interaction. After the interaction, they are embedded into a one stationary soliton with constant amplitude. 
\\[2mm]
We may have the opposite case where one stationary soliton 
unfolds to three different solitons, each keeping its amplitude.
\\[3mm]
In figure \ref{3solitonplot13}, we also have three-soliton,
two-soliton moving in opposite directions interacting with
an exponentially decreasing stationary soliton. In that case, 
after the interaction, they are embedded into one stationary decreasing soliton over the time due to effect of energy radiation.
\\[2mm]
In contrast, we can have one stationary increasing soliton that unfold to three-soltion, in which two of them are moving in opposite directions keeping their amplitude while the other one is stationary and increasing exponentially over the time.
\\[3mm]
In figure \ref{3solitonplot6}, we have three-soliton,
two-soliton moving in opposite directions interacting with
an exponentially decreasing stationary soliton. 
They are embedded into a one moving soliton that keeps its constant amplitude after the interaction where the stationary soliton vanish.
\\[2mm]
In contrary, one moving soliton can also unfold to three different solitons, where two are moving in opposite direct keeping the amplitude and the other one is increasing exponentially over the time \cite{Clarke2000}.
\subsection{The dynamice of the four-soliton solution}
\begin{figure}[H]
\begin{center}
\includegraphics[width=0.45\textwidth,height=4.5cm]{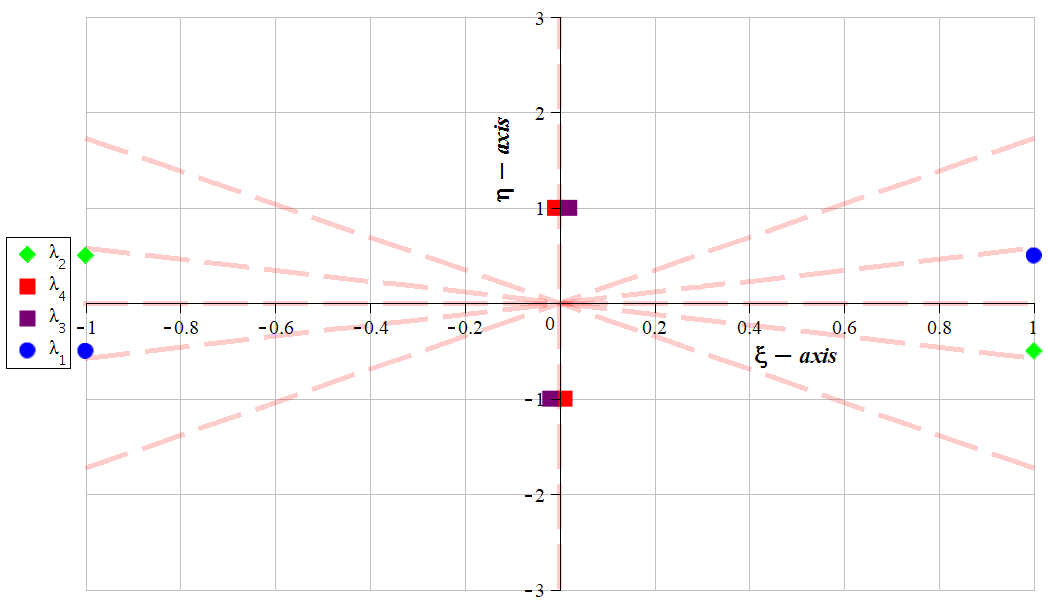}%
\includegraphics[width=0.45\textwidth,height=4.5cm]{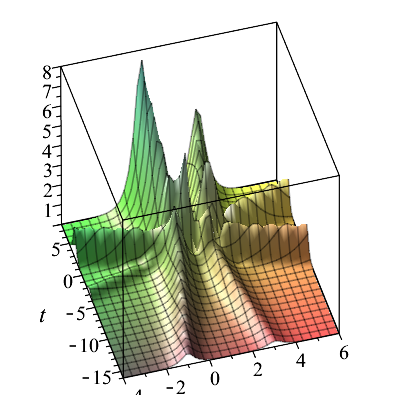}%
\\
\includegraphics[width=0.45\textwidth,height=4.5cm]{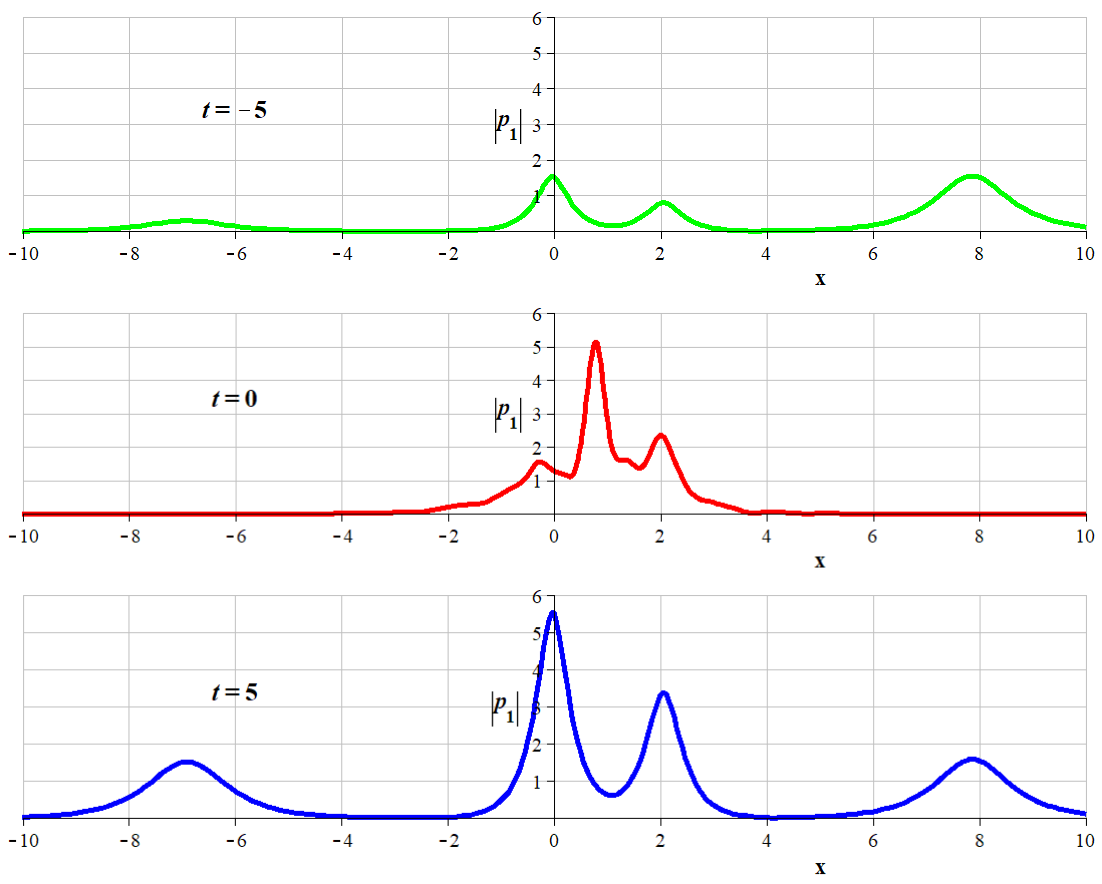}%
\includegraphics[width=0.45\textwidth,height=4.5cm]{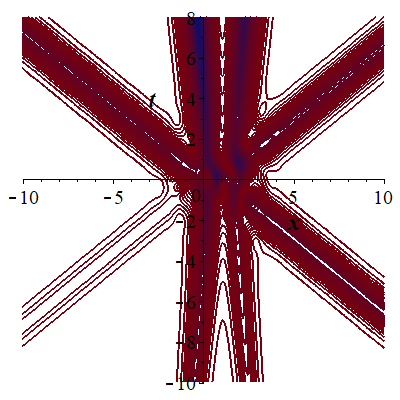}%
\caption{Spectral plane along with 3D, 2D and contour plots of $|p_{1}|$ of the four-soliton with parameters $(\rho_{1},\rho_{2},\rho_{3},\alpha_{1},\alpha_{2},\beta_{1},\beta_{2})=(1,-2,1,-2,1,-2,1)$, $(\lambda_{1},\lambda_{2},\lambda_{3},\lambda_{4})=(1+0.5i,-1+0.5i,0.02+i,-0.01+i)$, $w_{1}=(1-2i,1+3i,-i,1+i)$, $w_{2}=(-1+2i,1-3i,i,1-i)$, $w_{3}=(1+i,1+2i,0,2i)$, $w_{4}=(1,i,2+i,1)$.}%
\label{4solitonplot21}%
\end{center}
\begin{center}
\includegraphics[width=0.45\textwidth,height=4.5cm]{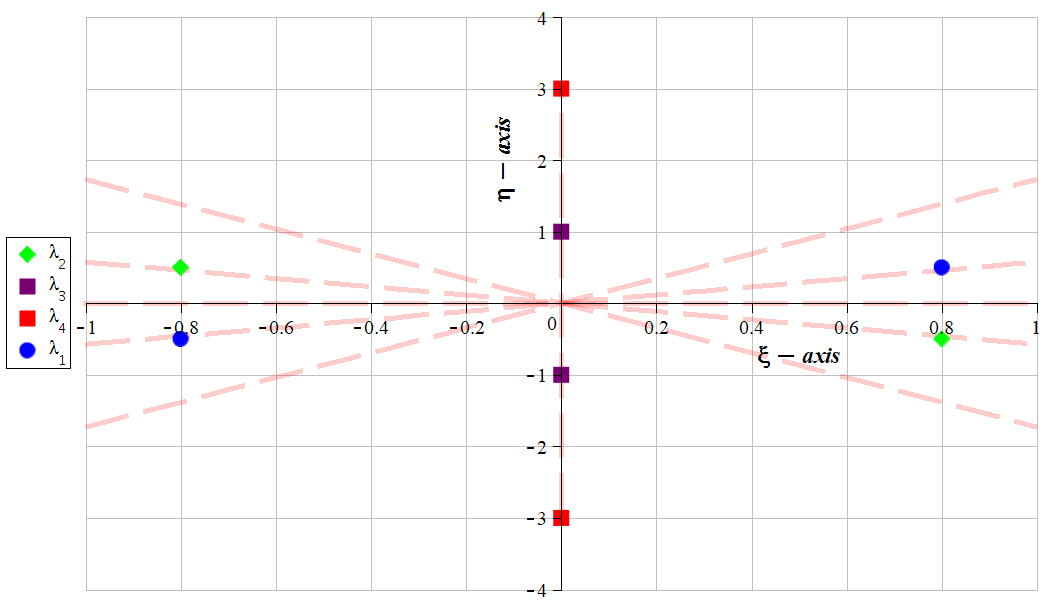}%
\includegraphics[width=0.45\textwidth,height=4.5cm]{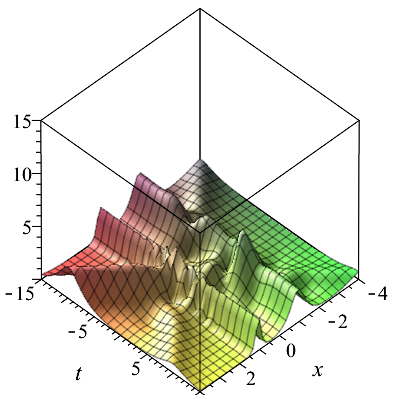}%
\\
\includegraphics[width=0.45\textwidth,height=4.5cm]{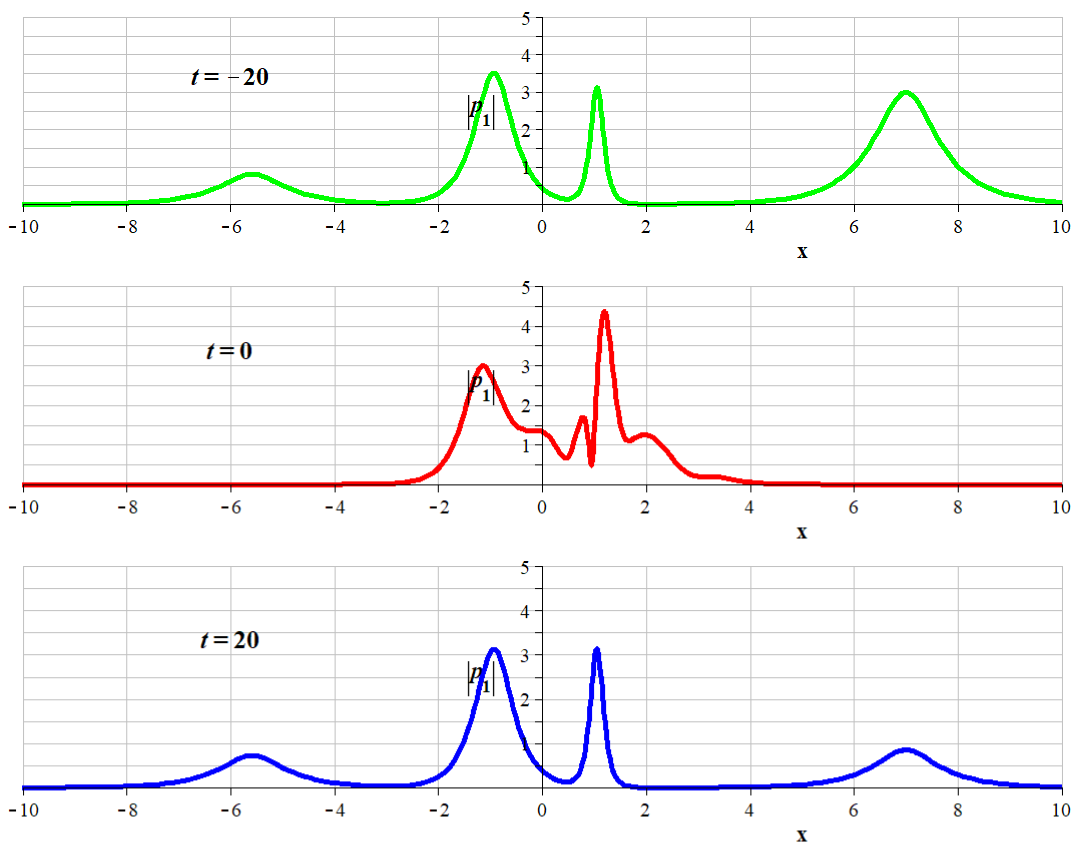}%
\includegraphics[width=0.45\textwidth,height=4.5cm]{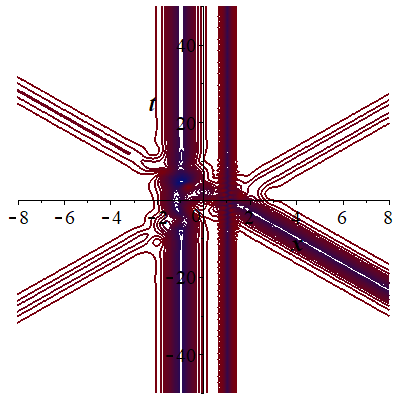}%
\caption{Spectral plane along with 3D, 2D and contour plots of $|p_{1}|$ of the four-soliton with parameters $(\rho_{1},\rho_{2},\rho_{3},\alpha_{1},\alpha_{2},\beta_{1},\beta_{2})=(1,1,1,-2,1,-2,1)$, $(\lambda_{1},\lambda_{2},\lambda_{3},\lambda_{4})=(0.8+0.5i,-0.8+0.5i,i,3i)$, $w_{1}=(1-0.5i,1+3i,-i,1+i)$, $w_{2}=(-1+2i,1-1.5i,i,1-i)$, $w_{3}=(30,i,2+i,1)$, $w_{4}=(-0.0005,1,2,1)$.}%
\label{4solitonplot23}%
\end{center}
\end{figure}

\begin{figure}[H]
\begin{center}
\includegraphics[width=0.45\textwidth,height=4.5cm]{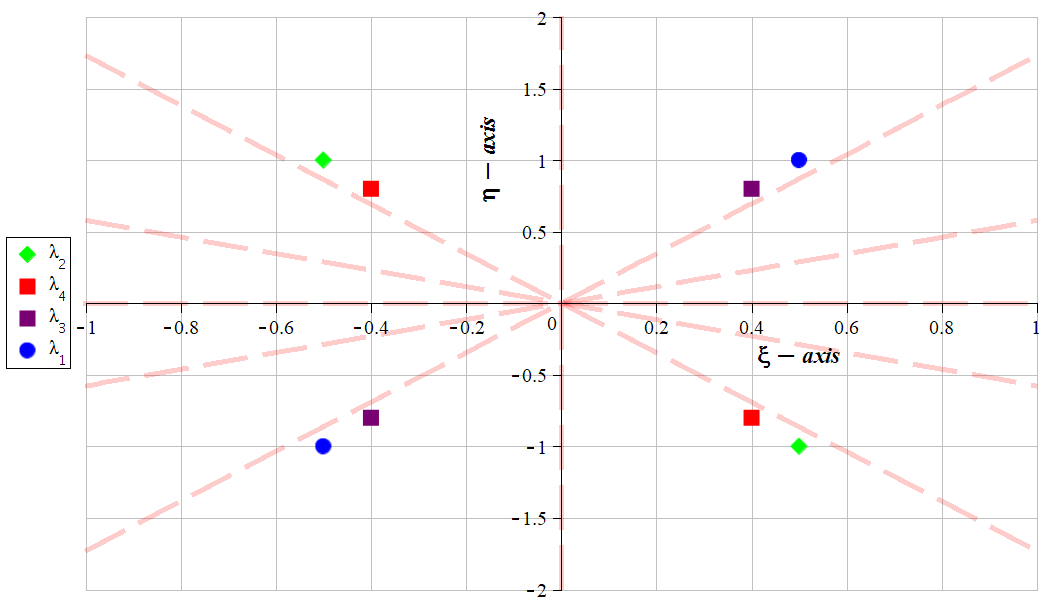}%
\includegraphics[width=0.45\textwidth,height=4.5cm]{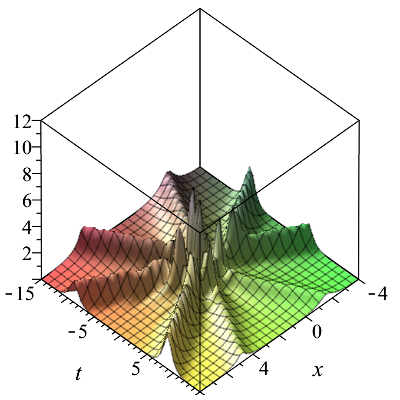}%
\\
\includegraphics[width=0.45\textwidth,height=4.5cm]{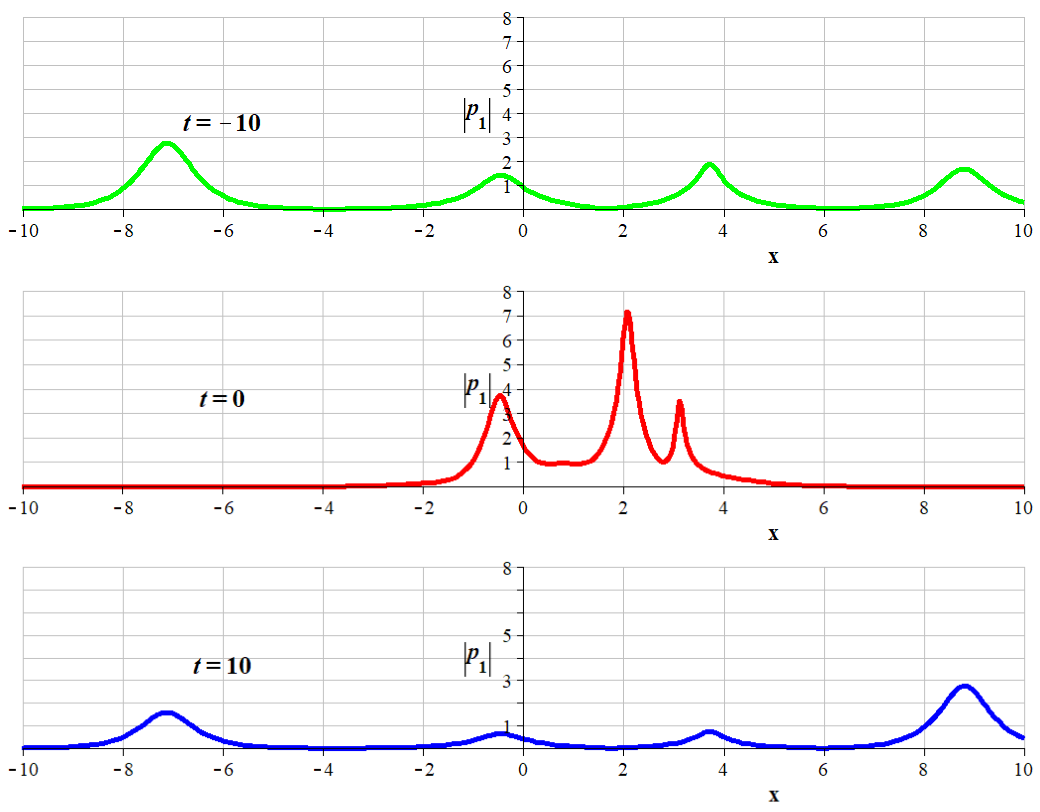}%
\includegraphics[width=0.45\textwidth,height=4.5cm]{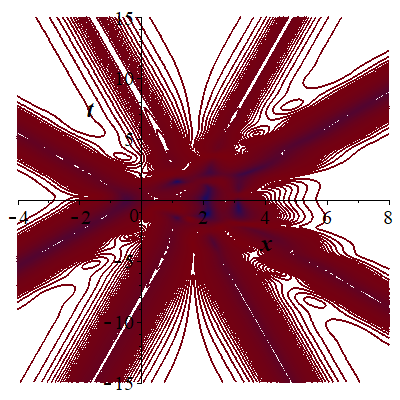}%
\caption{Spectral plane along with 3D, 2D and contour plots of $|p_{1}|$ in the focussing case of the four-soliton with parameters $(\rho_{1},\rho_{2},\rho_{3},\alpha_{1},\alpha_{2},\beta_{1},\beta_{2})=(-1,-1,-1,-1,1,-1,1)$, $(\lambda_{1},\lambda_{2},\lambda_{3},\lambda_{4})=(0.5+i,-0.5+i,0.4+0.8i,-0.4+0.8i,)$, $w_{1}=(-1.5+2i,2-3i,i,1-i)$, $w_{2}=(3+2i,-1+3i,-i,1+i)$, $w_{3}=(i,1,1-2i,1-i)$, $w_{4}=(-i,1-2i,1,1+i)$.}%
\label{4solitonplot30}%
\end{center}
\begin{center}
\includegraphics[width=0.45\textwidth,height=4.5cm]{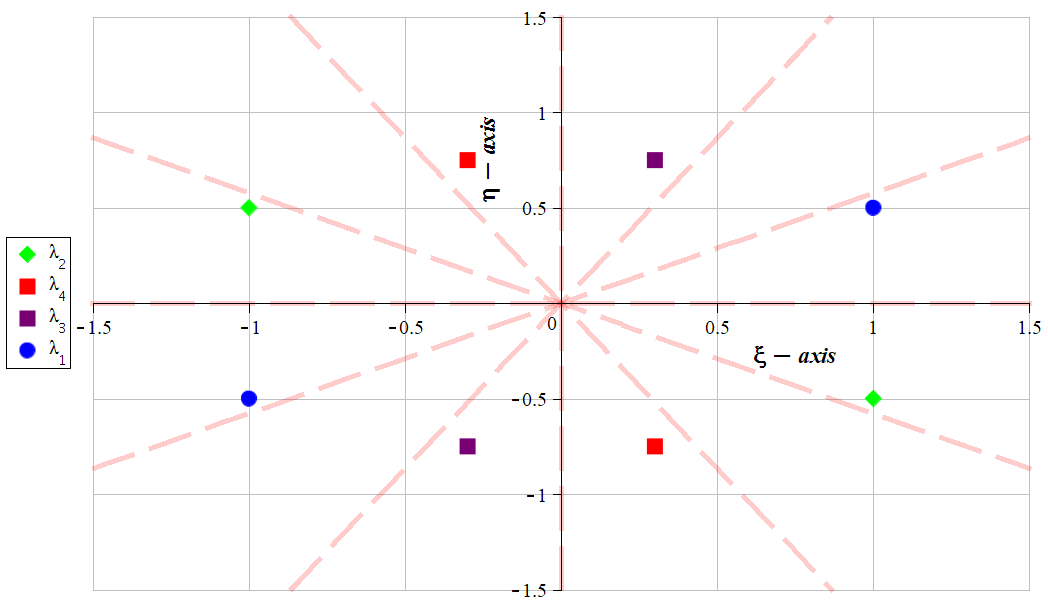}%
\includegraphics[width=0.45\textwidth,height=4.5cm]{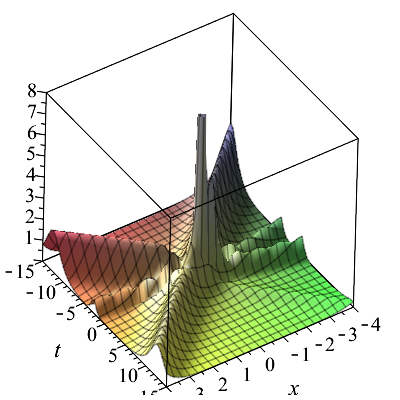}%
\\
\includegraphics[width=0.45\textwidth,height=4.5cm]{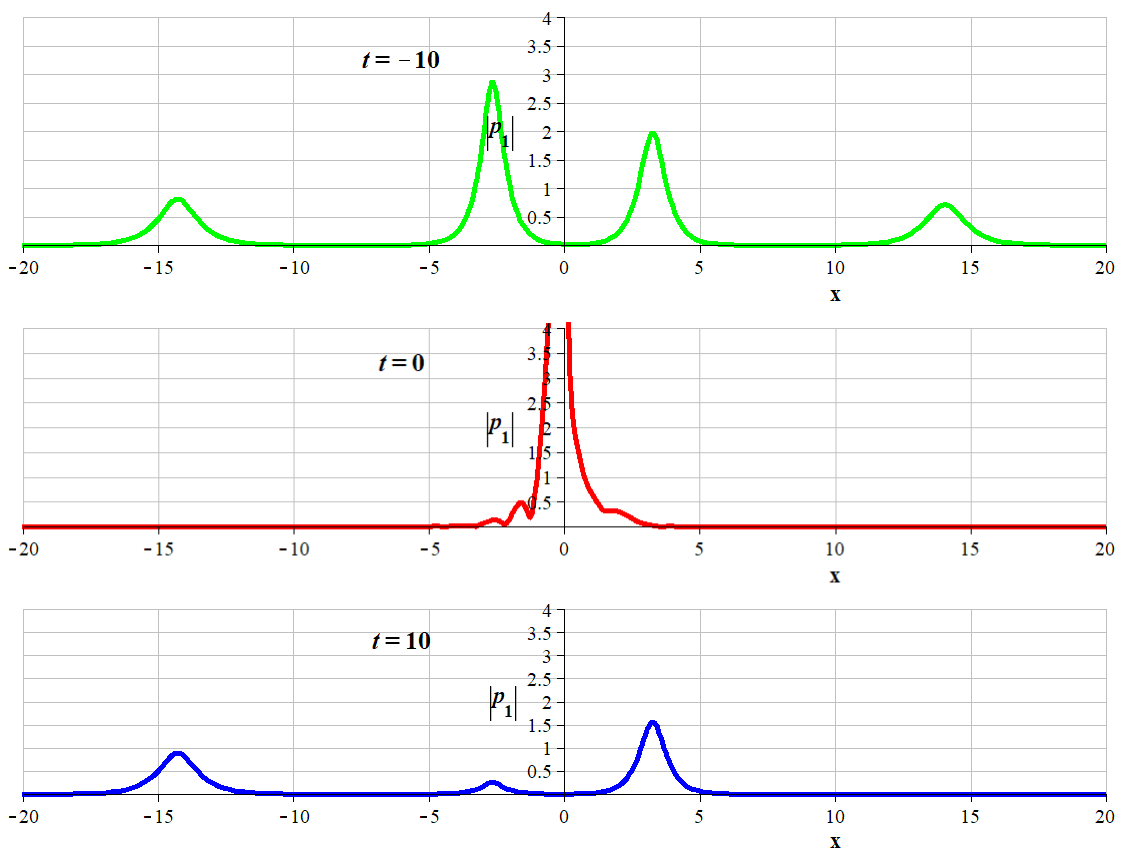}%
\includegraphics[width=0.45\textwidth,height=4.5cm]{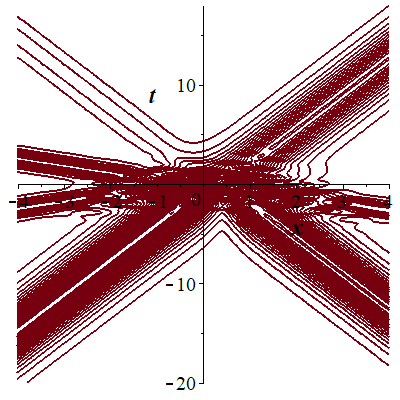}%
\caption{Spectral plane along with 3D, 2D and contour plots of $|p_{1}|$ in the focussing case of the four-soliton with parameters $(\rho_{1},\rho_{2},\rho_{3},\alpha_{1},\alpha_{2},\beta_{1},\beta_{2})=(-1,-1,-1,-2,1,-2,1)$, $(\lambda_{1},\lambda_{2},\lambda_{3},\lambda_{4})=(1+0.5i,-1+0.5i,0.3+0.75i,-0.3+0.75i)$, $w_{1}=(1+4i,0,2+i,i)$, $w_{2}=(-1+4i,i,1-2i,0)$, $w_{3}=(-2+i,0.5+i,-2+i,1)$, $w_{4}=(2+i,1-0.5i,1,1)$.}%
\label{4solitonplot24}%
\end{center}
\end{figure}

The four-soliton solution is given, for which $N=4$,
$w_{1}=(w_{11},w_{12},w_{13},w_{14})^{T}$,
$w_{2}=(w_{21},w_{22},w_{23},w_{24})^{T}$,
$w_{3}=(w_{31},w_{32},w_{33},w_{34})^{T}$,
$w_{4}=(w_{41},w_{42},w_{43},w_{44})^{T}$,
$(\lambda_{1},\lambda_{2},\lambda_{3},\lambda_{4}) \in \mathbb{C}^{4}$, and
$\hat{\lambda}_{1}=-\lambda_{1}$, $\hat{\lambda}_{2}=-\lambda_{2}$, $\hat{\lambda}_{3}=-\lambda_{3}$, 
$\hat{\lambda}_{4}=-\lambda_{4}$ by
\begin{align}
p_{1} = \alpha \sum\limits_{k,j=1}^{4} v_{k1} (M^{-1})_{kj}
\hat{v}_{j,2},
\\
p_{2} = \alpha \sum\limits_{k,j=1}^{4} v_{k1} (M^{-1})_{kj}
\hat{v}_{j,3},
\\
p_{3} = \alpha \sum\limits_{k,j=1}^{4} v_{k1} (M^{-1})_{kj}
\hat{v}_{j,4}.
\end{align}
Without loss of generality, for the four-soliton,
we take all four eigenvalues in the upper-half plane in such a way that 
$\lambda_{i} \neq \lambda_{j}$ for $i,j \in \{1,2,3,4\}$.
\\[3mm]
For the four-soliton dynamics, we have the interactions of four solitons. Two of them can be stationary or all the four solitons are moving.
\\[2mm]
Figure \ref{4solitonplot21} exhibits the interaction of two
exponentially increasing solitons moving in opposite directions
and interacting with two moving solitons with constant amplitude. After interaction the middle two solitons keep moving with increasing amplitude, while the two other solitons keep moving with constant amplitude.
\\[2mm]
We can notice that the middle two solitons can decrease exponentially while moving and interacting with the other two solitons.
\begin{remark}
The speed of the far right and left solitons are larger than the
speed of the middle two solitons, such that all four solitons collide together.
\end{remark}
Another behaviour is shown in figure \ref{4solitonplot23},
where two solitons moving in opposite directions interact with 
two stationary solitons with constant amplitudes. After the interaction,
the two stationary solitons remain stationary and the two moving solitons
continue to move in opposite directions, but their amplitudes can change to new constant amplitudes or it can stay unchanged. 
\\[3mm]
As for figure \ref{4solitonplot30}, we have the interaction of four moving solitons. Two waves are moving in the same direction and interacting with the other two waves coming from the opposite direction.
After the interaction, each of the four solitons can keep its amplitude unchanged or its amplitude can change to a new constant amplitude.
In the case that each soliton keeps its amplitude before and after the interaction, we have four travelling waves.
\\[2mm]
Finally in figure \ref{4solitonplot24}, four moving solitons are embedded
into three moving solitons. After the interaction each soliton keep its amplitude unchanged or it can be changed to a new constant amplitude
over the time.
\newpage
\newpage
\newpage
\section{Conclusion}
To summarize for fundamental soliton solutions, solitons interact elastically in a superposition manner, while for nonlocal solitons it is not always the case. Also, for the nonlocal soliton solutions, we can have singularities at a finite time. 
\newline
In reverse-time, the pairs of symmetric eigenvalues $(\lambda,-\lambda)$ make the Riemann-Hilbert problem simpler to solve \cite{Yang2019},
where $\lambda \in \mathbb{C}_{+}$ and $-\lambda \in \mathbb{C}_{-}$.
\\[2mm]
In addition, an interesting observation in this paper, is that 
we can explicit the one and two soliton solutions for any $n$-order ($n$ is even)
six-component AKNS integrable system for our spectral matrix 
$U(u,\lambda)$. 
\newline 
That is the general explicit one soliton solution for reverse-time $n$-order six-component system when
$\hat{\lambda}_{1}=-\lambda_{1}$ is given by
\begin{align}
p_{1}(x,t)= \frac{
2 \rho_{2} \rho_{3} \lambda_{1} (\alpha_{1}-\alpha_{2}) w_{11} w_{12}
e^{i \lambda_{1} (\alpha_{1}+\alpha_{2})x+i\lambda_{1}^{n}(\beta_{1}-\beta_{2})t}}{\rho_{1} \rho_{2} \rho_{3} w_{11}^{2} e^{2i\lambda_{1} \alpha_{1}x}
+(\rho_{2} \rho_{3} w_{12}^{2}+ \rho_{1} \rho_{3} w_{13}^{2}+ \rho_{1} \rho_{2} w_{14}^{2}) e^{2i\lambda_{1}\alpha_{2}x}},
\end{align}
similarly for $p_{2}(x,t)$ and $p_{3}(x,t)$.
\\[2mm]
As for the two-soliton, the general explicit solution 
when 
$\hat{\lambda}_{1}=-\lambda_{1}$, $\hat{\lambda}_{2}=-\lambda_{2}$,
if $\lambda_{1} \neq - \lambda_{2}$, is given by
\begin{align}
p_{1}(x,t) = 2 \rho_{2} \rho_{3} (\lambda_{1}+\lambda_{2}) (\alpha_{1}-\alpha_{2}) \frac{A(x,t)}{B(x,t)},
\end{align}
where
\begin{flalign}
&\begin{aligned}
A(x,t) = 
e^{i[\lambda_{2}^{n} (\beta_{1}-\beta_{2})  t 
+\lambda_{2} (\alpha_{1}+\alpha_{2}) x]} \cdot
& \bigg[ \bigg( w_{22} M (\lambda_{1}+\lambda_{2}) 
- 2 w_{12} K \lambda_{1} \bigg)
w_{21} \lambda_{2} e^{i 2\alpha_{2} \lambda_{1} x}
\\[-2mm]
& - \rho_{1} \rho_{2} \rho_{3} (\lambda_{1}-\lambda_{2}) w_{11}^{2} w_{21} w_{22} \lambda_{2}
e^{i 2 \alpha_{1} \lambda_{1} x} \bigg]
\\[-2mm]
+ e^{i[\lambda_{1}^{n} (\beta_{1}-\beta_{2})  t 
+\lambda_{1} (\alpha_{1}+\alpha_{2}) x]} \cdot
& \bigg[ \bigg( w_{12} N (\lambda_{1}+\lambda_{2}) 
- 2 w_{22} K \lambda_{2} \bigg)
w_{11} \lambda_{1} e^{i 2\alpha_{2} \lambda_{2} x}
\\[-2mm]
& + \rho_{1} \rho_{2} \rho_{3} (\lambda_{1}-\lambda_{2}) w_{11} w_{12} w_{21}^{2} \lambda_{1}
e^{i 2 \alpha_{1} \lambda_{2} x} \bigg],
\end{aligned}&
\end{flalign}
\newpage
\begin{flalign}
&\begin{aligned}
B(x,t) &= -4 \rho_{1} \rho_{2} \rho_{3} \lambda_{1} \lambda_{2} w_{11} w_{21} K
e^{i (\lambda_{1}+\lambda_{2})(\alpha_{1}+\alpha_{2}) x}
\cdot
\bigg[ 
e^{i (\lambda_{1}^{n}-\lambda_{2}^{n})(\beta_{1}-\beta_{2}) t}
+ e^{-i (\lambda_{1}^{n}-\lambda_{2}^{n})(\beta_{1}-\beta_{2}) t}
\bigg]
\\
&+ \rho_{1} \rho_{2} \rho_{3} w_{21}^{2} M (\lambda_{1}+\lambda_{2})^{2}
e^{i2 (\alpha_{1} \lambda_{2}+\alpha_{2} \lambda_{1}) x}
+ \rho_{1} \rho_{2} \rho_{3} w_{11}^{2} N (\lambda_{1}+\lambda_{2})^{2}
e^{i2 (\alpha_{1} \lambda_{1}+\alpha_{2} \lambda_{2}) x}
\\
&+
\rho_{1}^2 \rho_{2}^2 \rho_{3}^2 w_{11}^{2} w_{21}^{2} (\lambda_{1}-\lambda_{2})^{2}
e^{i2 \alpha_{1} (\lambda_{1}+\lambda_{2}) x}
+ \bigg[ (\lambda_{1}^{2}+\lambda_{2}^{2}) MN 
+ (2MN-4K^{2}) \lambda_{1} \lambda_{2}\bigg] 
e^{i2 \alpha_{2} (\lambda_{1}+\lambda_{2}) x},
\end{aligned}&
\end{flalign}
and
$M=\rho_{2} \rho_{3} w_{12}^{2}+ \rho_{1} \rho_{3} w_{13}^{2}+ \rho_{1} \rho_{2} w_{14}^{2}$, 
$N=\rho_{2} \rho_{3} w_{22}^{2}+ \rho_{1} \rho_{3} w_{23}^{2}+ \rho_{1} \rho_{2} w_{24}^{2}$ and
$K=\rho_{2} \rho_{3} w_{12} w_{22} + \rho_{1} \rho_{3} w_{13} w_{23} + \rho_{1} \rho_{2} w_{14} w_{24}$. Similarly for $p_{2}(x,t)$ and $p_{3}(x,t)$.
\\[2mm]
For higher order non-local reverse-time AKNS systems, 
the $n$-soliton exhibit similar dynamics (or combinations of dynamics) to the dynamics discussed in this paper and in previous work \cite{AlleAhmedMa}.
\newline
Solving integrable equations in reverse-space, reverse-time and reverse-spacetime using other techniques such as Darboux transformations and Hirota bilinear method is still an active 
investigation \cite{MatveevSalle1991}-\cite{SunMaYu}.


\end{document}